\documentclass[twocolumn]{aastex6}

\usepackage{color}

\newcommand{\ikome}{\textcolor{black}}

\newcommand{\kawashima}{\textcolor{black}}
\newcommand{\kawashimabf}{\textcolor{black}}
\newcommand{\kawashimar}{\textcolor{black}}
\newcommand{\ikomar}{\textcolor{black}}
\newcommand{\kawashimas}[1]{{\textcolor{black}{#1}}}

\newcommand{\ikomas}{\textcolor{black}}

\fullcollaborationName{The Friends of AASTeX Collaboration}

\newcommand{\pbar}{\mathrm{bar}}

\newcommand{\kelvin}{\mathrm{K}}

\newcommand{\mearth}{M_{\mathrm{\oplus}}}
\newcommand{\rearth}{R_{\mathrm{\oplus}}}

\newcommand{\rsun}{R_{\mathrm{\odot}}}

\newcommand{\rstar}{R_{\mathrm{s}}}
\newcommand{\mplanet}{M_{\mathrm{p}}}

\newcommand{\rtransit}{R_{\mathrm{tr}}}

\newcommand{\hydrogen}{\mathrm{H}_2}
\newcommand{\helium}{\mathrm{He}}
\newcommand{\water}{\mathrm{H}_2 \mathrm{O}}
\newcommand{\methane}{\mathrm{C}\mathrm{H}_4}
\newcommand{\carbonmonoxide}{\mathrm{C}\mathrm{O}}
\newcommand{\carbondioxide}{\mathrm{C}\mathrm{O}_2}
\newcommand{\ammonia}{\mathrm{N}\mathrm{H}_3}
\newcommand{\nitrogen}{\mathrm{N}_2}

\usepackage{amsmath}
\usepackage{longtable}

\begin{document}


\title{Theoretical transmission spectra of exoplanet atmospheres with hydrocarbon haze: Effect of creation, growth, and settling of haze particles. I. \\ Model description and first results}


\author{Yui Kawashima\altaffilmark{1, 2} and Masahiro Ikoma\altaffilmark{1, 3}}
\altaffiltext{1}{Department of Earth and Planetary Science, Graduate School of Science, The University of Tokyo, 7-3-1 Hongo, Bunkyo-ku, Tokyo 113-0033, Japan}
\altaffiltext{2}{ykawashima@eps.s.u-tokyo.ac.jp}
\altaffiltext{3}{Research Center for the Early Universe (RESCEU), Graduate School of Science, The University of Tokyo, 7-3-1 Hongo, Bunkyo-ku, Tokyo 113-0033, Japan}




\and




\begin{abstract}
Recently, 
properties of exoplanet atmospheres have been constrained via multi-wavelength transit observation, which measures an apparent decrease in stellar brightness during planetary transit in front of its host star (called transit depth). 
Sets of transit depths so far measured at different wavelengths (called transmission spectra) are somewhat diverse: Some show
steep \kawashima{spectral} slope features in the visible, 
some contain
featureless spectra in the near-infrared, 
some show distinct features from radiative absorption by gaseous species. These facts infer
the existence of haze in the atmospheres 
especially of warm, relatively low-density super-Earths and mini-Neptunes.
Previous studies that addressed theoretical modeling of transmission spectra of hydrogen-dominated atmospheres with 
haze used some assumed distribution and size of haze particles.
In this study, we \kawashima{model the atmospheric chemistry,} derive 
the spatial and size distributions of haze particles 
by simulating the creation, growth and settling of hydrocarbon haze particles directly, 
and develop 
transmission spectrum models 
of UV-irradiated, solar-abundance 
atmospheres of close-in warm ($\sim$~500~K) exoplanets. 
We find that the haze 
is distributed in the atmosphere much more broadly than previously 
assumed and consists of particles of various sizes. 
We also 
demonstrate that the observed diversity of transmission spectra can be explained by the difference in the production rate of haze monomers, which is related to the UV irradiation intensity from host stars.

\end{abstract}




\section{Introduction} \label{sec:intro}
\ikome{Composition of exoplanet atmospheres is} often measured by transit observation \citep[e.g.,][]{2010ARA&A..48..631S}.
\ikome{When transiting in front of its host star, a planet blocks} a fraction of the incident \ikome{stellar} light. 
\ikome{
The amount of blocked light relative to the original stellar light is called the transit depth. 
} 
Since a set of transit depths observed at different wavelengths (called transmission spectrum) depends on atmospheric constituents,
one can infer the atmospheric composition via multi-wavelength transit observations.

Recently, thanks to advance in observational techniques, 
\ikome{atmospheric} characterization \ikome{for} relatively small planets has become possible via transit observations. 
Typical examples are 
\ikome{GJ~1214b of mass 6.26~$\mearth$ and radius 2.80~$\rearth$ \citep{2009Natur.462..891C, 2013AA...551A..48A},}
GJ~3470b of 13.73~$\mearth$ and 3.88~$\rearth$ \citep{2014MNRAS.443.1810B}, 
\ikome{and} GJ~436b of 25.4~$\mearth$ and 4.10~$\rearth$ \citep{2014A&A...572A..73L}. 
\ikome{Interestingly,} 
transmission spectra of those planets observed so far cannot be explained \ikome{only} by absorption and scattering (i.e., extinction) of gaseous molecules in the atmospheres.

GJ~1214b is a super-Earth whose atmosphere has been probed most. 
\ikomar{Recent} multi-wavelength transit observations show a relatively featureless or flat spectrum from the optical \citep[e.g.][]{2013ApJ...773..144N, Nascimbeni:2015kk} to near-infrared \citep[e.g.][]{2014Natur.505...69K}\ikomar{, although \citet{2012A&A...538A..46D} reported a tentative increase in the transit depth in the optical.} 
This raises the possibility that particles such as clouds and hazes are present in the atmosphere, 
because those particles obscure molecular absorption features. 
\kawashima{\ikomar{(In this study}, we \ikomar{refer to thermochemical condensates as ``}clouds" and \ikomar{photochemical products as ``}hazes".\ikomar{)}}
Also, its transmission spectrum in the near-infrared is too flat to be explained even by a CO$_2$-dominated atmosphere \citep{2014Natur.505...69K}. 
GJ~436b is also reported as showing a featureless spectrum in the near-infrared by \citet{2014Natur.505...66K}, 
suggesting the presence of a cloudy/hazy layer.

GJ~3470b is reported to show a bit more complicated spectrum, which includes a steep \kawashima{spectral} slope\footnote{\ikomar{The steep slope in the optical is sometimes referred to as  the Rayleigh scattering slope in the literature. However, one can never conclude that the slope is due to Rayleigh scattering from an observed spectral slope alone \citep[see][]{2016ApJ...826L..16H}.}}  \ikome{in the optical} \citep{2013ApJ...770...95F, 2013A&A...559A..32N, 2014MNRAS.443.1810B, 2015ApJ...814..102D, Awiphan:2016ke} and 
is relatively featureless or flat in the near-infrared \citep{2013A&A...559A..33C, 2014A&A...570A..89E}. 
\ikome{A modest amount of cloud/haze particles,}
if present, tend to steepen the \kawashima{spectral} slope in the \ikome{optical} \citep{2008A&A...481L..83L}, \ikome{while a thick cloud/haze} obscures molecular and atomic absorption features, flattening the spectrum. 
Though the number of samples is still small, cloud/haze may be commonly present and bring about a diversity of spectra \kawashima{\citep{Sing:2016hi}}. 
\kawashima{\cite{2016ApJ...817L..16S} and \cite{2016ApJ...826L..16H} explored this diversity by quantifying the degree of cloudiness in atmospheres of transiting exoplanets from their spectra.
Both two studies reported the trend that \ikomar{cooler} planets were more likely to have \ikomar{cloudy/hazy} atmospheres.}

\ikome{As a candidate for the cloud/haze, we consider hydrocarbon haze in this study for the reason below, while some other constituents are assumed in previous studies.}
\ikome{The above three} exoplanets \ikome{are close-in super-Earths/mini-Neptunes} orbiting M stars.
\ikome{Their} atmospheric temperatures are typically $\sim$~500 to 1000~K. 
Also, \ikome{those} close-in planets are exposed to intense UV radiation from their host stars.
In such warm, highly-UV irradiated environments, hydrocarbon hazes are formed easily through photochemical reactions triggered by photo-dissociation of methane, provided the atmospheres are reducing enough that $\mathrm{CH_4}$ rather than CO dominates the atmospheric carbon chemistry \cite[e.g.][]{Yung:1984cv}.
Note that \ikome{a great number of exoplanets in the similar environments will be detected in near future, because MK dwarfs are most abundant in the solar neighborhood \citep[e.g.,][]{2013AJ....146...99C}. }
\ikome{Also,} transit exoplanet surveys \ikome{so far} have detected many low-density low-mass exoplanets, which indicates that there are abundant low-mass exoplanets with relatively hydrogen-rich, reducing atmospheres \citep[][and references therein]{Fortney:2007hf}.

\ikome{Some} studies \ikome{so far} addressed theoretical modeling of transmission spectra of \ikome{hydrogen-rich} atmospheres \ikome{in such environments}, considering the effect of \ikome{haze in the atmosphere.}
\cite{2012ApJ...756..176H} is the first to quantify the effects of haze on transmission spectrum of GJ~1214b's atmosphere. 
They assumed \ikome{an atmospheric layer that contained haze particles such as tholin.} 
\ikome{The haze layer} is characterized by four parameters, 
\ikome{including single values of} the number density and size of the haze particles and the \ikome{pressures of the} upper and lower \ikome{edges of the haze layer}.
(They also considered the existence of clouds, below which transmitted light is cut off completely, regardless of wavelength.) 
Comparing their theoretical spectra with various haze/cloud properties and molecular compositions to \ikome{the} observed transmission spectrum of GJ~1214b, 
they \ikome{demonstrated} that a hydrogen-rich atmosphere with \ikome{the} haze layer could explain the observed transmission spectrum, \ikome{provided appropriate sets of the haze parameters were chosen}.

The same way for incorporating the effect of a haze layer was adopted by \cite{2014A&A...570A..89E}, 
who modeled transmission spectra of GJ~3470b's atmosphere 
and \ikome{then} compare\ikome{d} them with \ikome{the} observed \ikome{one}, 
including their observations done with \textit{Hubble} Space Telescope (HST).
They found no solution that reproduced the observed steep \kawashima{spectral} slope in the optical, simultaneously with \ikome{the observed flat} spectrum in the near-infrared. 
Instead, they concluded that both of the observed features in the visible and the near-infrared could be matched by a hydrogen-rich atmosphere covered with clouds, which they modeled in the same way as \cite{2012ApJ...756..176H}.

In contrast to the above theoretical modelings that \textit{assume} the altitude \ikome{and thickness} of the haze layer,
\cite{2013ApJ...775...33M} tried to determine \ikome{those properties} by doing photochemical calculations.
They derived numerically the vertical distributions of the photochemically-produced hydrocarbons\kawashima{,} $\mathrm{HCN}$, $\mathrm{C_2 H_2}$, $\mathrm{C_2H_4}$, and $\mathrm{C_2H_6}$, which are precursors of haze particles. 
Assuming that haze particles formed from a given fraction of the precursors, 
which they regarded as a parameter \ikome{(called the haze-forming efficiency)}, 
they determined the distribution of haze particles and \ikome{then} modeled the transmission spectrum of GJ~1214b's atmosphere \ikome{with assumed particle size and number density}.
(In their modeling, they used the opacity data of soot \ikome{instead of those of tholin}.)
They found that the observed transmission spectra of GJ~1214b could be explained by the haze particles with the size \ikome{of} 0.01 to 0.25~$\mathrm{\mu}$m and the haze-forming efficiency of 1-5\%, although there remained a possibility of clouds composed of KCl and ZnS.

The above studies certainly demonstrated that theoretical transmission spectra of hazy atmospheres matched the corresponding observations for appropriate choices of the haze parameters.
However, they did not \ikome{access} the viability of those haze properties sufficiently from a physical point of view.
In addition, transmission spectra so far observed seem to be diverse \citep{Sing:2016hi}: 
In the visible, some show distinct \kawashima{spectral} slope features, some may not. 
Also, some show molecular and atomic features, some are featureless. 
Again, although \ikome{the previous studies} found that choice of various haze parameters resulted in variation in transmission spectra, it remains to be clarified what yields such a variety of haze properties.

Of special interest in this study is the distribution of the size and number density of haze particles and \ikome{its} impacts on transmission spectra, which have \ikome{not} \ikome{been} investigated \ikome{previously}.
\ikome{Therefore} we \ikome{develop} transmission spectrum \ikome{models with detailed calculations of} the creation, growth, and settling of hydrocarbon haze particles, \ikome{assuming} hydrogen-dominated atmosphere\ikome{s} of close-in warm ($\lesssim$~1000 K) exoplanets. 
\ikome{In this first paper, we focus on describing the methodology and demonstrating the sensitivity of transmission spectra to} the production rate of haze monomers, which relates to the \ikome{amount} of UV irradiation from the host star.
\ikome{In our forthcoming papers, we make detailed investigation of the} dependence of \ikome{transmission spectra} on model parameters, other than monomer production rate, such as \ikome{atmospheric} metallicity, \kawashima{C/O ratio, }eddy diffusion coefficient, atmospheric temperature profile, and monomer size. 
\ikome{Also, we explore in detail the composition of the atmospheres of known warm exoplanets by comparing the observed spectra with our theoretical ones, taking into account other possibilities of cloud/haze constituents.}

The rest of this paper is organized as follows.
In $\S$~\ref{sec:model}, we describe the assumptions, equations, and calculation methods for the size and number density distributions of haze particles and generating the transmission spectra. 
In $\S$~\ref{sec:result}, we investigate the vertical distribution of haze particles and its effects on the transmission spectra.
Also, we investigate the dependence of the spectra on the production rate of haze monomers, which is related to UV irradiation intensity from the host star.
In $\S$~\ref{sec:single}, to gain a deeper understanding of the effect of the haze particle distribution on transmission spectrum, we calculate the particle growth and transmission spectra with a characteristic-size approximation and then compare the results with those obtained in $\S$~\ref{sec:result}.
Finally, we conclude this paper in $\S$~\ref{sec:summary}.

\section{Model and Method Description} \label{sec:model}

As described in Introduction, 
we develop transmission spectrum models of warm transiting planets with hydrogen-rich atmospheres by incorporating the effects of the size and number density distributions of hydrocarbon haze 
that are determined through the production, growth, and settling processes of the particles.
First, precursor molecules of haze particles (i.e., higher-order hydrocarbons, which we call haze precursors, hereafter) are created through photochemical reactions triggered by UV photodissociation of $\mathrm{CH_4}$. 
Then, aggregation of the haze precursors results in haze particles of small size, which are called monomers. Note that the size of monomers in the atmosphere of Titan was reported as $40 \pm 10$~nm from observations \citep{2009Icar..204..271T}.
Those monomers diffuse and settle downward. Also, collisional growth of the haze particles takes place. Once the haze particles go down into hot, convective regions, they are likely to be thermodynamically broken and evaporated to be $\mathrm{CH_4}$ again, which can be diffused upward to be the source of haze precursors.

In this study, we first perform photochemical calculations to derive the vertical distribution of haze precursors in a similar way to \citet{2013ApJ...775...33M} ($\S$~\ref{method_photo}). Then, using the obtained vertical profiles of the precursors, we calculate the growth and settling of haze particles in the atmosphere to derive the steady-state distributions of the size and number density of the haze particles ($\S$~\ref{method_growth}). Finally, we calculate the extinction opacities of the gases and particles ($\S$~\ref{method_opacity}) and model transmission spectra of the atmospheres with obtained properties of haze ($\S$~\ref{method_transmission}).

Before explaining the details of the above three modules, we first describe the assumptions and treatments made in all the modules.
We make an reasonable assumption that the atmosphere is in hydrostatic equilibrium and composed of ideal gases. 
While considering the altitude variation of gravity for hydrostatic structure, we neglect the effect of curvature, which would yield only a small difference compared to other large uncertainties in model parameters, and assume plane-parallel structure in the photochemical and particle growth calculations.
In this paper, because our focus is on the effects of the size distribution of haze particles on transmission spectra, we assume, for simplicity, that the atmospheric structure is spherically symmetric.
In reality, since close-in exoplanets tend to be tidally locked, the structure may be far from spherically symmetric.
The details of \kawashima{this} effect will be explored in our forthcoming papers.

\subsection{Photochemical Model} \label{method_photo}
\kawashima{
Various photochemical models have been constructed for terrestrial \ikomar{and} gaseous \ikomar{planets}. 
\citet{1981JGR....86.3617A}\ikomar{'s model} \ikomar{is for studying} the vertical transport and photochemistry in the Earth's mesosphere and lower thermosphere (50-120~km).
Using their model, they derived the distributions of long-lived species and compared them with observations.
\citet{2011ApJ...738...32L} introduced their photochemical model to explore the chemistry of warm gaseous exoplanetary atmospheres \ikomar{for explaining the} observed depletion of methane in the atmosphere of GJ~436b.
\citet{2012A&A...546A..43V} released a large chemical network applying combustion models, which were validated over the temperature and pressure ranges relevant to hot Jupiter atmospheres.
After that, they expanded their networks to hydrocarbons up to six-order \citep{2015A&A...577A..33V}.
\citet{2012ApJ...761..166H} presented the photochemical model for terrestrial exoplanets applicable for all types of atmospheres, from reducing to oxidizing.
They presented the results for three benchmark cases of atmospheric scenarios from reducing to oxidizing for terrestrial exoplanets.
\citet{2017ApJS..228...20T} presented an open-source photochemical model for hot exoplanetary atmospheres, VULCAN, which they validated by reproducing the results of \citet{2011ApJ...737...15M}.
In this study, we newly develop a photochemical model to derive the vertical distribution of haze precursors.
}

\subsubsection{Model description}
The one-dimensional continuity-transport equation that governs the change in the number density of species $i$, $n_i$, is written as \citep{1999ppa..conf.....Y}
\begin{equation}
\frac{\partial n_i}{\partial t} = P_i - L_i - \frac{\partial \Phi_i}{\partial z},
\label{eq_cont}
\end{equation}
where $t$ and $z$ are the time and the altitude, respectively, $P_i$ and $L_i$ are the production and loss rates of species $i$ due to photochemical and thermochemical reactions, respectively, and $\Phi _i$ is the vertical transport flux of species $i$. We assume that the vertical transport occurs by eddy diffusion and ignore molecular diffusion. 
The eddy diffusion flux is given by \citep{1999ppa..conf.....Y}
\begin{equation}
\Phi_i = - K_{zz} N \frac{\partial f_i}{\partial z},
\label{eq_phi}
\end{equation}
where $K_{zz}$ is the eddy diffusion coefficient, $N$ is the total number density of the atmospheric gas molecules, and $f_i \equiv n_i / N$ is the mixing ratio of species $i$.
Here, we have used the definition of atmospheric scale height and the ideal gas law.

We include the following \kawashimas{29} chemical species composed of the five elements, C, H, O, N, and He: $\mathrm{O}$, $\mathrm{O_2}$, $\mathrm{H_2O}$, $\mathrm{H}$, $\mathrm{OH}$, $\mathrm{CO_2}$, $\mathrm{CO}$, $\mathrm{HCO}$, $\mathrm{CH_4}$, $\mathrm{CH_3}$, $\mathrm{CH_3O}$, $\mathrm{CH_3OH}$, $\mathrm{CH}$, $\mathrm{CH_2}$, $\mathrm{C}$, $\mathrm{C_2}$, $\mathrm{C_2H}$, $\mathrm{C_2H_2}$, $\mathrm{N}$, $\mathrm{N_2}$, $\mathrm{NH}$, $\mathrm{NH_2}$, $\mathrm{NH_3}$, $\mathrm{CN}$, $\mathrm{HCN}$, $\mathrm{H_2}$, $\mathrm{He}$, $\mathrm{O(^1D)}$, and $\mathrm{^1CH_2}$.
These species are the ones considered in the photochemical models of \cite{2012ApJ...745...77K}, who studied the atmosphere of the hot Jupiter WASP-12b\kawashimas{, except for $\mathrm{H_2CO}$ and $\mathrm{CH_2OH}$, which we do not consider}.
\ikomar{Since the main focus of this study is on calculating the size and  spatial distributions of haze particles and evaluating their impacts on resultant transmission spectra, we simply assume that the haze precursors form from $\rm C_2H_2$ and $\rm HCN$, in the same way as \citet{2013ApJ...775...33M}, and} 
do not include higher-order hydrocarbons such as $\mathrm{C_2 H_4}$ and $\mathrm{C_2H_6}$. 
\ikomar{They showed that} $\mathrm{HCN}$ and $\mathrm{C_2H_2}$ are the most dominant hydrocarbons phtochemically produced in solar-abundance atmospheres with temperature of 500-1000~K, 
\kawashima{
\ikomar{although} there remains uncertainties for the treatment of higher-order hydrocarbons \citep[see, e.g.,][]{2016ApJ...824..137Z}.
\ikomar{Also,} we do not consider sulphur compounds, because they 
\ikomar{are scarcely involved} in reactions 
\ikomar{with} hydrocarbons 
\ikomar{of interest here}.
As 
\ikomar{both the opacities of} $\mathrm{H_2S}$ and $\mathrm{OCS}$ are much smaller compared to \ikomar{those} of $\mathrm{H_2O}$ and $\mathrm{CH_4}$ 
\ikomar{according to} sulphur's 
\ikomar{small} elemental abundance, 
\ikomar{it is sure} that they 
\ikomar{have little} impact 
\ikomar{on} the transmission spectrum.
We do not consider Na and K because they condense as $\mathrm{Na_2S}$ and $\mathrm{KCl}$ clouds, respectively, and settle downward in the temperature range of interest ($\lesssim$ 1000~K) \citep{2013ApJ...775...33M}.
}

We adopt 154 thermochemical reactions from the reaction list of \citet{2012ApJ...761..166H}.
All the thermochemical reactions and their \kawashima{rate coefficients} are listed in Table~\ref{tab:thermo_reactions}. 
We have chosen the reactions that involve only some of the above 31 species, although the reaction list of \citet{2012ApJ...761..166H} contains more reactions.
We also consider their reverse reactions using the method described in \citet{Visscher:2011jf}. 
Thus, in total, we consider 308 thermochemical reactions. 
For the calculation of the Gibbs free energy of each species, which is needed to calculate the equilibrium constants (the ratios of forward to reverse reaction rate coefficients), we use the polynomial coefficients for calculating enthalpies of formation, entropies, and heat capacities from the Third Millennium Ideal Gas and Condensed Phase Thermochemical Database for Combustion \footnote{http://garfield.chem.elte.hu/Burcat/burcat.html}.
\kawashima{Although some rate coefficients are invalid in the temperature range considered in this study, we use them outside their temperature range due to the lack of data and/or theory.}

For photochemistry, we consider \kawashimas{16} reactions listed in Table~\ref{tab:photo_reactions}. 
Likewise, all the reactions are extracted from the reaction list of \citet{2012ApJ...761..166H} if the reaction involves only some of the above 31 species.
Photodissociation rate of species $i$ \ikomar{(i.e., the number of atoms or molecules dissociated per unit time)} at altitude $z$, $J_i \left( z \right)$, is written as
\begin{equation}
J_i \left( z \right) = \kawashimabf{\frac{1}{2}} \int_{0}^{\infty} q_i \left( \lambda \right) \sigma_i \left( \lambda \right) F \left( z, \lambda \right) d\lambda,
\end{equation}
where $\lambda$ is the wavelength, $q_i \left( \lambda \right)$ and $\sigma_i \left( \lambda \right)$ are the \kawashima{dimensionless} quantum yield \ikomar{of species $i$,} \ikomar{the} absorption cross section \kawashima{(\ikomar{its} physical \ikomar{dimension being} area)} of species $i$, and $F \left( z, \lambda \right)$ is the \kawashima{actinic photon} flux \kawashima{\ikomar{per unit area, unit time, and unit wavelength}}.
\kawashima{The factor 1/2 is \ikomar{needed} to account for diurnal variation \citep[see][]{2012ApJ...761..166H}.}
The references from which we take the data of the quantum yields and absorption cross sections are tabulated in Table~\ref{tab:photo_cross-sections} and \ref{tab:photo_reactions}, respectively, most of which can be downloaded from the website of the MPI-Mainz UV/VIS Spectral Atlas of Gaseous Molecules of Atmospheric Interest\footnote{http://satellite.mpic.de/spectral\_atlas}.
Temperature dependences of absorption cross sections are known for some of the species, but measured only in a temperature range between 200 and 300~K.
Thus, following \citet{2012ApJ...761..166H},
we calculate the absorption cross sections at 300~K by a linear interpolation with the use of the measured data and use them for temperatures higher than 300~K, namely $\sigma (\lambda, T) = \sigma (\lambda, 300~\mathrm{K})$, instead of extrapolating beyond 300~K.
We consider the attenuation of the \kawashima{actinic} flux as
\begin{equation}
F \left( z, \lambda \right) = F \left( \infty, \lambda \right) e^{-\tau \left( z, \lambda \right) / \mu},
\end{equation}
where $F \left( \infty, \lambda \right)$ is the \kawashima{actinic} flux at the top of the atmosphere at wavelength $\lambda$ and $\mu$ is the cosine of the zenith angle of the star. 
$\tau \left( z, \lambda \right)$ is the optical depth defined by
\begin{equation}
\tau \left( z, \lambda \right) = \sum_i^{\kawashimabf{\mathcal{N}}} \int_z^{\infty} n_i \left( z' \right) \sigma_i \left( \lambda \right) dz',
\end{equation}
\kawashima{where $\mathcal{N}$ is the number of the species.}
We assume the zenith angle to be $57.3^{\circ}$, as done in \citet{2012ApJ...761..166H}. 
They found that the mean zenith angle differed depending on the optical depth of interest and concluded that the assumption of $\mu$ = $57^{\circ}$-$48^{\circ}$, which corresponded to $\tau$ = 0.1-1.0, was appropriate for the one-dimensional photochemical models. 

\kawashima{
\ikomar{For the boundary conditions, we set the diffusion flux $\Phi_i$ as zero for all the species at the upper boundary, while we fix the volume mixing ratios $f_i$ of all the species at the thermochemical equilibrium values at the lower boundary. The exact conditions are, however, uncertain, so that}
\ikomar{previous studies chose different conditions at} 
\ikomar{both boundaries.} 
As for \ikomar{the} upper boundary condition, while some studies set 
\ikomar{the diffusion flux equal to the assumed} atmospheric escape 
\ikomar{flux} \citep[e.g.,][]{2012ApJ...761..166H},
\ikomar{some studies} set 
$\Phi_i = 0$ 
for all the species 
\citep[\ikomar{e.g.}][]{2011ApJ...737...15M, 2012A&A...546A..43V, 2017ApJS..228...20T}. 
\ikomar{In this study, we choose the latter because}
\kawashimar{the atmospheric escape rate is unknown for exoplanets.}
As for \ikomar{the} lower boundary condition, 
\ikomar{photochemical modeling of terrestrial planet atmospheres often sets the flux of surface emission and/or deposition at the lower boundary}
 \citep[e.g.,][]{2012ApJ...761..166H}. 
However,  
\ikomar{gas-rich} planets\ikomar{, which we consider in this study,} have no \ikomar{rigid} surfaces. 
While some studies adopted zero flux \citep[e.g.,][]{2011ApJ...737...15M, 2012A&A...546A..43V, 2017ApJS..228...20T}, we fix the volume mixing ratios $f_i$ of all the species at thermochemical equilibrium values 
\ikomar{in a similar way to,} for example, \citet{2011ApJ...738...32L} and \citet{2014ApJ...797...41Z}.
This is because 
\ikomar{the gases at} deep levels would be \ikomar{in} thermochemical equilibrium.
While \citet{2011ApJ...737...15M} \kawashimar{reported that they did not find any} differences in the results between 
\ikomar{the two types of inner boundary condition}, 
\citet{2017ApJS..228...20T} found 
that \ikomar{only the} minor (\ikomar{$f_i$} $\lesssim 10^{-9}$) molecules, CO and $\mathrm{CO_2}$, deviated from their thermochemical equilibrium values at \kawashimas{relatively cool (1000~K)} lower boundary (1000~bar)\ikomar{, but} 
major molecules are in thermochemical equilibrium.}

\subsubsection{Calculation method} \label{method_photo2}
\kawashima{The calculation method \ikomar{we use in this study is basically the same as that used in} previous works \citep[e.g.,][]{2012A&A...546A..43V, 2012ApJ...761..166H, 2017ApJS..228...20T}.}
We discretize Eq.~(\ref{eq_cont}) as
\begin{equation}
\frac{\partial n_{i, j}}{\partial t} = P_{i, j} - L_{i, j} - \frac{\Phi_{i, j+1/2} - \Phi_{i, j-1/2}}{\Delta z_j},
\label{eq_cont_disc}
\end{equation}
where the subscript $j$ represents the physical quantities in the $j$th layer and $\Delta z_j$ is the thickness of the $j$th layer.
We prepare layers with the same thickness $\Delta z$ and set the pressure at the mid-point altitude of the lowest layer as the lower boundary pressure.
From Eq.~(\ref{eq_phi}), we approximate $\Phi_{i, j+1/2}$ as \kawashima{\citep[e.g.,][]{2012A&A...546A..43V, 2012ApJ...761..166H, 2017ApJS..228...20T}}
\begin{equation}
\Phi_{i, j+1/2} = - K_{zz} N_{j+1/2} \frac{f_{i, j+1} - f_{i, j}}{\Delta z}.
\end{equation}

To obtain a steady-state solution, 
we solve Eq.~(\ref{eq_cont_disc}) implicitly with the use of the solver DLSODES \citep{Hindmarsh1982}, which is suitable to solve stiff ODE systems such as chemical network calculations \kawashima{\citep[e.g.,][]{2014MNRAS.439.2386G}. It is based on \ikomar{a backward differentiation formula} (BDF), which is also called Gear's method. 
The most suitable order is chosen within the solver. We set the maximum order allowed to be five.}
We adopt the values of relative (RTOL) and absolute (ATOL) tolerances as $10^{-4}$ and $N_{j} \times 10^{-15}$, respectively; 
the value of ATOL differs from layer to layer.

The initial number densities of the species are set to their thermochemical equilibrium values, which we calculate \kawashima{in the following way. 
A system composed of $\mathcal{N}$ gaseous species being considered, the Gibbs free energy of the system is minimized at equilibrium.
The Gibbs free energy at fixed temperature $T$, pressure $P$, and composition $\mathbf{\xi}$ is written as \citep{RefWorks:6}
\begin{equation}
G \left(T, P, \mathbf{\xi} \right) = \sum_{i = 1}^\mathcal{N} \xi_i \phi_i,
\end{equation}
where $\xi_i$ and $\phi_i$ are the \ikomar{molar} number and chemical potential of species $i$, respectively, and $\mathbf{\xi} = \{ \xi_1, \xi_2, \cdot \cdot \cdot, \xi_\mathcal{N} \}$. 
The chemical potential of an ideal gas is given by \citep{RefWorks:6}
\begin{equation}
\phi_i \left( T, P \right) = \phi^\circ_i \left( T \right) + \mathcal{R}T \ln{\frac{p_i}{p_\mathrm{ref}}}.
\end{equation}
Here $\phi^\circ$ is the standard chemical potential that is a function of $T$ only, $p_i$ is the partial pressure of gaseous species $i$, $p_\mathrm{ref}$ is the reference pressure, and $\mathcal{R}$ is the molar gas constant.
If a collection of species in the system is given, theoretically permissible chemical reactions can be derived from the law of conservation of mass:
\begin{equation}
\sum_{i = 1}^\mathcal{N} a_{ki} \xi_i = b_k,
\end{equation}
where $a_{ki}$ is the number of the $k$th element contained in species $i$ and $b_k$ is the total number of moles of the $k$th element. The composition that gives the minimum value of the Gibbs free energy is searched for to determine the equilibrium values of the mole fractions of the elements in the system. We assume vertically constant elemental abundance ratios and use the same Gibbs free energy data as that we use for calculation of reverse rate coefficients.}

We time-integrate Eq.~(\ref{eq_cont_disc}) until the system becomes in a steady state.
We adopt the criteria of convergence such that all the species of $f_i$ $> 10^{-10}$ vary in mixing ratio by less than 1\% in all the layers.
The integration is done over a period longer than the eddy diffusion timescale, which we assume as the maximum value of $H_j^2/K_{zz}$ among all the layers at the initial condition. Here, $H_j$ is the atmospheric scale height for layer $j$.
The time step is self-adjusted within the solver so that the estimated local error in $n_{i, j}$ is not larger by an order of magnitude than that of $\mathrm{RTOL} \times n_{i, j} + \mathrm{ATOL}_{j}$ ($\equiv$ $\mathrm{EWT}_{i, j}$).
At each time after calling the solver, for the atmosphere to be in hydrostatic equilibrium and the total mixing ratio to be unity, we set the output negative number densities to be zero, renormalize the volume mixing ratio of each species, recalculate the total number density at each layer assuming hydrostatic equilibrium, and calculate the number density of each species at each layer. 
Note that the output negative number densities are not larger by an order of magnitude than EWT.

\kawashimas{We compare our \ikomas{photochemical} model with \ikomas{the} previous thermochemical models for \ikomas{the} atmospheres of HD~189733b and HD~209458b \ikomas{presented by} \citet{2017ApJS..228...20T} in APPENDIX~\ref{tsai} and \ikomas{the} photochemical models for \ikomas{the} WASP-12b's atmosphere \ikomas{presented by} \citet{2012ApJ...745...77K} in  APPENDIX~\ref{kopparapu}.
We have confirmed that the abundances of most of the species match those of the previous works within one order of magnitude and the profiles of the molecules are similar except for absolute value.
And the differences in abundances for some molecules would not affect our results \ikomas{regarding haze distributions and transmission spectra}.}
We have also confirmed the major trend found in those for GJ~1214b's atmosphere \citep{2012ApJ...745....3M, 2013ApJ...775...33M} and other low temperature ($\lesssim 1000$~K) atmospheres \citep{2013ApJ...777...34M, 2014A&A...562A..51V}.

\subsection{Particle Growth Model} \label{method_growth}
We simulate the growth and settling of hydrocarbon haze particles after the production of monomers in the upper atmosphere and determine their steady-state distribution.
We assume that monomers form in situ from the precursor molecules of haze particles. 
We assume $\mathrm{HCN}$ and $\mathrm{C_2H_2}$ as the precursor molecules.
While higher-order hydrocarbons may have to be also included as the precursors, previous studies \citep[e.g.,][]{2013ApJ...775...33M} showed that $\mathrm{HCN}$ and $\mathrm{C_2H_2}$ are the most dominant hydrocarbons photochemically produced in solar abundance atmospheres with temperature of 500-1000~K, as mentioned in \S~\ref{method_photo}.

\subsubsection{Model description}
\kawashima{We follow the classical formalism \ikomar{for} cloud particle growth \citep[see][]{RefWorks:16}, which has also been used to simulate haze particle growth in Titan's atmosphere \citep[e.g.,][]{1980Icar...43..260T, 1992Icar...95...24T}.
Note that the same formalism has been also used \ikomar{for dust} particle growth in the field of planet formation \citep[see, \ikomar{e.g.,}][]{RefWorks:13}.
\ikomar{Also,} as for dust particle growth for brown dwarf atmospheres, there is a series of work \citep{2003A&A...399..297W, 2004A&A...414..335W, 2006A&A...455..325H},
\ikomar{which} is different from 
\ikomar{ours} in the point that they considered particle growth due to chemical surface reactions \ikomar{but did} not consider the growth due to coagulation.}

Adopting a discrete volume grid, one can write the one-dimensional continuity-transport equation for the number density of particles with volume $v_i$, $n \left( v_i \right)$, as \citep[e.g.][]{2010Icar..210..832L}
\begin{eqnarray}
\frac{\partial n \left( v_i \right)}{\partial t} =&& \frac{1}{2} \sum_{k = 1}^{i-1} K \left(v_k, v_i-v_k \right) n \left( v_k \right) n \left( v_i-v_k \right) \nonumber \\
&& - n \left( v_i \right) \sum_{k = 1}^{\mathcal{N}} K \left(v_i, v_k \right) n\left( v_k \right) \nonumber \\
&& - \frac{\partial \Phi \left( v_i \right)}{\partial z} + p \left( v_i \right),
\label{eq_cont2}
\end{eqnarray}
where the subscript denotes the volume grid, $K \left(v_i, v_k \right)$ is the coagulation kernel between two particles with volumes $v_i$ and $v_k$, and $\mathcal{N}$ is the total number of volume bins used in the calculation.
The first and second terms on the right-hand side describe the production and loss of the particles of volume $v_i$ (hereafter, the $i$th particles, for simplicity) due to the coagulation.
$\Phi \left( v_i \right)$ is the vertical transport flux and $p \left( v_i \right)$ is the photochemical production rate of the $i$th particles, which takes a non-zero value only for $v_1$, namely monomers.

Assuming that the vertical transport occurs by sedimentation and eddy diffusion, 
one can write $\Phi \left( v_i \right)$
as \citep[e.g.][]{2010Icar..210..832L}
\begin{equation}
\Phi \left( v_i \right) = - V_{\mathrm{s}, i} n \left( v_i \right) - K_{zz} N \frac{\partial \left(n \left( v_i \right) / N \right)}{\partial z},
\label{eq_phi2}
\end{equation}
where $V_{\mathrm{s}, i}$ is the sedimentation velocity of the $i$th particles written as \citep[e.g.][]{2010Icar..210..832L}
\begin{equation}
V_{\mathrm{s}, i} = \frac{2 s_i^2 \rho_p g}{9 \eta_a} f_{\mathrm{slip}, i}.
\label{eq_sedimentation}
\end{equation}
Here, $s_i$ is the radius of the $i$th particle, 
$\rho_p$ is the particle internal density, 
\kawashima{and} $g$ is the local gravitational acceleration\kawashima{.} 
$\eta_a$ is the dynamic viscosity defined as
\begin{equation}
\eta_a = \frac{1}{3} \rho_a \overline{V}_\mathrm{th} \lambda_a,
\end{equation}
where $\rho_a$ is the mass density of the gas, $\overline{V}_\mathrm{th}$ is the thermal velocity of the gaseous molecules defined as 
\begin{equation}
\overline{V}_\mathrm{th} = \sqrt{\frac{8 k_\mathrm{B} T}{\pi m_{a}}}
\end{equation}
with the Boltzmann constant $k_\mathrm{B}$, and the temperature $T$, \kawashima{and} the mean mass of gaseous molecules $m_{a}$\kawashima{.} $\lambda_a$ is the atmospheric mean free path defined as
\begin{equation}
\lambda_a = \frac{k_\mathrm{B} T}{\pi \sqrt{2} P d^2}
\end{equation}
with the pressure $P$ and the diameter of the gas molecule $d$. Because $\mathrm{H_2}$ is the most abundant gas species in the atmosphere of interest in this study, we use the diameter of $\mathrm{H_2}$ for the value of $d$, taken from CRC Handbook of CHEMISTRY and PHYSICS \citep{RefWorks:4}.
$f_\mathrm{slip}$ is the Cunningham slip-flow correction factor given by \citep{Davies:1945ci}
\begin{equation}
f_{\mathrm{slip}, i} = 1 + 1.257 K_{n, i} + 0.400K_{n, i} \exp{\left( -1.10/K_{n, i} \right)}, 
\label{eq_slip}
\end{equation}
where $K_{n, i}$ is the Knusden number defined as $K_{n, i} \equiv \lambda_a / s_i$.

As for coagulation, we consider two rate-controlling processes 
which include the Brownian diffusion and gravitational collection. The latter is the collisional process that occurs as a result of difference in sedimentation velocity between different size particles.
The total kernel is assumed to be the sum of the two kernels, namely
\begin{equation}
K \left( v_i, v_k \right) = K_\mathrm{BD} \left( v_i, v_k \right) + K_\mathrm{GC} \left( v_i, v_k \right) \kawashima{.}
\end{equation}
The Brownian collision kernel for the $i$th and $k$th particles, $K_\mathrm{BD} \left( v_i, v_k \right)$, can be written as \citep{RefWorks:16}
\begin{equation}
K_\mathrm{BD} \left( v_i, v_k \right) = \frac{4 \pi \left(s_i + s_k \right) \left(D_{p, i} + D_{p, k} \right)}{\frac{s_i + s_k}{s_i + s_k + \sqrt{\delta_i^2 + \delta_k^2}} + \frac{4 \left(D_{p, i} + D_{p, k} \right)}{\sqrt{\overline{v}_{\mathrm{th}, i}^2 + \overline{v}_{\mathrm{th}, k}^2} \left(s_i + s_k \right)}}
\end{equation}
with
\begin{equation}
\delta_i = \frac{ \left( 2s_i + \lambda_{p, i} \right)^3 - \left( 4 s_i^2 + \lambda_{p, i}^2 \right)^{3/2}}{6 s_i \lambda_{p, i}}.
\end{equation}
$D_{p, i}$ and $\overline{v}_{\mathrm{th}, i}$ are the diffusion coefficient and thermal velocity for the $i$th particle, respectively.
These parameters are given as
\begin{equation}
D_{p, i} = \frac{k_\mathrm{B} T}{6 \pi s_i \eta_a} f_{\mathrm{slip}, i}
\end{equation}
and
\begin{equation}
\overline{v}_{\mathrm{th}, i} = \sqrt{\frac{8 k_\mathrm{B} T}{\pi m_{p, i}}}
\label{eq_vth}
\end{equation}
with the particle mass $m_{p, i}$.
$\lambda_{p, i}$ is the particle's mean free path written as
\begin{equation}
\lambda_{p, i} = \frac{8D_{p, i}}{\pi \overline{v}_{\mathrm{th}, i}}.
\end{equation}
The gravitational collection kernel for the $i$th and $k$th particles, $K_\mathrm{GC} \left( v_i, v_k \right)$, can be written as \citep{RefWorks:16}
\begin{equation}
K_\mathrm{GC} \left( v_i, v_k \right) = E_{\mathrm{coll}, i, k} \pi \left(s_i + s_k \right)^2 |V_{s, i} - V_{s, k}|,
\end{equation}
where $E_{\mathrm{coll}, i, k}$ is a collision efficiency given by
\begin{equation}
E_{\mathrm{coll}, i, k} = \frac{60 E_{\mathrm{V}, i, k} + E_{\mathrm{A}, i, k} \mathrm{Re}_i}{60 + \mathrm{Re}_i} \; \left( s_i \geq s_k \right)
\end{equation}
\begin{eqnarray}
E_{\mathrm{V}, i, k} = \left\{ \begin{array}{ll}
\left[ 1 + \frac{0.75 \ln{\left( 2 \mathrm{St}_{i, k}\right)}}{\mathrm{St}_{i, k} - 1.214} \right]^{-2} & \left( \mathrm{St}_{i, k} > 1.214 \right) \\
0 & \left( \mathrm{St}_{i, k} \leq 1.214 \right)\\
\end{array} \right .
\end{eqnarray}
\begin{equation}
E_{\mathrm{A}, i, k} = \frac{\mathrm{St}_{i, k}^2}{\left( \mathrm{St}_{i, k} + 0.5 \right)^2}.
\end{equation}
Here, $\mathrm{Re}_i$ is the Reynolds number written as
\begin{equation}
\mathrm{Re}_i = \frac{2 s_i V_{s, i}}{\nu_a}
\end{equation}
with the kinematic viscosity
\begin{equation}
\nu_a = \frac{\eta_a}{\rho_a}
\end{equation}
and $\mathrm{St}_{i, k}$ is the Stokes number written as
\begin{equation}
\mathrm{St}_{i, k} = \frac{V_{s, k} |V_{s, i} - V_{s, k}|}{s_i g} \; \left( s_i > s_k \right).
\end{equation}

When we simulate the particle growth with the discretized size distribution, 
we face the problem that the coagulation between the $i$th and $k$th particles ($v_i > v_k$) produces particles of an intermediate volume,
\begin{equation}
v_{i, k} = v_i + v_k.
\end{equation}
To satisfy the conservations of the mass and the particle numbers at the same time, 
we partition this intermediate-volume particle into the two volume bins, $v_l$ and $v_{l + 1}$ ($v_l < v_{i, k} < v_{l + 1}$), with fractions $\gamma_l$ and $\gamma_{l + 1}$, respectively.
Unless $v_{l}$ is the largest volume bin, these fractions can be written as
\begin{equation}
\gamma_l = \frac{v_{l + 1} - v_{i, k}}{v_{l + 1} - v_l}
\end{equation}
and
\begin{equation}
\gamma_{l + 1} = 1 - \frac{v_{l + 1} - v_{i, k}}{v_{l + 1} - v_l}. 
\end{equation}
If $v_{l}$ is the largest volume bin, we cannot partition the intermediate particle but just put it into the largest volume bin $v_{l + 1}$, although the mass conservation is not satisfied.
We specify the volume ratio of two adjacent bins in $\S$~\ref{method_procedure}.

\subsubsection{Monomer production rate}
As described above, we assume that monomers are formed in situ from the precursor molecules $\mathrm{HCN}$ and $\mathrm{C_2H_2}$. 
Thus, we calculate the vertical profile of the mass production rate of monomers according to the distribution of the two molecules.
We consider that the mass production rate of monomers\kawashima{, which \ikomar{means the total mass of monomers produced per unit volume per unit time},} at altitude $z$ is given by
\begin{equation}
p \left( v_1, z \right) = 
\frac{\left[ f_\mathrm{HCN} \left( z \right) + f_\mathrm{C_2H_2} \left( z \right) \right] N \left( z \right)}
{\int_0 ^{\infty} \left[ f_\mathrm{HCN} \left( z' \right) + f_\mathrm{C_2H_2} \left( z' \right) \right] N \left( z' \right) dz'} \dot{M},
\end{equation}
where $f_\mathrm{HCN}$ and $f_\mathrm{C_2H_2}$ are the volume mixing ratios of HCN and $\mathrm{C_2H_2}$, respectively, and $\dot{M}$ is the total mass production rate of monomers throughout the atmosphere \kawashima{and \ikomar{its} physical unit \ikomar{is} mass/area/time}.

We assume that $\dot{M}$ is proportional to the incident stellar Lyman-alpha (Ly$\alpha$) flux at the planet's orbital distance, $I_\mathrm{Ly\alpha}$, because monomer production is relevant to UV photodissociation.
\kawashima{Thus, we assume the photochemistry of monomer formation to be driven entirely by $\mathrm{Ly}\alpha$.}
For the reference, we use the observed values of the incident solar Ly$\alpha$ flux, $I_\mathrm{Ly\alpha, Titan}$, and mass production rate, $\dot{M}_\mathrm{Titan}$, in the present Titan's atmosphere; \kawashima{\sout{N}n}amely, \begin{equation}
\label{eq_monomer_total}
\dot{M} = \beta \frac{I_\mathrm{Ly\alpha}}{I_\mathrm{Ly\alpha, Titan}} \dot{M}_{\mathrm{Titan}}.
\end{equation}
\kawashima{\ikomar{This} is a simpler version of Eq.~(8) of \citet{2006PNAS..10318035T}\ikomar{, which} they derived empirically.}
Although both linear and quadratic dependences of $\dot{M}$ on $I_\mathrm{Ly\alpha}$ are proposed, there is still room for discussion to determine which relationship is appropriate \citep{2006PNAS..10318035T}.
The linear relationship would be valid when haze monomers are produced predominantly by photodissociation of hydrocarbon intermediate molecules, which is the product of photodissociation of $\mathrm{CH_4}$, while the quadratic relationship would be valid when haze monomers are produced mainly by thermochemical reactions between multiple intermediates (see \cite{2006PNAS..10318035T} for details).
Because the relationship is totally uncertain for exoplanet atmospheres, we have adopted the linear relationship for simplicity, and added a numerical parameter $\beta$ in the above equation.
We adopt $1 \times$~$10^{-14}$~g~$\mathrm{cm^{-2}}$~$\mathrm{s^{-1}}$ for $\dot{M}_{\mathrm{Titan}}$, since microphysical models, photochemical models, and laboratory simulations all imply that the production rate of the monomers on Titan is in the range between $0.5 \times 10^{-14}$ and $2 \times 10^{-14}$~g~$\mathrm{cm^{-2}}$~$\mathrm{s^{-1}}$ \citep{2001P&SS...49...79M}.
Also, we use $6.2 \times 10^9$~photons~$\mathrm{cm^{-2}}$~$\mathrm{s^{-1}}$ for $I_\mathrm{Ly\alpha, Titan}$ \citep{2006PNAS..10318035T}.
\kawashima{When we vary $\beta$, we \ikomar{also} vary the intensities of the actinic flux at all the wavelengths according to $\beta$ (i.e., the Ly$\alpha$ intensity).}

Finally, the boundary conditions for Eq.~(\ref{eq_cont2}) are given as follows. As the lower boundary conditions, we consider that all the particles are lost with the larger of the sedimentation velocity and the downward velocity imposed by the atmospheric mixing, following \cite{2010Icar..210..832L}.
As the upper boundary conditions, we set zero fluxes for all the particle sizes.

\subsubsection{Calculation method}
We divide the atmosphere into layers with the same thickness $\Delta z$ and discretize Eq.~(\ref{eq_cont2}) as
\begin{eqnarray}
\frac{\partial n_j \left( v_i \right)}{\partial t} =&& \frac{1}{2} \sum_{k = 1}^{i-1} K_j \left(v_k, v_i-v_k \right) n_j \left( v_k \right) n_j \left( v_i-v_k \right) \nonumber \\
&& - n_j \left( v_i \right) \sum_{k = 1}^{\mathcal{N}} K_j \left(v_i, v_k \right) n_j \left( v_k \right) \nonumber \\
&& - \frac{\Phi_{j + 1/2} \left( v_i \right) - \Phi_{j - 1/2} \left( v_i \right)}{\Delta z} + p_j \left( v_i \right),
\label{eq_cont_disc2}
\end{eqnarray}
where the subscript $j$ represents the physical quantities in the $j$th layer.
We set the pressure at the mid-point altitude of the lowest layer as the lower boundary pressure.
For Eq.~(\ref{eq_phi2}), we use the upwind difference scheme instead of the central difference scheme for the calculation of sedimentation flux, because of numerical stability, and approximate $\Phi_{j + 1/2} \left( v_i \right)$ as
\begin{eqnarray}
\Phi_{j + 1/2} \left( v_i \right) =&& - V_{s, i, j +1} n_{j + 1} \left( v_i \right) \nonumber \\
&& - K_{zz} N_{j + 1/2} \frac{n_{j + 1} \left( v_i \right) / N_{j + 1} - n_j \left( v_i \right) / N_j}{\Delta z}. \nonumber \\
\end{eqnarray}

To obtain a steady-state solution, 
we solve the continuity Eq.~(\ref{eq_cont_disc2}) implicitly with the same solver DLSODES \citep{Hindmarsh1982} that we use in the photochemical calculations ($\S$~\ref{method_photo}).
We adopt the values of relative (RTOL) and absolute (ATOL) tolerances as $10^{-4}$ and $10^{-20}$, respectively.
The initial number densities of all the sizes are set to zero.
We adopt the criteria of convergence such that the volume-averaged sizes of particles in all the layers, which we calculate as
\begin{equation}
\label{eq_s_vol}
s_\mathrm{vol} = \frac{\sum_{i = 1}^{\mathcal{N}} n \left(s_i \right) s_i^4}{\sum_{i = 1}^{\mathcal{N}} n \left(s_i \right) s_i^3},
\end{equation}
are different by less than 1\%.
\subsection{Opacity} \label{method_opacity}
\subsubsection{Haze particles}
\label{method_opacity_haze}
We calculate the extinction cross sections of haze particles based on the Mie theory \citep{1908AnP...330..377M}.
In the limit where the particle radius, $s$, is large compared to the radiation wavelength, $\lambda$, the Mie theory agrees with geometric optics.
On the other hand, the Mie theory reduces to the Rayleigh theory in the limit of $s \ll \lambda$.

From the Mie theory, the extinction cross section of a homogeneous spherical particle of radius $s$, $\sigma_\mathrm{ext}$, can be written as \citep{RefWorks:9} 
\begin{equation}
\sigma_\mathrm{ext} = \pi s^2 \frac{2}{x^2} \sum_{n=1}^{\infty} \left( 2n+1 \right) \mathrm{Re} \left(a_n + b_n \right),
\label{eq_cs}
\end{equation}
where Re denotes the real part.
Here, $x$ is the size parameter defined as
\begin{equation}
x \equiv \frac{2 \pi s}{\lambda}.
\end{equation}
Coefficients $a_n$ and $b_n$ are calculated as
\begin{equation}
a_{n} = \frac{m \phi_{n} \left( mx \right) \phi_{n}' \left( x \right) - \phi_{n} \left( x \right) \phi_{n}' \left( mx \right)}{m \phi_{n} \left( mx \right) \zeta_{n}' \left( x \right) - \zeta_{n} \left( x \right) \phi_{n}' \left( mx \right)},
\end{equation}
and
\begin{equation}
b_{n} = \frac{\phi_{n} \left( mx \right) \phi_{n}' \left( x \right) - m \phi_{n} \left( x \right) \phi_{n}' \left( mx \right)}{\phi_{n} \left( mx \right) \zeta_{n}' \left( x \right) - m \zeta_{n} \left( x \right) \phi_{n}' \left( mx \right)},
\label{eq_bn}
\end{equation}
where $m$ is the ratio of the complex refractive indices of the particle to the surrounding atmosphere.
$\phi$ and $\zeta$ are the so-called Ricatti-Bessel functions and the prime indicates differentiation with respect to the argument in parentheses.

We use the bhmie code \citep{RefWorks:9} to calculate Eqs.~(\ref{eq_cs})-(\ref{eq_bn}). Complex refractive indices of haze are taken from \cite{1984Icar...60..127K}, which reports laboratory experiment results for production of tholin hazes in a simulated Titan's atmosphere (0.9 $\nitrogen$/0.1 $\methane$ gas mixture at 0.2~mb).

\begin{figure}[ht!]
\plotone{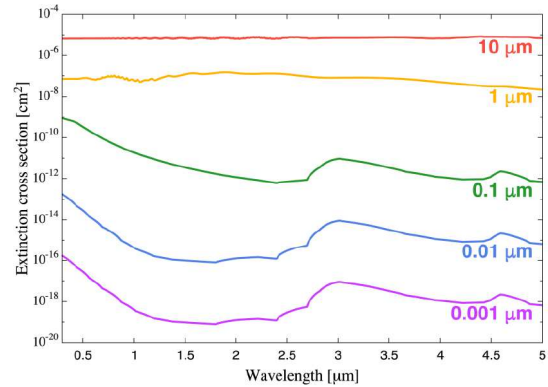}
\caption{Extinction cross sections of the tholin-like haze particles of five different particle sizes of 0.001~$\mu\mathrm{m}$, 0.01~$\mu\mathrm{m}$, 0.1~$\mu\mathrm{m}$, 1~$\mu\mathrm{m}$, and 10~$\mu\mathrm{m}$.}
\label{fig_opacity_haze}
\end{figure}

In Figure~\ref{fig_opacity_haze}, we show the extinction cross sections of the haze particles of five different particle sizes, namely, 0.001~$\mu\mathrm{m}$, 0.01~$\mu\mathrm{m}$, 0.1~$\mu\mathrm{m}$, 1~$\mu\mathrm{m}$, and 10~$\mu\mathrm{m}$. 
When the particle size is sufficiently small relative to the wavelength, the scattering is approximated by the Rayleigh scattering. 
More specifically, the cross sections for $s =$ 0.001~$\mu\mathrm{m}$, 0.01~$\mu\mathrm{m}$, and 0.1~$\mu\mathrm{m}$ show the behavior due to the Rayleigh scattering in the visible wavelength region; namely, $\sigma_\mathrm{ext} \propto \lambda^{-4}$.
Also, the dependence on the particle radius is such that $\sigma_\mathrm{ext} \propto s^3$ \citep[e.g.,][]{RefWorks:10}. 
In contrast,
for larger particles of 1~$\mu$m and 10~$\mu$m, no such feature is found, and the cross sections are relatively independent of wavelength.
Note that the bumps found around $3.0$~$\mu\mathrm{m}$ and $4.6$~$\mu\mathrm{m}$ come from the vibrational transitions of the C-H bond and C$\equiv$N bond of the tholin-like haze particles, respectively \citep{1984Icar...60..127K}.

\subsubsection{Gaseous species}
For another source of radiative extinction in the atmosphere, we consider line absorption by $\water$, $\carbondioxide$, $\carbonmonoxide$, $\methane$, $\mathrm{O_2}$, $\ammonia$, $\mathrm{OH}$, $\nitrogen$, $\mathrm{HCN}$, $\mathrm{C_2H_2}$, and $\hydrogen$.
We ignore the extinction by Na and K because they condense as $\mathrm{Na_2S}$ and $\mathrm{KCl}$ clouds, respectively, and settle downward in a temperature range of interest ($\lesssim$ 1000~K) \citep{2013ApJ...775...33M}.

\kawashima{The extinction cross section of species $i$, at wavenumber $\nu$, $\sigma_i \left( \nu \right)$, is written as
\begin{equation}
\sigma_i \left( \nu \right) = \sum_{\eta, \eta'} \sigma_{i, \eta \eta'} \left( \nu \right)
\end{equation}
where $\sigma_{i, \eta \eta'}$ is the line absorption cross section for the transition from lower state $\eta$ to upper state $\eta'$.}

\kawashima{For briefly, we omit the subscript $i$ hereafter.
The line absorption cross section, $\sigma_{\eta \eta'}$, is given as
\begin{equation}
\sigma_{\eta \eta'} \left( \nu \right)= S_{\eta \eta'} \left( T \right) f\left( \nu - \nu_{\eta \eta'} \right),
\end{equation}
where $\nu_{\eta \eta'}$ is the spectral line transition wavenumber, $S_{\eta \eta'}$ is the spectral line intensity, and $f$ is the line profile function.
We calculate $\sigma_{\eta \eta'}$, using the line data from HITRAN2012 \citep{2013JQSRT.130....4R}.
When summing the absorption cross section for each transition, we do not consider the cross sections whose spectral line intensities are less than $10^{-40}$ $\mathrm{cm}^{-2}$ because of the computational cost.}

\kawashima{According to \cite{2007ApJS..168..140S} and \cite{1998JQSRT..60..665R}, the spectral line intensity at temperature $T$, $S_{\eta \eta'} \left( T \right)$, is written as
\begin{eqnarray}
S_{\eta \eta'} \left( T \right) &=& \frac{\pi e^2 g_\eta f_{\eta \eta'}}{m_e c} \frac{\exp{\left( - h c E_\eta / k_\mathrm{B} T \right)}}{Q \left( T \right)} \nonumber \\
&& \left[ 1 - \exp{\left( - h c \left( E_{\eta'} - E_{\eta} \right) / k_\mathrm{B} T \right)} \right] \\
&=& S_{\eta \eta'} \left( T_{\mathrm{ref}} \right) \frac{Q \left( T_{\mathrm{ref}} \right)}{Q \left( T \right)} \frac{\exp{\left( - h c E_\eta / k_\mathrm{B} T \right)}}{\exp{\left( - h c E_\eta / k_\mathrm{B} T_{\mathrm{ref}} \right)}} \nonumber \\
&& \frac{\left[ 1 - \exp{\left( - h c \left( E_{\eta'} - E_{\eta} \right) / k_\mathrm{B} T \right)} \right]}{\left[ 1 - \exp{\left( - h c \left( E_{\eta'} - E_{\eta} \right) / k_\mathrm{B} T_{\mathrm{ref}} \right)} \right]},
\end{eqnarray}
where $g_\eta$ is the statistical weight of the lower state $\eta$, $f_{\eta \eta'}$ is the oscillator strength for the transition between the lower and upper states, $E_{\eta}$ and $E_{\eta'}$ are the lower-state and upper-state energy, respectively, and $Q \left( T \right)$ is the total internal partition function at temperature $T$.
$e$ is the elementary charge, $m_e$ is the electron mass, and 
$h$, $c$, and $k_\mathrm{B}$ are the Planck constant, the speed of light, and the Boltzmann constant, respectively. 
$S_{\eta \eta'} \left( T_{\mathrm{ref}} \right)$ is the spectral line intensity at the reference temperature $T_{\mathrm{ref}}$ and written as
\begin{eqnarray}
S_{\eta \eta'} \left( T_{\mathrm{ref}} \right) &=& \frac{\pi e^2 g_\eta f_{\eta \eta'}}{m_e c} \frac{\exp{\left( - h c E_\eta / k_\mathrm{B} T_{\mathrm{ref}} \right)}}{Q \left( T_{\mathrm{ref}} \right)} \nonumber \\
&& \left[ 1 - \exp{\left( - h c \left( E_{\eta'} - E_{\eta} \right) / k_\mathrm{B} T_{\mathrm{ref}} \right)} \right].
\end{eqnarray}}

\kawashima{The HITRAN2012 database provides the values of $E_\eta$, $E_{\eta'}$, $S_{\eta \eta'} \left( T_{\mathrm{ref}} \right)$, and $Q \left( T_{\mathrm{ref}} \right)$, where $T_{\mathrm{ref}} = 296 \kelvin$.
We calculate $Q \left( T \right)$ with the total internal partition sums (TIPS) code \citep{2003JQSRT..82..401F} in the HITRAN database.
This code calculates $Q \left( T \right)$ for given temperature (the temperature range is 70-3000~K) and molecular species in the HITRAN database.}

\kawashima{We consider the air-broadened pressure-shift in the following way. The shifted spectral line transition wavenumber $\nu_{\eta \eta'}^*$ can be written as
\begin{equation}
\nu_{\eta \eta'}^* = \nu_{\eta \eta'} + \delta \left( P_{\mathrm{ref}} \right) P,
\end{equation}
where $\delta \left( P_{\mathrm{ref}} \right)$ is the air-broadened pressure shift, provided that the shift, $\delta \left( P_{\mathrm{ref}} \right)$, is small relative to $\nu_{\eta \eta'}$.
Here, $P_{\mathrm{ref}}$ is the reference pressure.
The HITRAN2012 database provides the values of $\delta \left( P_{\mathrm{ref}} \right)$, which we use in calculating the line absorption cross sections.}

\kawashima{As for line broadening, we consider pressure broadening and Doppler broadening.
The line profile for pressure broadening is given by the Lorentz profile \citep{RefWorks:10},
\begin{equation}
f_{\mathrm{L}} \left( \nu - \nu_{\eta \eta'} \right) = \frac{\Gamma_{\mathrm{P}}}{\pi \left[ \left( \nu - \nu_{\eta \eta'} \right)^2 + {\Gamma_{\mathrm{P}}}^2 \right]},
\end{equation}
where $\Gamma_{\mathrm{P}}$ is the line half width of the pressure broadening.
On the other hand, the line profile for Doppler broadening is given by the Gaussian profile \citep{RefWorks:10},
\begin{equation}
f_{\mathrm{D}} \left( \nu - \nu_{\eta \eta'} \right) = \frac{1}{\Delta \nu_{\mathrm{D}} \pi^{1/2}} \exp{\left[ - \frac{\left( \nu - \nu_{\eta \eta'} \right)^2}{{\Delta \nu_{\mathrm{D}}}^2} \right]},
\end{equation}
where $\nu_{\mathrm{D}}$ is the line half width of the Doppler broadening.}

\kawashima{To consider both line profiles, the convolution of the Lorentz and Gaussian profiles, which is called the Voigt profile, is used:
\begin{eqnarray}
f_{\mathrm{V}} \left( \nu - \nu_{\eta \eta'} \right) &=& \int_{- \infty}^{\infty} f_{\mathrm{L}} \left( \nu' - \nu_{\eta \eta'} \right) f_{\mathrm{D}} \left( \nu - \nu' \right) d\nu' \\
&=& \int_{- \infty}^{\infty} \frac{\Gamma_{\mathrm{L}}}{\pi \left[ \left( \nu' - \nu_{\eta \eta'} \right)^2 + {\Gamma_{\mathrm{L}}}^2 \right]} \frac{1}{\Delta \nu_{\mathrm{D}} \pi^{1/2}} \nonumber \\
&& \exp{\left[ - \frac{\left( \nu - \nu' \right)^2}{{\Delta \nu_{\mathrm{D}}}^2} \right]} d\nu' \\
&=& \frac{1}{\Delta \nu_{\mathrm{D}} \pi^{1/2}} \mathcal{H} \left( \frac{\Gamma_{\mathrm{L}}}{\Delta \nu_{\mathrm{D}}}, \frac{\nu - \nu'}{\Delta \nu_{\mathrm{D}}} \right),
\end{eqnarray}
where  $\mathcal{H} \left( a, y \right)$ is called the Voigt function and defined as
\begin{equation}
\mathcal{H} \left( a, y \right) \equiv \frac{a}{\pi} \int_{- \infty}^{\infty} \frac{e^{-x^2}}{\left( y - x \right)^2 + a^2} dx.
\end{equation}
For the calculation of the Voigt function, we use the polynomial expansion of this function \citep{1997JQSRT..57..819K, 2004JQSRT..86..231R}. We adopt any cut-off in the line wings.
}

\kawashima{In the HITRAN2012 database, the line half width of the pressure broadening is calculated as
\begin{eqnarray}
\Gamma_{\mathrm{P}} \left(P, T \right) &=& \left( \frac{T_{\mathrm{ref}}}{T} \right)^n \nonumber \\
&& \left[ \Gamma_{\mathrm{air}} \left( P_{\mathrm{ref}}, T_{\mathrm{ref}} \right) \left( P - P_s \right) + \Gamma_{\mathrm{self}} \left( P_{\mathrm{ref}}, T_{\mathrm{ref}} \right) P_{\mathrm{s}} \right],
\end{eqnarray}
where $\Gamma_{\mathrm{air}}$ and $\Gamma_{\mathrm{self}}$ are, respectively, the air-broadened halfwidth and the self-broadened halfwidth at half maximum (HWHM) at $T_{\mathrm{ref}} = 296$~K and $P_{\mathrm{ref}} = 1$~atm and $P_s$ is the partial pressure.
The line half width of the Doppler broadening is given by
\begin{equation}
\Delta \nu_{\mathrm{D}} = \nu \left( \frac{2 k T}{m c^2} \right)^{1/2},
\end{equation}
where $m$ is the mass of the molecule \citep{RefWorks:10}.}

We also consider the Rayleigh scattering by those molecules except $\mathrm{OH}$ and the collision-induced absorption by $\hydrogen$-$\hydrogen$ and $\hydrogen$-$\helium$.
We have confirmed that the Rayleigh scattering by OH is negligible for the total extinction by all the molecules because of its low abundance in the atmosphere.
The Rayleigh scattering cross section is given by \citep{RefWorks:12}
\begin{equation}
\sigma_{\mathrm{Rayleigh}} = \frac{128 \pi^5}{3 \lambda^4} \alpha^2,
\end{equation}
where $\alpha$ is the polarizability.
We use the value of the polarizability for each molecule from CRC Handbook of CHEMISTRY and PHYSICS \citep{RefWorks:4}.
The collision-induced absorption cross sections are taken from HITRAN2012 \citep{2013JQSRT.130....4R}.

\subsection{Transmission Spectrum Model} \label{method_transmission}
\kawashima{We model transmission spectra following \citet{2001ApJ...553.1006B}.}
The transit depth at wavelength $\lambda$, $D \left( \lambda \right)$, can be defined as
\begin{equation}
D \left( \lambda \right) = \frac{L_\mathrm{s} \left( \lambda \right) - L_\mathrm{obs} \left( \lambda \right)}{L_\mathrm{s} \left( \lambda \right)}.
\label{eq_transit-depth0}
\end{equation}
Here, $L_\mathrm{s}$ is the disk-integrated \kawashima{luminosity} from the host star given by
\begin{equation}
L_\mathrm{s} \left( \lambda \right) = \int_0^{R_\mathrm{s}} F_\mathrm{s} \left( \lambda \right) \, 2 \pi r dr,
\end{equation}
where $R_\mathrm{s}$ and $F_\mathrm{s}$ are the stellar radius and flux, respectively, 
and $r$ is the impact parameter measured from the disk center.
$L_\mathrm{obs}$ is the disk-integrated \kawashima{luminosity} of the host star during transit.
Here, we assume that the incident stellar light rays are parallel and thus $F_\mathrm{s}$ is constant through the stellar disk, 
because the orbital distances of planets of interest are much larger (by a factor of 10-100) than the host star's radius. 
With this assumption, $L_\mathrm{obs}$ is expressed as
\begin{equation}
L_\mathrm{obs} \left( \lambda \right) = \int_0^{R_\mathrm{s}} F_\mathrm{s} \left( \lambda \right) \mathrm{e}^{- \tau \left( r, \lambda \right)} \, 2 \pi r dr,
\label{eq_luminosity-obs}
\end{equation}
where $\tau \left( r, \lambda \right)$ is the so-called chord optical depth defined by
\begin{equation}
\tau \left( r, \lambda \right) = 2 \int^{\infty}_0 \sum_{i = 1}^{\mathcal{N}} \sigma_i \left( r, s, \lambda \right) N_i \left( r, s \right) ds.
\end{equation}
Here, $\sigma_i$ and $N_i$ are the extinction cross section and number density of species $i$, $\mathcal{N}$ is the number of species whose extinction is considered, and $ds$ is the line element along the line of sight.

In this study, we assume that all the parts inside the sphere of radius $R_0$ are optically thick enough to block the incident stellar light completely. 
The radius $R_0$ may be defined as that of a solid surface or an optically thick cloud deck in the atmosphere, if present. 
However, some exoplanets may have no such well-defined boundary.
Even if there is such a boundary, its radius is unknown in advance.
According to our numerical results, $\tau$ is sufficiently larger than unity below the pressure level of 10~$\pbar$ in the atmosphere considered in this study. 
Thus, we define $R_0$ as the radial distance from the planetary center at which the pressure is 10~bar. 

With the above assumption and from Eqs.~(\ref{eq_transit-depth0}) to (\ref{eq_luminosity-obs}), 
the transit depth $D \left( \lambda \right)$ can be written as
\begin{equation}
D \left( \lambda \right) = \frac{ {R_0}^2 + \int^{{\rstar}^2}_{{R_0}^2} \left[ 1 - \mathrm{e}^{- \tau \left( r, \lambda \right)}\right] {dr}^2}{{\rstar}^2}.
\end{equation}
The so-called transit radius, $\rtransit \left( \lambda \right)$, is defined as
\begin{equation}
\rtransit \left( \lambda \right) \equiv \rstar \sqrt{D \left( \lambda \right)}.
\end{equation}

\subsection{Calculation Procedure and Model Parameters} \label{method_procedure}
Finally, we summarize the calculation procedure 
and the model parameters and their values that we use in our simulations.

First, we derive the vertical profiles of volume mixing ratios of the gaseous species, $f_i$, from the photochemical calculations ($\S$~\ref{method_photo}). 
Then, from the sum of $f_\mathrm{HCN}$ and $f_\mathrm{C_2H_2}$, which corresponds to the distribution of the haze precursors, 
we simulate the particle growth and calculate the number density distribution of each haze volume $n(v_i, z)$ ($\S$~\ref{method_growth}).
After that, with the obtained size and number density distributions of haze particles and the vertical distribution of the gaseous species, 
we model transmission spectrum of the atmosphere ($\S$~\ref{method_transmission}) with calculations of opacities of gaseous species and haze particles ($\S$~\ref{method_opacity}).
The \kawashima{opacity and} transit depth is calculated every wavenumber grid with width of 0.1~$\mathrm{cm}^{-1}$.

In this study, we model the transmission spectra assuming the properties of the super-Earth GJ~1214b.
Among super-Earths found so far, 
the atmosphere of GJ~1214b has been probed most by transit observations at multiple wavelengths. 
The model parameters and their values we use are listed in Table~\ref{tab:parameters}.
We will explore dependence of results on model parameters other than monomer production rate such as metallicity, \kawashima{C/O ratio, }eddy diffusion coefficient, atmospheric temperature profile, and monomer size in our forthcoming papers.

\kawashima{We adopt the value of the radius at the 1000-bar pressure level 
\ikomar{(simply called the 1000-bar radius, hereafter)} 
as 2.07}~$\rearth$, which is \kawashima{74}\% of the planet radius reported by \cite{2013AA...551A..48A}; 
We have found that this value of \kawashima{\ikomar{the 1000-bar} radius} 
can roughly match the observed transit radii of GJ~1214b when we assume a \kawashima{clear} solar composition atmosphere.
\kawashima{Note that when we infer the molecular abundance from observational transmission spectrum, \ikomar{we suffer from} degeneracy \ikomar{among} the reference radius, 
\ikomar{1000-bar} radius, 
and inferred molecular abundance \citep[see][]{2017MNRAS.470.2972H}.}

\kawashima{
For the temperature-pressure profile, we use the analytical formula of \citet{2010A&A...520A..27G}, because its smooth and simple function 
\ikomar{suits} computationally-heavy photochemical calculations. 
With Eq.~(29) of \citet{2010A&A...520A..27G}, we calculate the temperature-pressure profile averaging over the whole planetary surface (i.e., $f = 1/4$ in the equation).
We choose the parameters, namely, \ikomar{the intrinsic temperature} $T_\mathrm{int}$, \ikomar{equilibrium temperature} $T_\mathrm{irr}$, \ikomar{averaged opacity in the optical} $k_\mathrm{v}$, and \ikomar{averaged opacity in the infrared} $k_\mathrm{th}$, so as to match the temperature-pressure profile of GJ~1214b that \cite{2010ApJ...716L..74M} derived for a solar composition atmosphere \ikomar{under the assumption of} efficient 
\ikomar{heat} redistribution 
\ikomar{from the day and night sides}.
This yieds $T_\mathrm{int} = 120$~K, $T_\mathrm{irr} = 790$~K, $k_\mathrm{v} = 10^{-4.0}$~$\mathrm{g}$~$\mathrm{cm^{-2}}$, and $k_\mathrm{th} = 10^{-2.6}$~$\mathrm{g}$~$\mathrm{cm^{-2}}$.
We have confirmed that \ikomar{our} profile agrees with that of \cite{2010ApJ...716L..74M} within 86~K for the grids we adopt.}
We adopt the value of eddy diffusion coefficient $K_{zz}$ as $1 \times 10^7$~$\mathrm{cm^2}$~$\mathrm{s^{-1}}$.
\kawashima{We will explore 
\ikomar{the sensitivity of transmission spectrum to} eddy diffusion coefficient in our forthcoming papers.}
As for the elemental abundance ratios, we assume that of the solar system abundance, which we take from Table 2 of \citet{2003ApJ...591.1220L}\kawashima{, \ikomar{corresponding} to C/O, O/H, and N/H \ikomar{of} $5.010 \times 10^{-1}$, $5.812 \times 10^{-4}$, and $8.021 \times 10^{-5}$, respectively.}

As for the stellar spectrum used in the photochemical model, we use that of GJ~1214 \kawashima{constructed} by the MUSCLES Treasury Survey \citep{2016ApJ...820...89F, Youngblood:2016ib, 2016ApJ...824..102L}, the wavelength coverage of which is from 0.55~nm to 5500~nm.
\kawashima{The spectrum for X-rays is constructed from Chandra/XMM-Newton and APEC models \citep{2001ApJ...556L..91S}, that for EUV from empirical scaling relation based on Ly$\alpha$ flux \citep{2014ApJ...780...61L}, that for Ly$\alpha$ from model fit to line wings \citep{Youngblood:2016ib}, and that for visible--IR from synthetic photospheric spectra from PHOENIX atmosphere models \citep{2013A&A...553A...6H}.}
We use the version 1.1 of the panchromatic SED binned to a constant 1~{\AA} resolution and downsampled in low signal-to-noise regions to avoid negative flux, the data of which is taken from the MUSCLES team's website\footnote{https://archive.stsci.edu/prepds/muscles/}. \kawashima{We adopt 1 angstrom as the spectral resolution we use.}
The Ly$\alpha$ flux of GJ~1214, which is located at 14.6~pc far away from the Sun, was observed as $1.3 \substack{+ 1.4 \\ - 0.5} \times 10^{-14}$~erg~$\mathrm{cm^{-2}}$~$\mathrm{s^{-1}}$ at the Earth \citep{Youngblood:2016ib}.
From this value, we calculate the Ly$\alpha$ flux at the planet's orbit as $3.3 \times 10^{13}$~photons~$\mathrm{cm^{-2}}$~$\mathrm{s^{-1}}$ using the value of GJ~1214b's semi-major axis, 0.0148~AU \citep{2013AA...551A..48A}, and the Ly$\alpha$ wavelength of 121.6~nm. 
Note that when considering the effects of the mass production rate of haze monomers (i.e., the Ly$\alpha$ flux), we vary the intensities \kawashima{of the actinic flux} at all the wavelengths according to the Ly$\alpha$ intensity.

As for the monomer radius $s_1$, we adopt $1 \times 10^{-3}$~$\mu$m.
We prepare 40 volume bins, 
setting the volume ratio of two adjacent bins to be 3 \citep{2010Icar..210..832L}, 
and cover from $1 \times 10^{-3}$~$\mu$m (monomer size) to 1600~$\mu$m.
As for the value of haze particle internal density $\rho_p$, we adopt $1.0$~g~$\mathrm{cm^{-3}}$ , which is adopted by most of the particle growth models for hydrocarbon hazes in Titan's atmosphere \citep[e.g.][]{1992Icar...95...24T, 2010Icar..210..832L}.

In the photochemical calculations, the atmosphere is vertically divided into \kawashima{165} layers with thickness of \kawashima{45}~km, placing the lower boundary pressure at 10\kawashima{00}~bar.
This thickness is sufficiently smaller relative even to the \kawashima{minimum} atmospheric scale hight \kawashima{in the atmosphere}, which is \kawashima{177}~km.
In the case of the particle growth model, we consider the pressure range from 10~bar to $10^{-10}$~bar with 200 same thickness layers.

Simplified version of our transmission spectrum models are used for WASP-80b in \cite{2014ApJ...790..108F} and for HAT-P-14b in \cite{Fukui:2016ky}.
In the spectrum models of \cite{2014ApJ...790..108F}, we ignored the photochemical reactions and regarded the particle size, particle number density, altitude and thickness of the haze layer as input parameters.
In \cite{Fukui:2016ky}, we did not consider the effects of haze on the spectra.

\begin{deluxetable*}{llll}
\tablecaption{Model parameters and their values used in the simulations \label{tab:parameters}}
\tablehead{
\colhead{Parameter} & \colhead{Description} & \colhead{Value} & \colhead{Reference} \\
}
\startdata
$\rstar$ & Host star radius & 0.201~$\rsun$ & \citet{2013AA...551A..48A} \\
$\mplanet$ & Planet mass & $6.26$~$\mearth$ & \citet{2013AA...551A..48A} \\
$R_{1000}$~$_\mathrm{bar}$ & 1000-bar radius & \kawashima{2.07}~$\rearth$ & \\
$K_{zz}$ & Eddy diffusion coefficient & $1.00 \times 10^7$~$\mathrm{cm^2}$~$\mathrm{s^{-1}}$ & \\
$s_1$ & Monomer radius & $1.00 \times 10^{-3}$~$\mu$m & \\
$\rho_p$ & Particle internal density & $1.00$~g~$\mathrm{cm^{-3}}$ & \\
$I_\mathrm{Ly\alpha}$ & Ly$\alpha$ flux at the planet's orbit & $3.30 \times 10^{13}$~photons~$\mathrm{cm^{-2}}$~$\mathrm{s^{-1}}$ & \cite{Youngblood:2016ib} \\
\enddata
\end{deluxetable*}

\section{Results} \label{sec:result}
In this section, we show results of our numerical simulations. First, we investigate the fiducial monomer production case (i.e., $\beta = 1$) in \S~\ref{result_photo}-\ref{result_transmission} and then explore the dependence on the monomer production rate by changing $\beta$ in \S~\ref{result_dependence}.
\subsection{Photochemical Calculations} \label{result_photo}
First we outline the photochemistry of the atmosphere.
Although the results we show below are basically the same as those from the previous studies, we show them because they are helpful in interpreting our later results. We note that our photochemical models of GJ~1214b's atmosphere are the first ones that use the observed GJ~1214's UV spectrum \citep{2016ApJ...820...89F, Youngblood:2016ib, 2016ApJ...824..102L}.

\begin{figure}[ht!]
\includegraphics[width=0.48\textwidth]{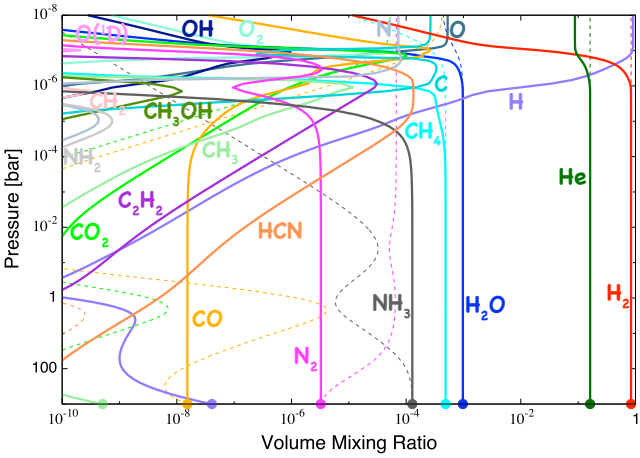}
\caption{Vertical distributions of gaseous species in the photochemical equilibrium atmosphere. Filled circles represent the thermochemical equilibrium values at the lower boundary. The thermochemical equilibrium abundances are shown with dashed lines for reference. Note that the eddy diffusion transport is not included in the thermochemical equilibrium calculations.
\label{fig_photo}}
\end{figure}

Figure~\ref{fig_photo} shows the calculated vertical distributions of gaseous species in the photochemical equilibrium state (solid lines).
We also present the distributions obtained by thermochemical equilibrium calculations (dashed lines) that ignore photochemical processes and eddy diffusion.
In the lower atmosphere ($P \gtrsim 10^{-4}$~bar), the eddy diffusion mixing, which tends to smooth out compositional gradients, is found to yield constant abundances of $\mathrm{H_2O}$, $\mathrm{CH_4}$, $\mathrm{NH_3}$, $\mathrm{N_2}$\kawashimar{, and CO} equal to the lower boundary values.
In the upper atmosphere ($P \lesssim 10^{-4}$~bar), it turns out that many species (i.e., $\mathrm{H}$, $\mathrm{O}$, $\mathrm{C}$, $\mathrm{HCN}$, $\mathrm{N}$, $\mathrm{O_2}$, $\mathrm{C_2H_2}$, $\mathrm{CH_3}$, $\mathrm{OH}$, $\mathrm{CH_3OH}$, $\mathrm{NH_2}$, \kawashimar{$\mathrm{CH_2}$, and $\mathrm{O (^1D)}$}) that are quite rare in thermochemical equilibrium states are produced photochemically and H is the most abundant species.
\kawashima{\ikomar{The} H is \ikomar{known to act as a} reactive radical in reducing atmospheres \citep{2012ApJ...761..166H}.}

\begin{figure*}[ht!]
\gridline{\fig{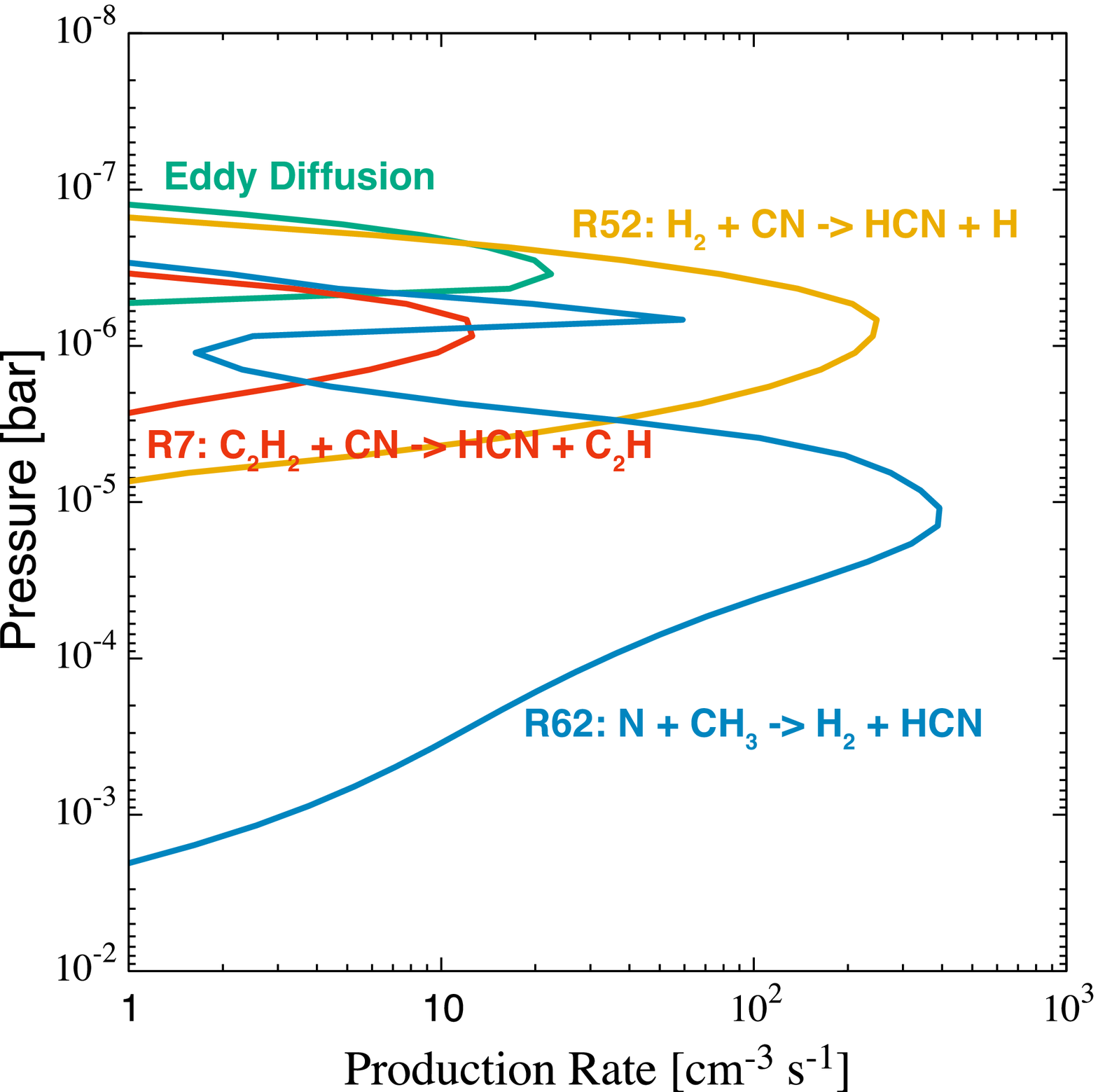}{0.4\textwidth}{(a) Production Rate of HCN}\fig{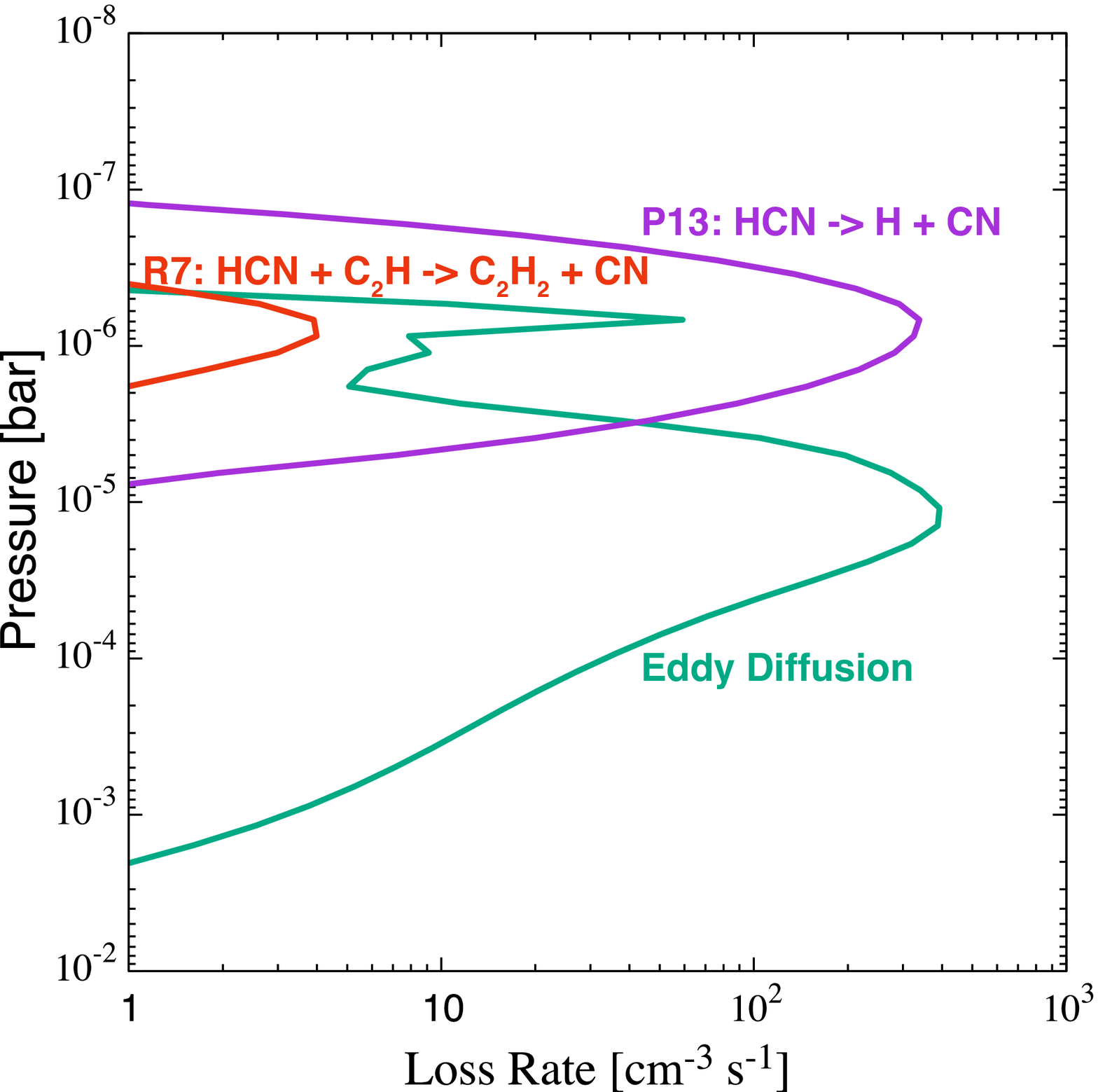}{0.4\textwidth}{(b) Loss Rate of HCN}}
\caption{Distributions of (a) the production and (b) loss rates of HCN due to thermochemical and photochemical reactions, and transport by eddy diffusion for the steady-state distribution of HCN.}
\label{fig_rate_HCN}
\end{figure*}

As for the haze precursors, HCN and $\mathrm{C_2H_2}$, $f_\mathrm{HCN}$ is always greater than $f_\mathrm{C_2H_2}$. This means that in our simulations, the profile of the production rate of monomers is determined mainly by that of $f_\mathrm{HCN}$.
The ratio $f_\mathrm{HCN}$ is constant in the pressure range of $1 \times 10^{-6}$~bar to $1 \times 10^{-5}$~bar because HCN is the most stable N-bearing species in this range.

\kawashima{\ikomar{The details of} the production and loss mechanisms of HCN and $\mathrm{C_2H_2}$ \ikomar{was discussed in} \citet{2011ApJ...737...15M} for the cases of two hot Jupiters, HD 189733b and HD 209458b.
\ikomar{Nevertheless, below, we also explore how the steady-state abundances of HCN and $\rm C_2H_2$ are maintained,} since the atmospheric temperature considered in this study is lower than HD 189733b ($T_\mathrm{eq} = 1100$~K) and HD 209458b ($T_\mathrm{eq} = 1316$~K\footnote{http://www.openexoplanetcatalogue.com}).}

In Figure~\ref{fig_rate_HCN}, we plot the distributions of the production and loss rates of HCN due to thermochemical and photochemical reactions, and transport by eddy diffusion for the steady-state distribution of HCN.
In the pressure range of $1 \times 10^{-7}$~bar to $3 \times 10^{-6}$~bar, the steady-state is maintained almost by the production process via the thermochemical reaction,
\begin{equation}
\mathrm{R52:\;} \mathrm{H_2} + \mathrm{CN} \rightarrow \mathrm{HCN} + \mathrm{H}, \nonumber
\end{equation}
and the loss process via photodissociation, 
\begin{equation}
\mathrm{P13:\;} \mathrm{HCN} \rightarrow \mathrm{H} + \mathrm{CN}. \nonumber
\end{equation}
On the other hand, in the pressure range of $3 \times 10^{-6}$~bar to $2 \times 10^{-3}$~bar, the steady-state is maintained by a balance between the production process via the thermochemical reaction,
\begin{equation}
\mathrm{R62:\;} \mathrm{N} + \mathrm{CH_3} \rightarrow \mathrm{H_2} + \mathrm{HCN}, \nonumber
\end{equation}
and the loss process via eddy diffusion transport to the upper atmosphere.

\begin{figure*}[ht!]
\gridline{\fig{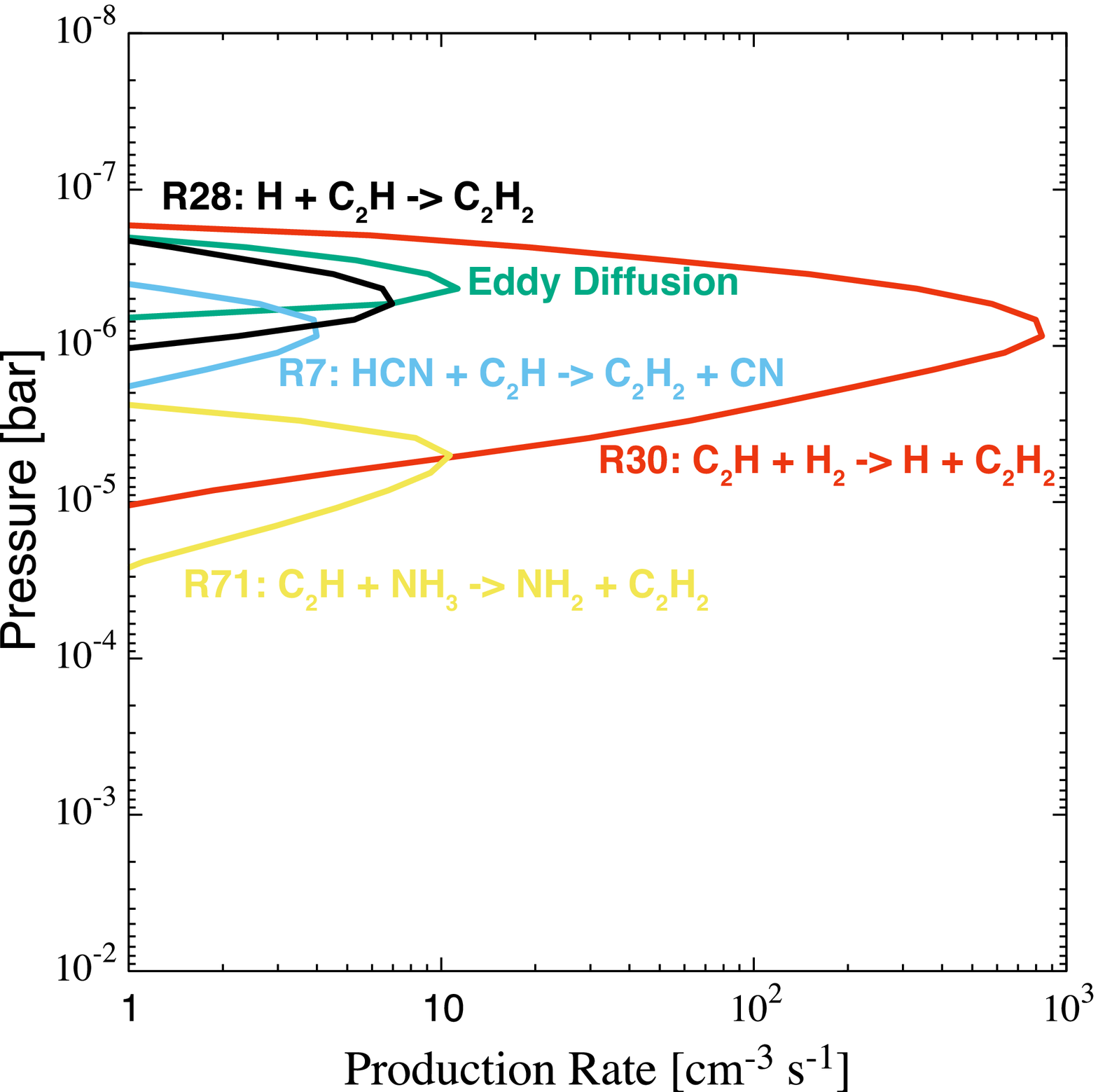}{0.4\textwidth}{(a) Production Rate of $\mathrm{C_2H_2}$}\fig{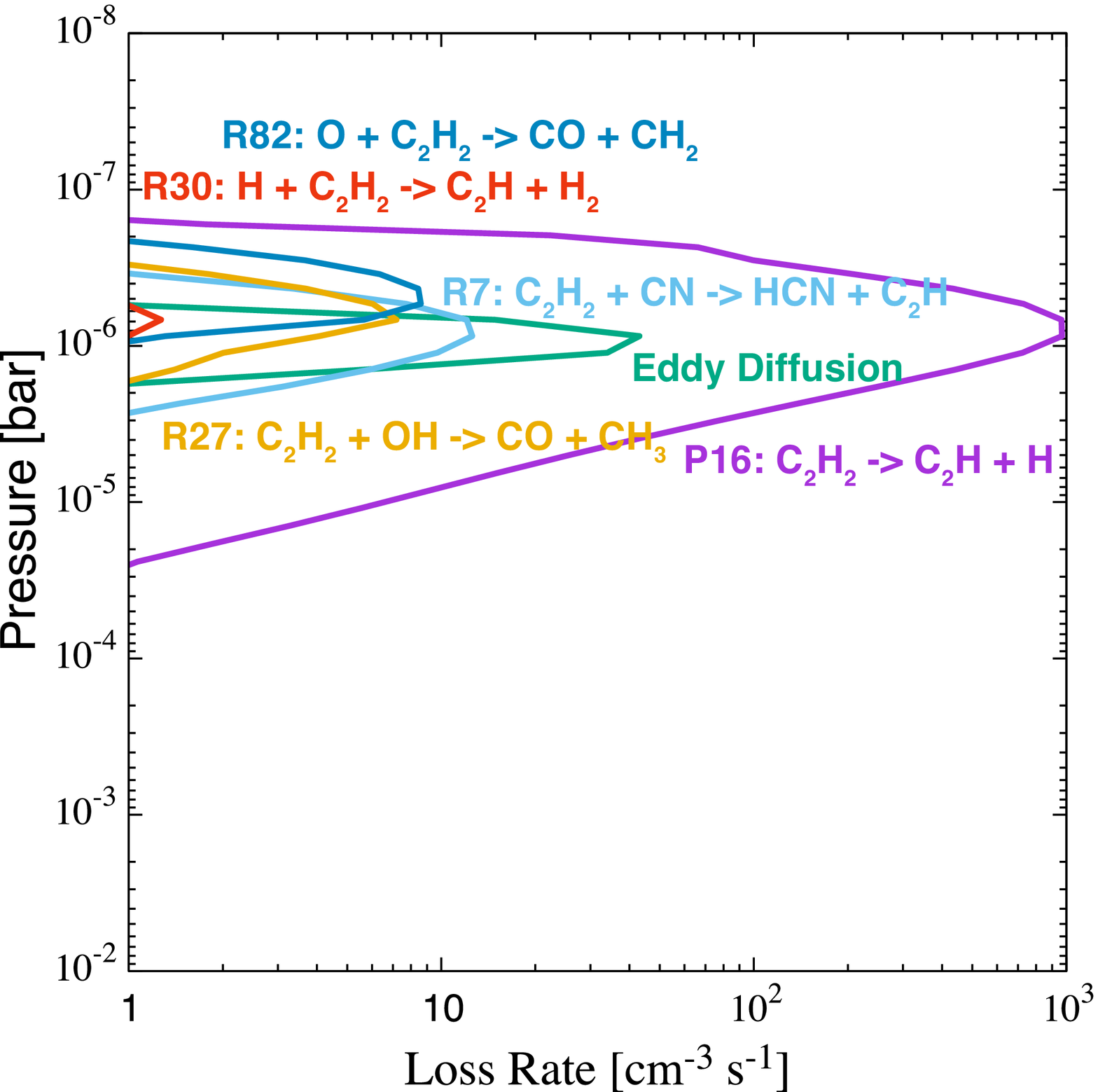}{0.4\textwidth}{(b) Loss Rate of $\mathrm{C_2H_2}$}}
\caption{Distributions of (a) the production and (b) loss rates of $\mathrm{C_2H_2}$ due to thermochemical and photochemical reactions, and transport by eddy diffusion for the steady-state distribution of $\mathrm{C_2H_2}$.}
\label{fig_rate_C2H2}
\end{figure*}

Figure~\ref{fig_rate_C2H2} is the same as Fig.~\ref{fig_rate_HCN} but for $\mathrm{C_2H_2}$.
In the pressure range of $1 \times 10^{-7}$~bar to $5 \times 10^{-6}$~bar, the steady-state is determined by the production process via the thermochemical reaction,
\begin{equation}
\mathrm{R30:\;} \mathrm{C_2H} + \mathrm{H_2} \rightarrow \mathrm{H} + \mathrm{C_2H_2}, \nonumber
\end{equation}
and the loss process via photodissociation, 
\begin{equation}
\mathrm{P16:\;} \mathrm{C_2H_2} \rightarrow \mathrm{C_2H} + \mathrm{H}. \nonumber
\end{equation}
On the other hand, in the pressure range of $5 \times 10^{-6}$~bar to $2 \times 10^{-5}$~bar, the steady-state is determined by production process via the thermochemical reaction,
\begin{equation}
\mathrm{R71:\;} \mathrm{C_2H} + \mathrm{NH_3} \rightarrow \mathrm{NH_2} + \mathrm{C_2H_2}, \nonumber
\end{equation}
and the loss process, via photodissociation, 
\begin{equation}
\mathrm{P16:\;} \mathrm{C_2H_2} \rightarrow \mathrm{C_2H} + \mathrm{H}. \nonumber
\end{equation}

\subsection{Particle Growth Calculations} \label{result_growth}

\begin{figure}[ht!]
\plotone{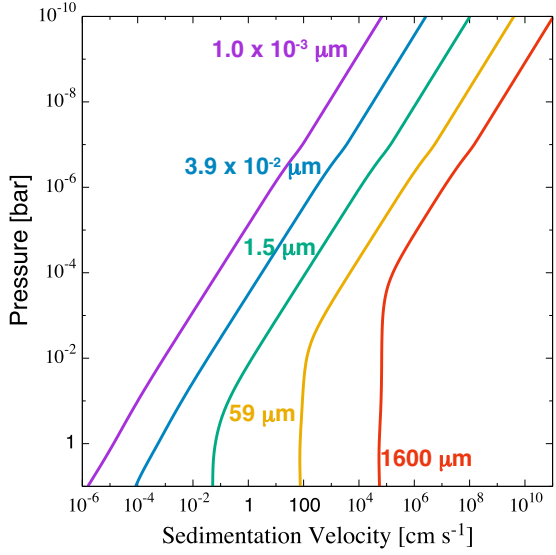}
\caption{Sedimentation velocity for five different particle radii, $1.0 \times 10^{-3}$~$\mu$m (purple line), $3.9 \times 10^{-2}$~$\mu$m (blue line), $1.5$~$\mu$m (green line), $59$~$\mu$m (orange line), and $1600$~$\mu$m (red line) along pressure.}
\label{fig_velocity}
\end{figure}

The growth of haze particles occurs via competition among coagulation, sedimentation, and diffusion.
Knowledge of the sedimentation velocity is therefore helpful in understanding the particle growth.
Figure~\ref{fig_velocity} shows the sedimentation velocity along pressure for five different particle radii, $1.0 \times 10^{-3}$~$\mu$m, $3.9 \times 10^{-2}$~$\mu$m, $1.5$~$\mu$m, $59$~$\mu$m, and $1600$~$\mu$m.
Change of the trend found at $P \sim 10^{-2}$~bar for the 59~$\mu$m particle and $P \sim 10^{-3}$~bar for the 1600~$\mu$m particle, respectively, results from the transition from slip flow ($K_{n, i} = \lambda_a / s_i > 1$) to Stokes flow ($K_{n, i} = \lambda_a / s_i < 1$).
In the slip flow regime, the sedimentation velocity is proportional to the particle radius (see Eqs.~(\ref{eq_sedimentation}) and (\ref{eq_slip})).
On the other hand, in the Stokes flow regime, the sedimentation velocity is proportional to the square of the particle radius (see Eqs.~(\ref{eq_sedimentation}) and (\ref{eq_slip})).

\begin{figure}[ht!]
\plotone{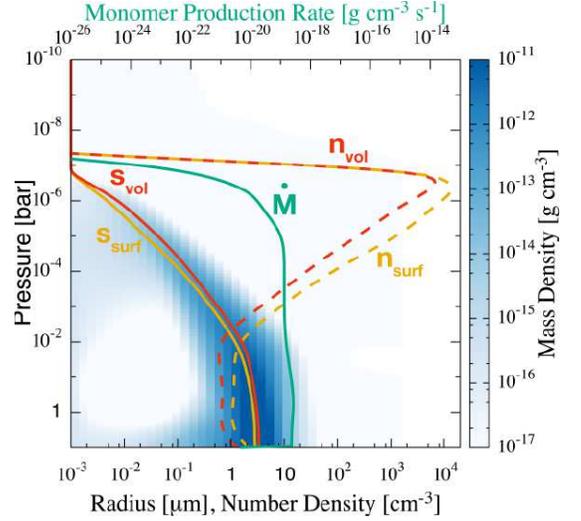}
\caption{Vertical profiles of the surface average radius $s_\mathrm{surf}$ (yellow solid line) and number density $n_\mathrm{surf}$ (yellow dashed line), and the volume average radius $s_\mathrm{vol}$ (red solid line) and number density $n_\mathrm{vol}$ (red dashed line) along with that of the monomer mass production rate (green solid line). See the text for the definition of each quantity. Also, the mass densities for all the size bins at each pressure level are plotted with the blue color contour.}
\label{fig_growth}
\end{figure}

Figure~\ref{fig_growth} shows the vertical profiles of haze properties. Here, we define the surface average radius $s_\mathrm{surf}$ (yellow solid line) as
\begin{equation}
s_\mathrm{surf} = \frac{\sum_{i = 1}^{\mathcal{N}} n \left(s_i \right) s_i^3}{\sum_{i = 1}^{\mathcal{N}} n \left(s_i \right) s_i^2},
\end{equation}
and the volume average radius $s_\mathrm{vol}$ (red solid line) by Eq.~(\ref{eq_s_vol}).
If the two average sizes agree with each other at a certain altitude, the size distribution is unimodal at the altitude.
The surface average number density $n_\mathrm{surf}$ (yellow dashed line) and the volume average number density $n_\mathrm{vol}$ (red dashed line) are calculated as
\begin{equation}
n_\mathrm{surf} = \frac{\sum_{i = 1}^{\mathcal{N}} n \left(s_i \right) s_i^3}{s_\mathrm{surf}^3}
\end{equation}
and
\begin{equation}
n_\mathrm{vol} = \frac{\sum_{i = 1}^{\mathcal{N}} n \left(s_i \right) s_i^3}{s_\mathrm{vol}^3},
\end{equation}
respectively.
Also, the mass densities for all the size bins at each pressure level are plotted with the blue color contour and the vertical profile of the monomer mass production rate is plotted with the green solid line.

From Fig.~\ref{fig_growth}, it is demonstrated that the average radii change dramatically with altitude.
In the upper atmosphere, particles grow little because they settle faster than coagulational growth proceeds.
The number densities become larger as altitude decreases (or the pressure increases) and they take the peak value at $P \sim 10^{-7}$~bar.
Coagulational growth occurs significantly below this pressure level.
As altitude decreases, the average radii increase from $1 \times 10^{-3}$~$\mu$m to 2-3~$\mu$m because of coagulational growth, and the number densities decrease by several orders of magnitude from the peak values.
Again, change of the trend found at $P \sim 10^{-2}$~bar results from the transition from the slip flow to Stokes flow regimes.
A significant increase in the sedimentation velocity due to the regime transition of drag force (see Fig.~\ref{fig_velocity}) inhibits the collision between particles.

The slight difference between $s_\mathrm{surf}$ and $s_\mathrm{vol}$ means that the haze contains different size particles at each altitude.
The color contour indicates that particles in some narrow range of size are abundant at each altitude and the monomer size particles exist broadly below the level of $10^{-7}$~bar because monomer production occurs in this region.

\begin{figure}[ht!]
\plotone{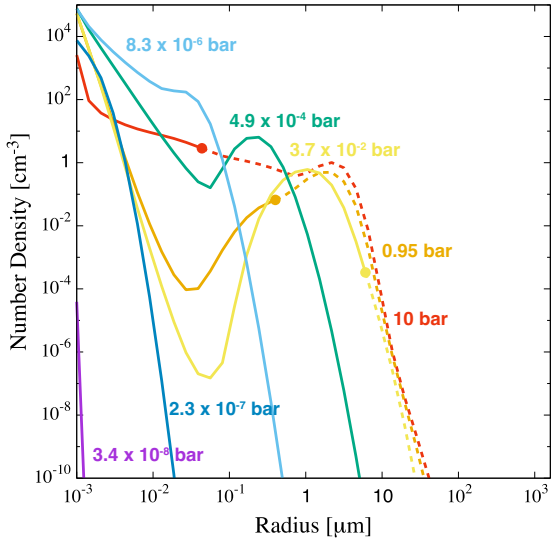}
\caption{Distributions of number density of haze particles for all the size bins at seven different pressure levels, $3.4 \times 10^{-8}$~bar (purple line), $2.3 \times 10^{-7}$~bar (blue line), $8.3 \times 10^{-6}$~bar (light blue line), $4.9 \times 10^{-4}$~bar (green line), $3.7 \times 10^{-2}$~bar (yellow line), 0.95~bar (orange line), and 10~bar (red line). The Stokes regime is indicated by dashed lines, while the slip flow regime is indicated by solid lines; the transition points are marked by filled circles.}
\label{fig_distribution-p_n}
\end{figure}

In Figure~\ref{fig_distribution-p_n}, we plot the distributions of number density of haze particles for all the size bins at seven different pressure levels, $3.4 \times 10^{-8}$~bar, $2.3 \times 10^{-7}$~bar, $8.3 \times 10^{-6}$~bar, $4.9 \times 10^{-4}$~bar, $3.7 \times 10^{-2}$~bar, 0.95~bar, and~10 bar.
First it is found that the number density of monomer size, $10^{-3}$~$\mu$m, is the largest among all the sizes at all the pressure levels because of the large monomer production rate.
At low pressures of $P \lesssim 10^{-5}$~bar, the coagulation due to brownian diffusion is the dominant process, whereas that due to gravitational collection hardly occurs.
On the other hand, at high pressures of $P \gtrsim 10^{-5}$~bar, both coagulation mechanisms contribute to the particle growth.
The coagulation due to gravitational collection makes a second peak of number density for the pressure levels higher than $8.3 \times 10^{-6}$~bar, because it occurs in a runaway fashion much more rapidly compared to that due to brownian diffusion.

The change of size distribution can be understood as follows: 
The particles grow through the frequent collisions with the abundant small particles. The collision timescale $\tau_\mathrm{coll}$ between a large particle and monomer size particles can be written as $\tau_\mathrm{coll} = \left( n_1 \sigma \Delta v \right)^{-1}$, where $n_1$ is the number density of monomers, $\sigma$ is the collision cross section of the large particle, and $\Delta v$ is the relative velocity between the particles.
The relative velocity due to sedimentation is proportional to particle radius $s$ in the slip flow regime and $s^2$ in the Stokes flow regime (see Eqs.~(\ref{eq_sedimentation}) and (\ref{eq_slip})), while the relative velocity due to brownian diffusion is proportional to $s^{-\frac{3}{2}}$ (see Eq.~(\ref{eq_vth})).
Thus, $\tau_\mathrm{coll} \propto s^{-3}$ (slip flow) and $\propto s^{-4}$ (Stokes flow) for gravitational collection, while $\tau_\mathrm{coll} \propto s^{-1/2}$ for brownian diffusion. This means the particle growth is always a runaway process: The larger the particle, the faster the growth proceeds. Also, the gravitational collection is much faster than the brownian diffusion especially for large size particles.
Therefore, from $P \sim 10^{-5}$~bar on, the second peak grows rapidly and a valley-shaped distribution develops (see yellow and orange lines), because gravitational collection contributes predominantly to the particle growth above this pressure.

At $P = 10$~bar, however, the valley is found to disappear.
This is because the drag law for the large particles shifts from the slip flow regime to the Stokes regime.
In Fig~\ref{fig_distribution-p_n}, the Stokes regime is indicated by dashed lines, while the slip flow regime is indicated by solid lines; the transition points are marked by filled circles.
Since the sedimentation velocity is so high in the Stokes regime (see Fig.~\ref{fig_velocity}) that the particles settle faster than they grow, the largest-size group ($\gtrsim$ 2~$\mu$m) stops growing (see the orange lines).
Then, small particles, which are still in the slip flow regime, grow and are catching up with the largest particles.

\begin{figure}[ht!]
\plotone{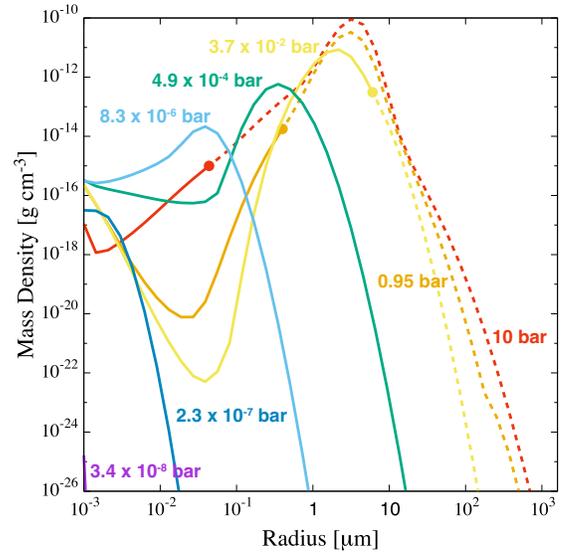}
\caption{Same as Fig.~\ref{fig_distribution-p_n} but the distribution of mass density.}
\label{fig_distribution-p_rho}
\end{figure}

In Figure~\ref{fig_distribution-p_rho}, we plot the distributions of mass density for all the size bins at the same set of seven different pressure levels as shown in Fig.~\ref{fig_distribution-p_n}.
It can be noticed that there are dominant sizes that account for most of the total haze mass for all the seven pressure levels.
And the dominant size becomes larger, as pressure increases, because of the coagulational growth.

\subsection{Transmission Spectrum Models} \label{result_transmission}
\begin{figure}[ht!]
\includegraphics[width = 0.48\textwidth]{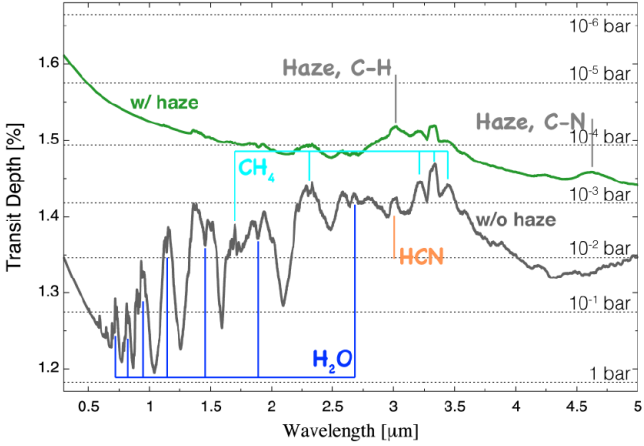}
\caption{Transmission spectrum models for the atmosphere with haze (green line) and without haze (black line). Horizontal dotted lines represent the transit depths corresponding to the pressure levels from $1 \times 10^{-6}$~bar to $1$~bar for the atmosphere without haze.
Note that the transmission spectrum models are smoothed for clarity by averaging over the nearest 633 wavenumber points, namely 63.2~$\mathrm{cm}^{-1}$, for each point.}
\label{fig_spectra}
\end{figure}

Figure~\ref{fig_spectra} shows the transmission spectrum models for the atmosphere with haze (green line) and without haze (black line).
\kawashima{\ikomar{Also,} 
\ikomar{the relative cross section of the planetary disk with radius} 
corresponding to a certain pressure level, 
\ikomar{which is defined as}}
\begin{equation}
\kawashimabf{D_P = \frac{R_P^2}{R_s^2},}
\label{DP}
\end{equation}
\ikomar{is presented by h}orizontal dotted lines from \ikomar{$P =$} $1 \times 10^{-6}$~bar to $1$~bar for the atmosphere without haze. 
\ikomar{In equation~(\ref{DP}), $R_P$ and $R_s$ are the radius at the pressure level $P$ and the stellar radius, respectively.}
Roughly at these pressure levels, there exist the molecules accountable for the spectral features. We have confirmed that the chord optical depth at the pressure that corresponds to the transit radius is between 0.1 and 1, depending on wavelength.
Note that the transmission spectrum models are smoothed for clarity by averaging over the nearest 633 wavenumber points, namely 63.2~$\mathrm{cm}^{-1}$, for each point. We use the same smoothing method for the results of spectrum models hereafter.

In the spectrum model for the atmosphere without haze (black line), several characteristic spectral features can be seen.
For example, prominent features of $\mathrm{H_2O}$ are found around $\lambda =$ 0.7~$\mu$m, 0.8~$\mu$m, 0.9~$\mu$m, 1.2~$\mu$m, 1.3-1.6~$\mu$m, 1.9~$\mu$m, and 2.5-3.0~$\mu$m, those of $\methane$ around 1.7~$\mu$m, 2.2-2.4~$\mu$m, and 3.3~$\mu$m, and that of $\mathrm{HCN}$ around 3.0~$\mu$m. 
The Rayleigh scattering feature mainly due to $\mathrm{H_2}$ can be seen in the optical wavelength region.

The spectrum for the atmosphere with haze (green line) is relatively featureless, compared to that for the atmosphere without haze (black line).
This is because the haze particles in the upper atmosphere ($P \lesssim 10^{-4}$~bar) makes the atmosphere optically thick and prevent the molecules in the lower atmosphere ($P \gtrsim 10^{-4}$~bar) from showing their absorption features.
However, the small features of $\methane$ above $10^{-4}$~bar can be seen at 2.2-2.4~$\mu$m and 3.3~$\mu$m because of their large extinction cross sections at these wavelengths.
Also, the spectral features due to the C-H and C$\equiv$N bonds of the haze particles appear at 3.0 and 4.6~$\mu$m, respectively.

In the wavelength region of 0.3-1~$\mu$m (green line), the \kawashima{spectral} slope due to \kawashima{Rayleigh scattering by} small ($\lesssim 0.1$~$\mu$m) haze particles in the upper atmosphere ($P \lesssim 10^{-4}$~bar) can be seen. 
Previous studies demonstrated that the existence of two separate cloud layers were needed to explain both the \kawashima{spectral} slope in the optical and the lack of the absorption features in the near-infrared simultaneously; A layer composed of small size ($\lesssim 0.1$~$\mu$m) particles in the upper atmosphere responsible for the \kawashima{spectral} slope \kawashima{due to Rayleigh scattering} and the dense cloud layer that prevents the molecules from showing their absorption features \citep{2014A&A...570A..89E, 2015MNRAS.446.2428S, 2015ApJ...814..102D}.
This study is the first to produce the transmission spectrum that has the \kawashima{spectral} slope, but no distinct molecular absorption features, without assuming such cloud layers, by calculating the distribution of the size and number density of haze particles in the atmosphere directly.

\subsection{Dependence on Monomer Production Rate} \label{result_dependence}
\begin{figure*}[ht!]
\gridline{\fig{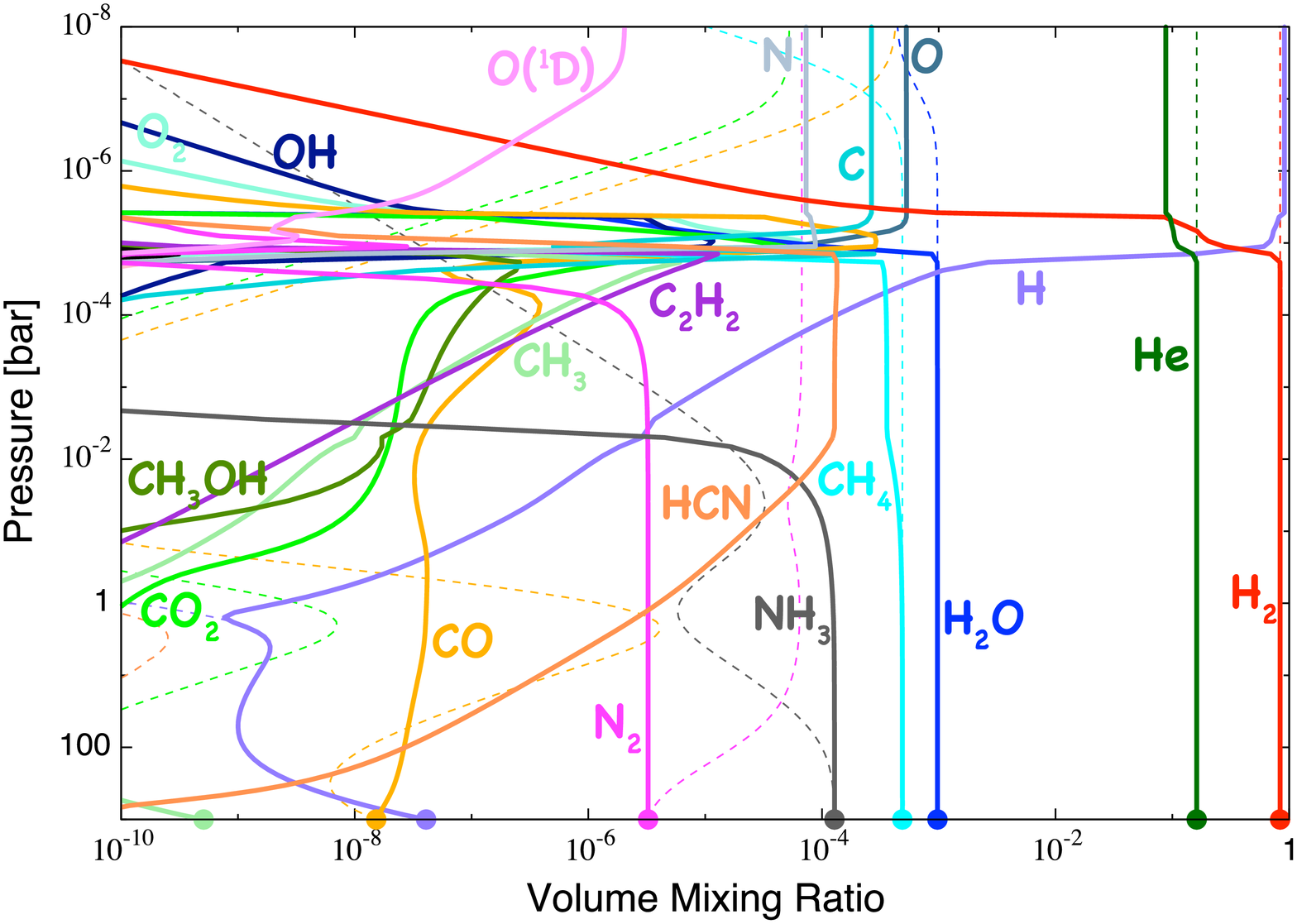}{0.5\textwidth}{(a) $\beta = 10^5$}\fig{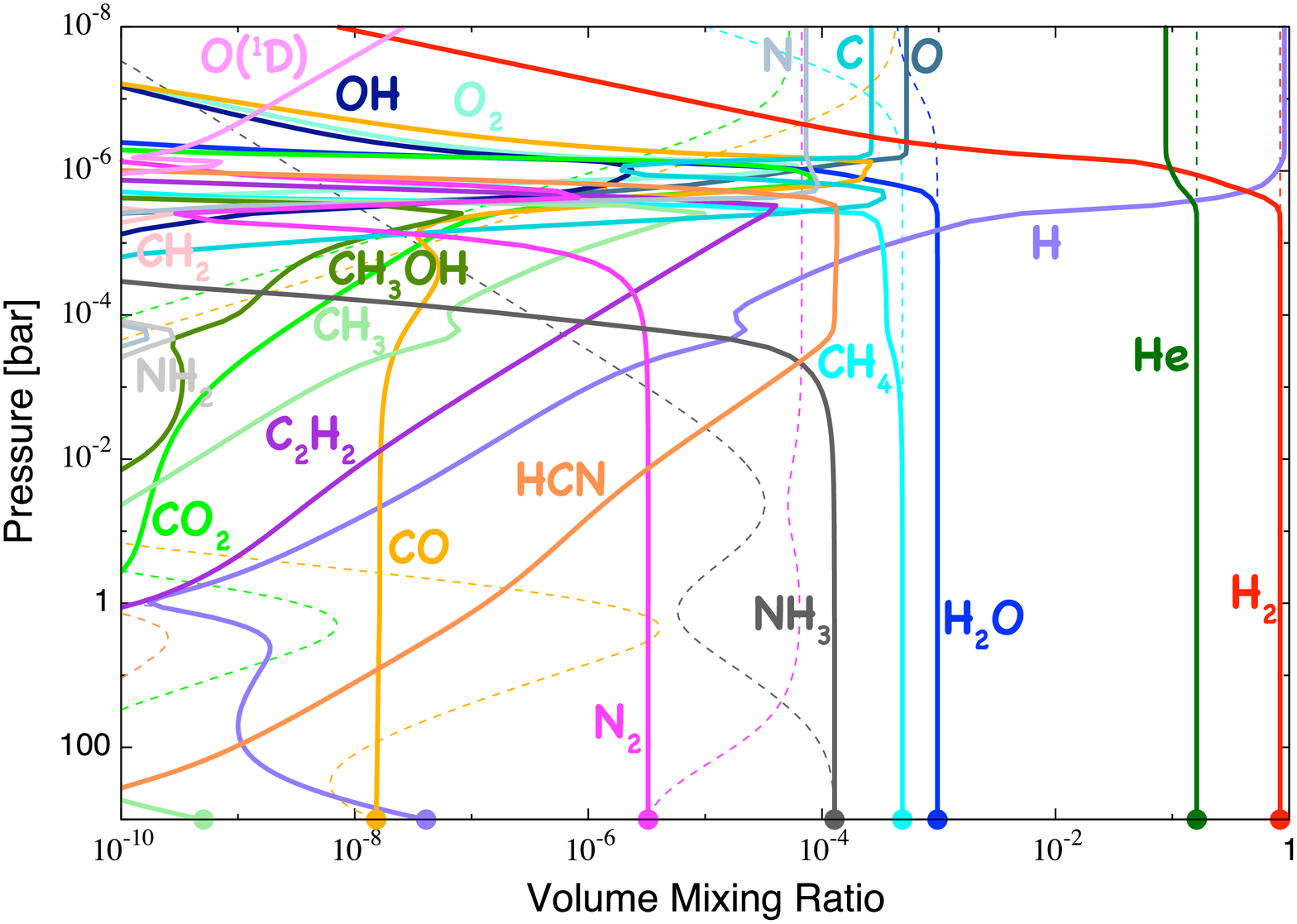}{0.5\textwidth}{(b) $\beta = 10^{2.5}$}}
\gridline{\fig{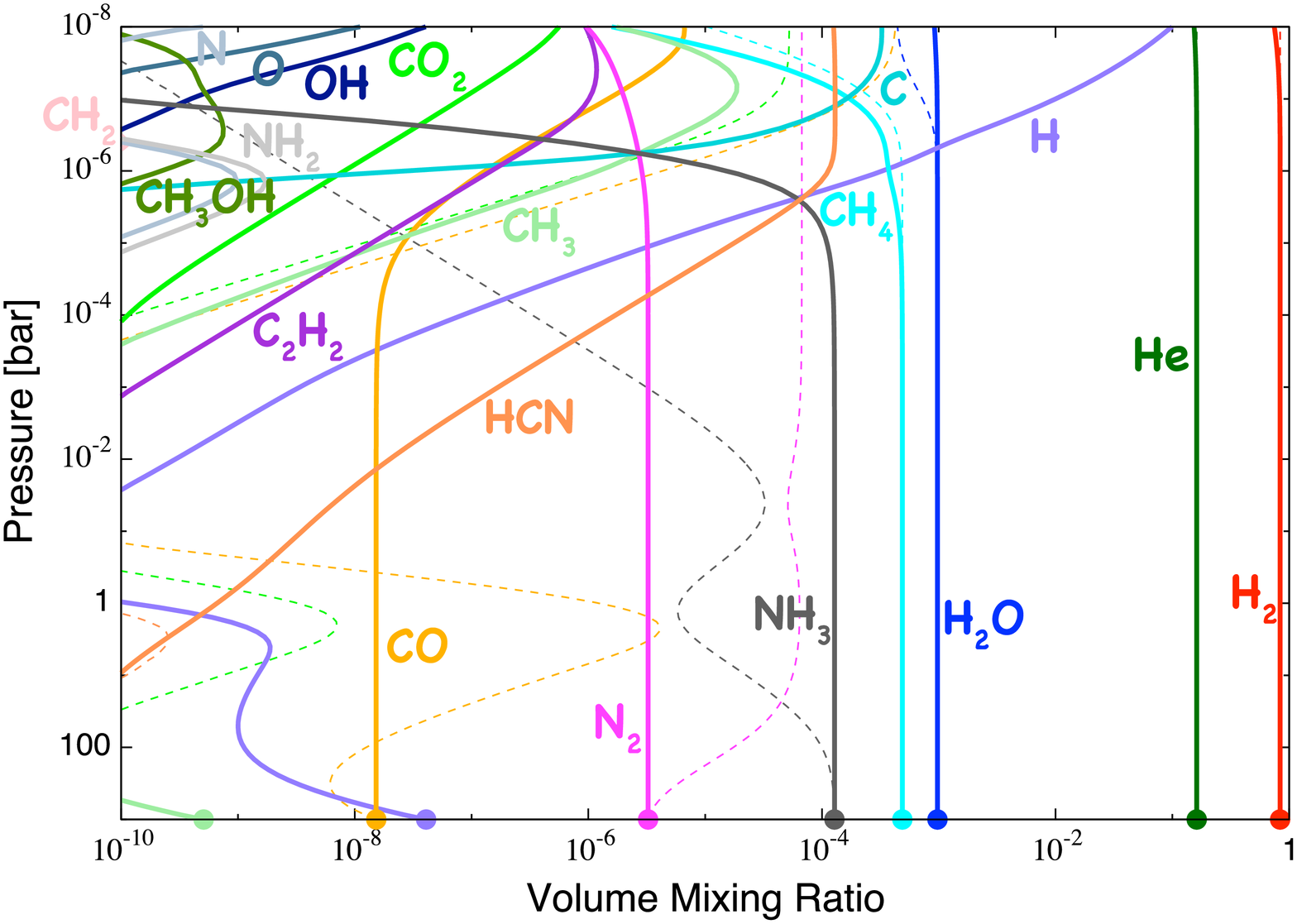}{0.5\textwidth}{(c) $\beta = 10^{-2.5}$}\fig{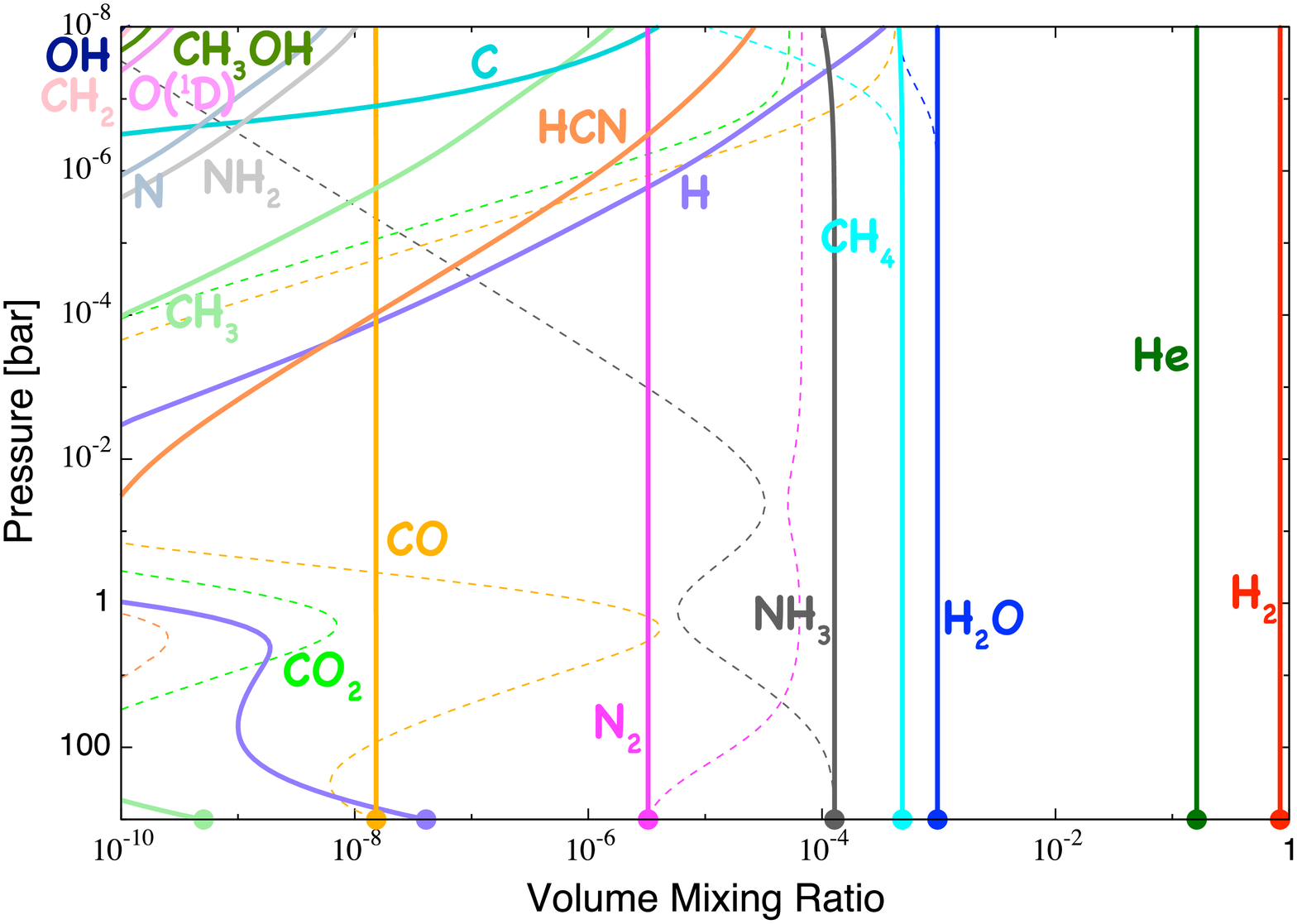}{0.5\textwidth}{(d) $\beta = 10^{-5}$}}
\caption{Same as Fig. \ref{fig_photo} but for haze monomer production parameter $\beta =$ (a) $10^{5}$, (b) $10^{2.5}$, (c) $10^{-2.5}$, and (d) $10^{-5}$. See Eq. (\ref{eq_monomer_total}) for the definition of $\beta$.}
\label{fig_dep_photo}
\end{figure*}

Here, we explore the dependence of the transmission spectrum on monomer production rate $\dot{M}$ by changing the haze monomer production parameter $\beta$ (see Eq.~(\ref{eq_monomer_total})).
As mentioned in \S~\ref{method_procedure}, when we vary the value of $\dot{M}$, we also vary the \kawashima{actinic flux at all the wavelengths} according to the change in the incident stellar Ly$\alpha$ flux.

Figure~\ref{fig_dep_photo} shows the calculated vertical distributions of gaseous species for four different values of $\beta$, (a) $10^5$, (b) $10^{2.5}$, (c) $10^{-2.5}$, and (d) $10^{-5}$, respectively.
We have confirmed the dependence of the molecular vertical distributions on the incident UV flux reported by previous works \citep[e.g.,][]{2014ApJ...780..166M, 2014A&A...562A..51V}, as shown in Fig.~\ref{fig_dep_photo}. 
In the high UV cases ($\beta =$ $10^5$ and $10^{2.5}$), the photodissociation of the molecules such as $\mathrm{H_2}$, $\mathrm{H_2O}$, $\mathrm{CH_4}$, and $\mathrm{NH_3}$ occurs and produces $\mathrm{H}$, $\mathrm{O}$, $\mathrm{C}$, $\mathrm{HCN}$, $\mathrm{N}$, $\mathrm{O_2}$, $\mathrm{C_2H_2}$, $\mathrm{CH_3}$, $\mathrm{OH}$, $\mathrm{O (^1D)}$, and $\mathrm{CH_3OH}$ at deeper levels than in the fiducial case (Fig.~\ref{fig_photo}).
On the other hand, in the low UV cases ($\beta =$ $10^{-2.5}$ and $10^{-5}$), the photodissociation does not occur effectively and the eddy diffusion evens out the abundance of the molecules such as $\mathrm{H_2O}$, $\mathrm{CH_4}$, and $\mathrm{NH_3}$ up to higher altitudes.

As for the haze precursors, HCN is always more abundant than $\mathrm{C_2H_2}$, irrespective of UV flux. \kawashima{Note that assumed values of C/O, O/H, and N/H are $5.010 \times 10^{-1}$, $5.812 \times 10^{-4}$, and $8.021 \times 10^{-5}$, respectively.}
It can be seen that the higher (lower) the incident UV flux is, the lower (higher) the region where the precursors are produced photochemically becomes, because of the effective photodissociation.

\begin{figure*}[ht!]
\gridline{\fig{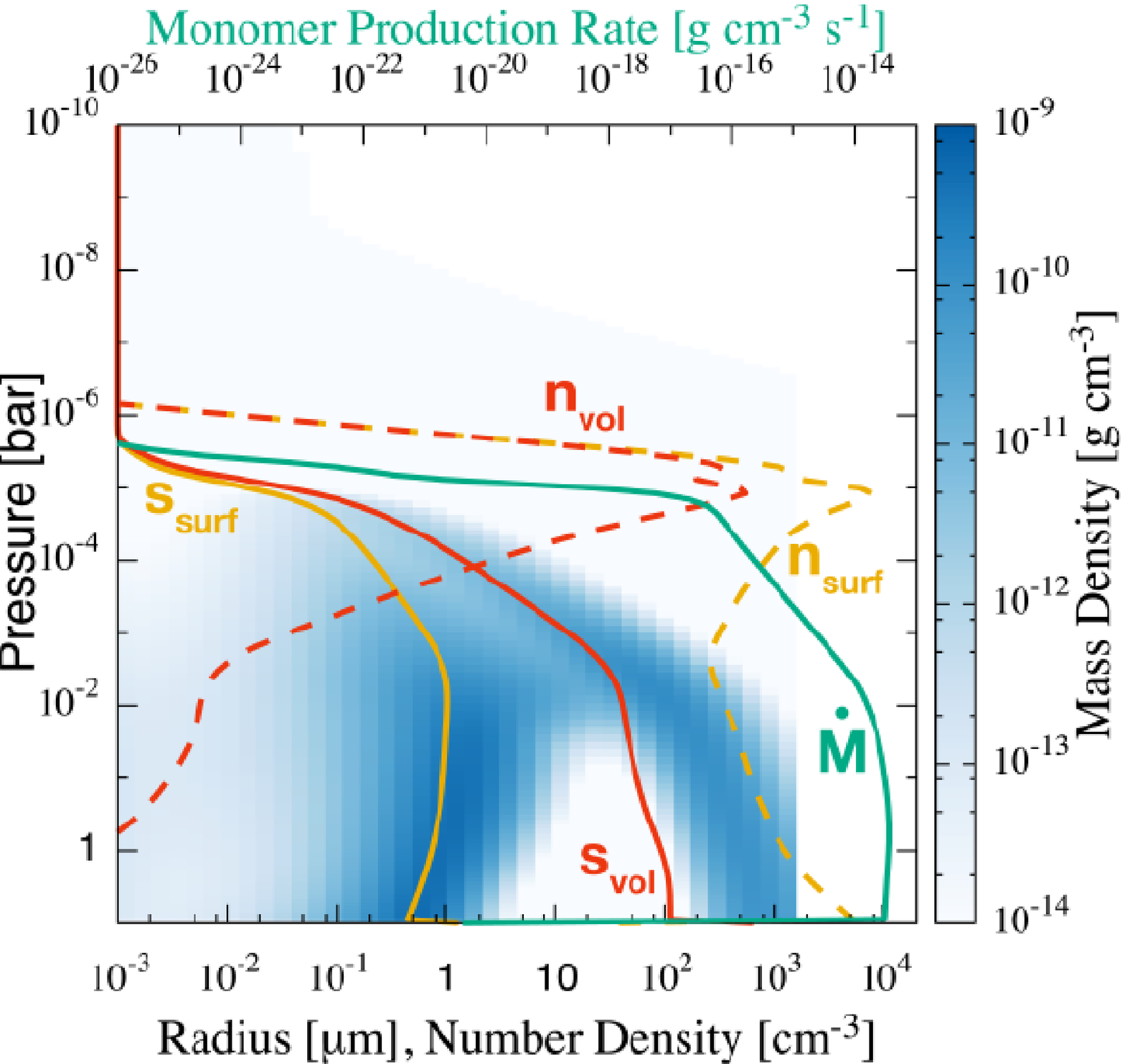}{0.4\textwidth}{(a) $\beta = 10^5$}\fig{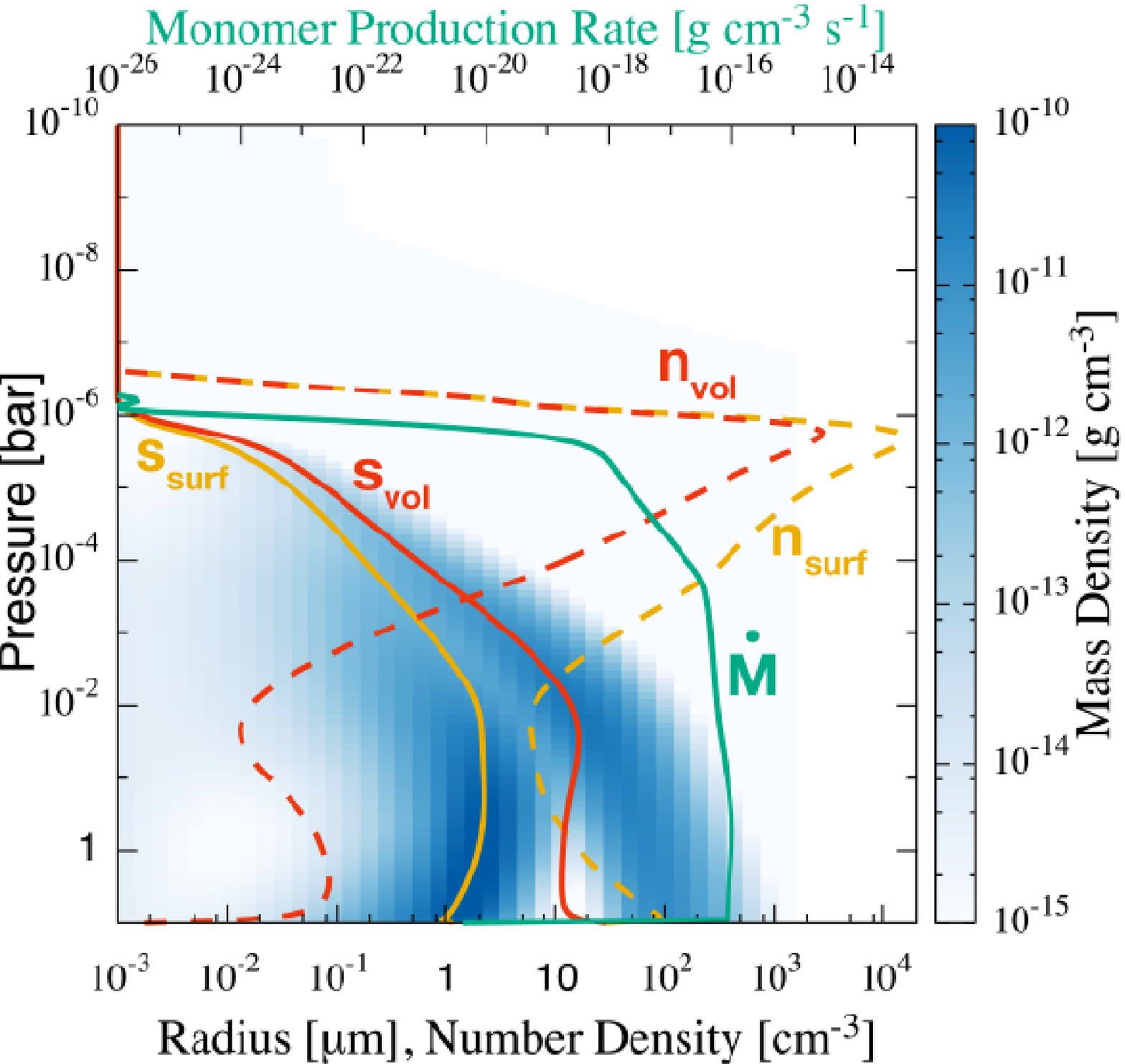}{0.4\textwidth}{(b) $\beta = 10^{2.5}$}}
\gridline{\fig{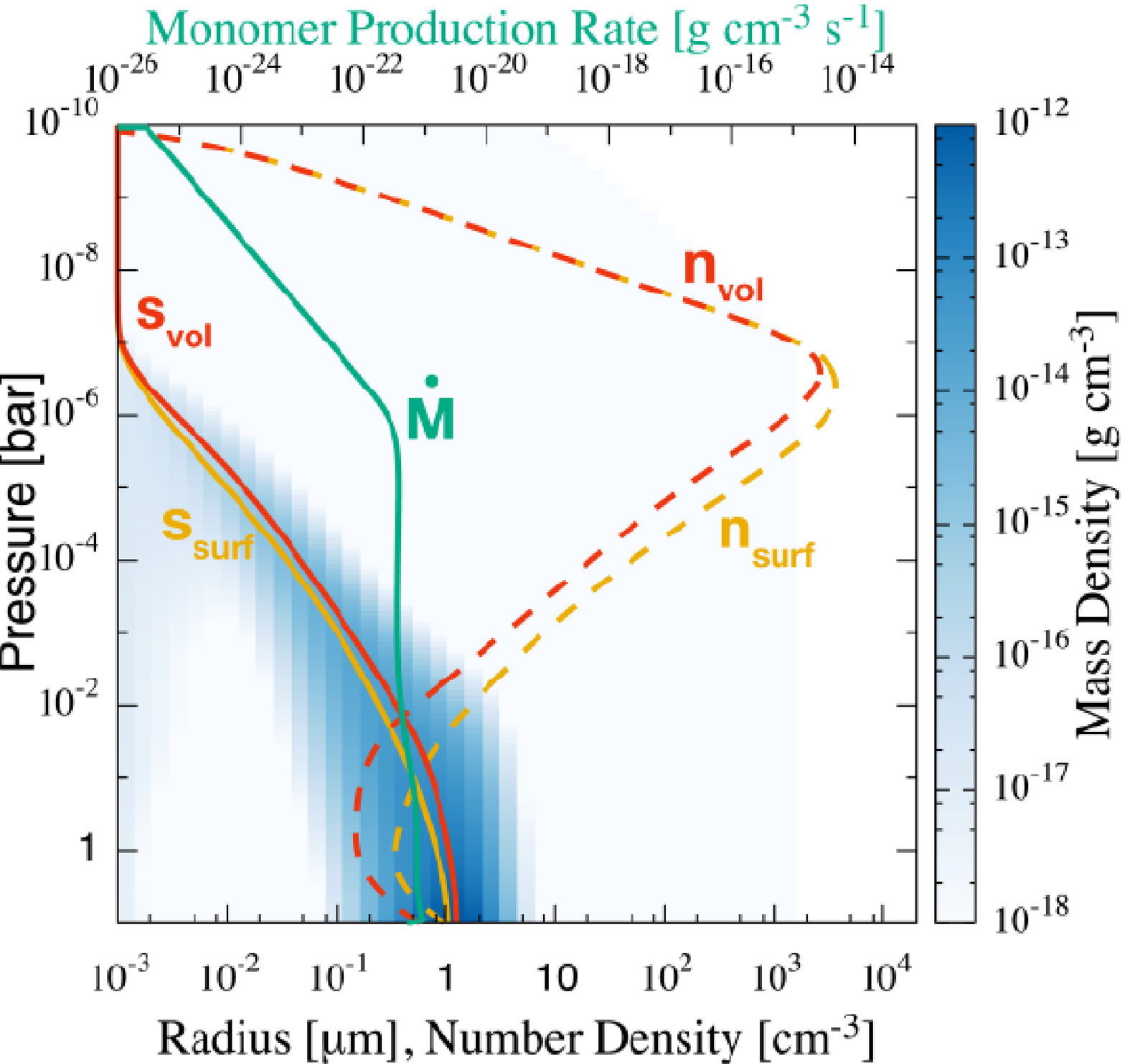}{0.4\textwidth}{(c) $\beta = 10^{-2.5}$}\fig{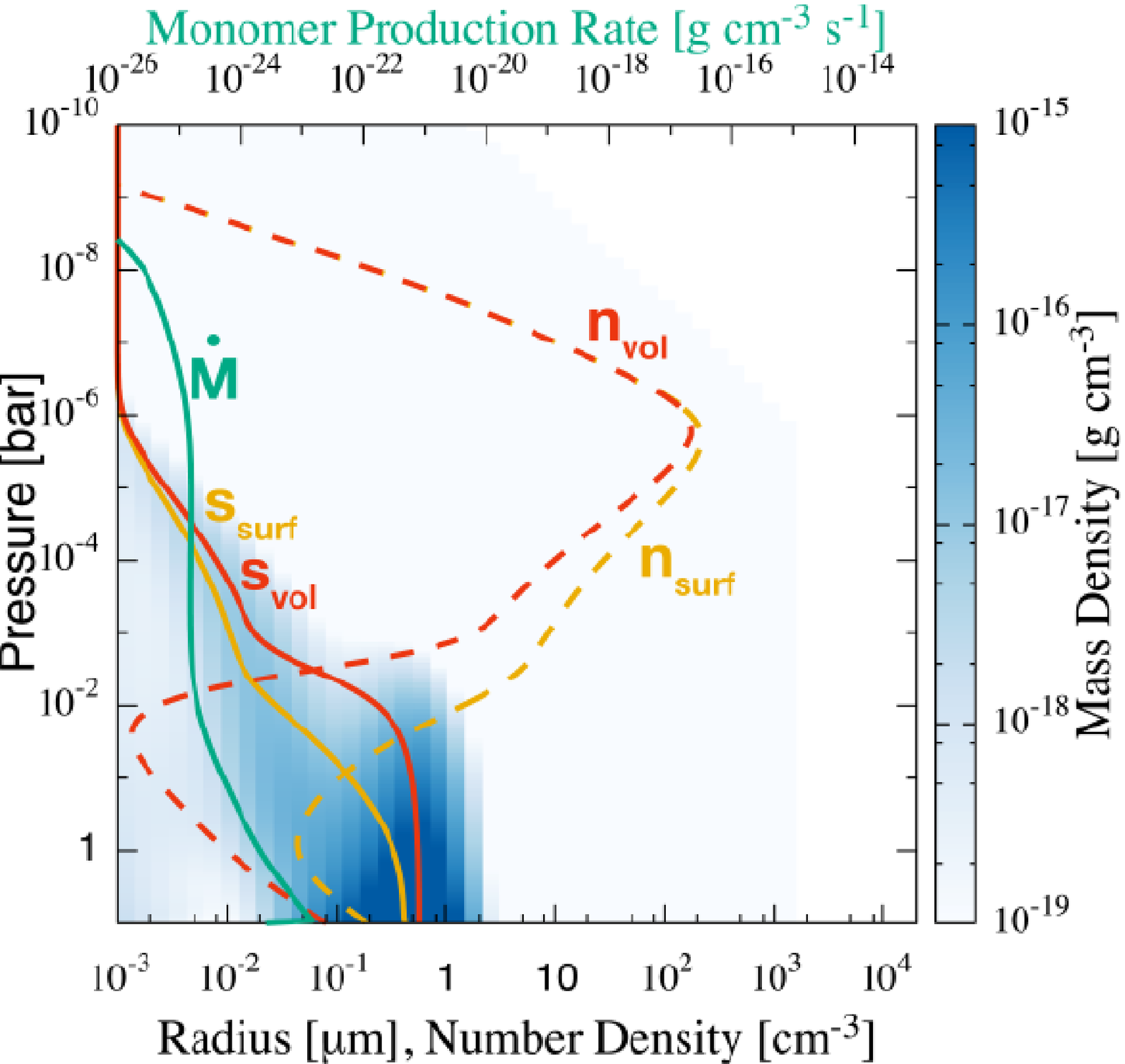}{0.4\textwidth}{(d) $\beta = 10^{-5}$}}
\caption{Same as Fig.~\ref{fig_growth} but for $\beta =$ (a) $10^{5}$, (b) $10^{2.5}$, (c) $10^{-2.5}$, and (d) $10^{-5}$. See Eq.~(\ref{eq_monomer_total}) for the definition of $\beta$.}
\label{fig_dep_growth}
\end{figure*}

Figure~\ref{fig_dep_growth} shows the vertical profiles of the surface average radius $s_\mathrm{surf}$ (yellow solid line) and number density $n_\mathrm{surf}$ (yellow dashed line), and the volume average radius $s_\mathrm{vol}$ (red solid line) and number density $n_\mathrm{vol}$ (red dashed line) along with that of the monomer mass production rate $\dot{M}$ (green solid line) for four different values of $\beta$, (a) $10^5$, (b) $10^{2.5}$, (c) $10^{-2.5}$, and (d) $10^{-5}$.
The mass densities for all the size bins at each pressure level are also plotted with the blue color contour.
The average radii are found to depend on the value of $\beta$ dramatically: $s_\mathrm{vol}$ becomes as large as $10^3$~$\mu$m in the case of $\beta = 10^5$, while it grows only to less than 1~$\mu$m in the case of $\beta = 10^{-5}$ at the lower boundary where the pressure is 10~bar. 
For the high UV cases ($\beta =$ $10^5$, and $10^{2.5}$), the disagreement between $s_\mathrm{surf}$ and $s_\mathrm{vol}$ is significantly larger compared to that in the fiducial case (Fig.~\ref{fig_growth})  and one clearly finds bimodal distributions due to the large monomer production rate, as explained in detail below.

\begin{figure}[ht!]
\plotone{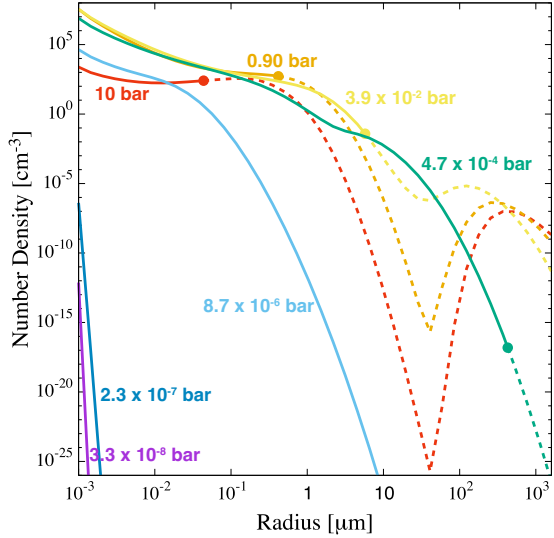}
\caption{Distributions of number density for all the size bins at seven different pressure levels, $3.3 \times 10^{-8}$~bar (purple line), $2.3 \times 10^{-7}$~bar (blue line), $8.7 \times 10^{-6}$~bar (light blue line), $4.7 \times 10^{-4}$~bar (green line), $3.9 \times 10^{-2}$~bar (yellow line), 0.90~bar (orange line), and 10~bar (red line) for the case of $\beta = 10^{5}$. The Stokes regime is indicated by dashed lines, while the slip flow regime is indicated by solid lines; The transition points are marked by filled circles.}
\label{fig_distribution-p_n_uv5}
\end{figure}

In Figure~\ref{fig_distribution-p_n_uv5}, we plot the distributions of number density for all the size bins at seven different pressure levels, $3.3 \times 10^{-8}$~bar, $2.3 \times 10^{-7}$~bar, $8.7 \times 10^{-6}$~bar, $4.7 \times 10^{-4}$~bar, $3.9 \times 10^{-2}$~bar, 0.90~bar, and 10~bar for the case of $\beta = 10^{5}$.
Same as in Fig.~\ref{fig_distribution-p_n}, the slip flow and Stokes regimes are indicated by solid and dashed lines, respectively, and the transition points are marked by filled circles.
First, similarly to the case of $\beta = 1$ (Fig.~\ref{fig_distribution-p_n}), the number density of the monomer size, $10^{-3}$~$\mu$m, is the largest at all the pressure levels, because of the large monomer production rate.
Like in the fiducial case, for $P \lesssim 10^{-5}$~bar, the coagulation due to brownian diffusion is the dominant process, whereas that due to gravitational collection hardly occurs.
On the other hand, at high pressures of $P \gtrsim 10^{-5}$~bar, both coagulation mechanisms contribute to the particle growth.
One finds a bimodal distribution with a wide gap whose center is around $40$~$\mu$m for $3.9 \times 10^{-2}$~bar, 0.90~bar, and 10~bar (note that the vertical range of Fig.~\ref{fig_distribution-p_n_uv5} differs greatly from that of Fig.~\ref{fig_distribution-p_n}).
In contrast to the fiducial case, the particle growth proceeds rapidly as a whole and, then, the large-size particles ($\gtrsim$ 400~$\mu$m) enter to the Stokes regime (see the green line) before development of any peak like ones observed in Fig.~\ref{fig_distribution-p_n}. Thus, the largest-size ($\gtrsim$ 400~$\mu$m) group stops growing and the small particles in the slip flow regime (40~$\mu$m $\lesssim s \lesssim$ 400~$\mu$m) grow and catch up with the largest ($\gtrsim$ 400~$\mu$m) particles in the Stokes regime.
However, in this case, even relatively small ($\lesssim$ 40~$\mu$m) particles are already in the Stokes regime at $3.9 \times 10^{-2}$~bar, $0.90$~bar, and 10~bar.
Thus, the transition points \kawashima{place limits on growth for these relatively small particles}.
The reason why the gap continues to deepen is that smaller particles settle more slowly than larger ones in the Stokes regime.

\begin{figure}[ht!]
\plotone{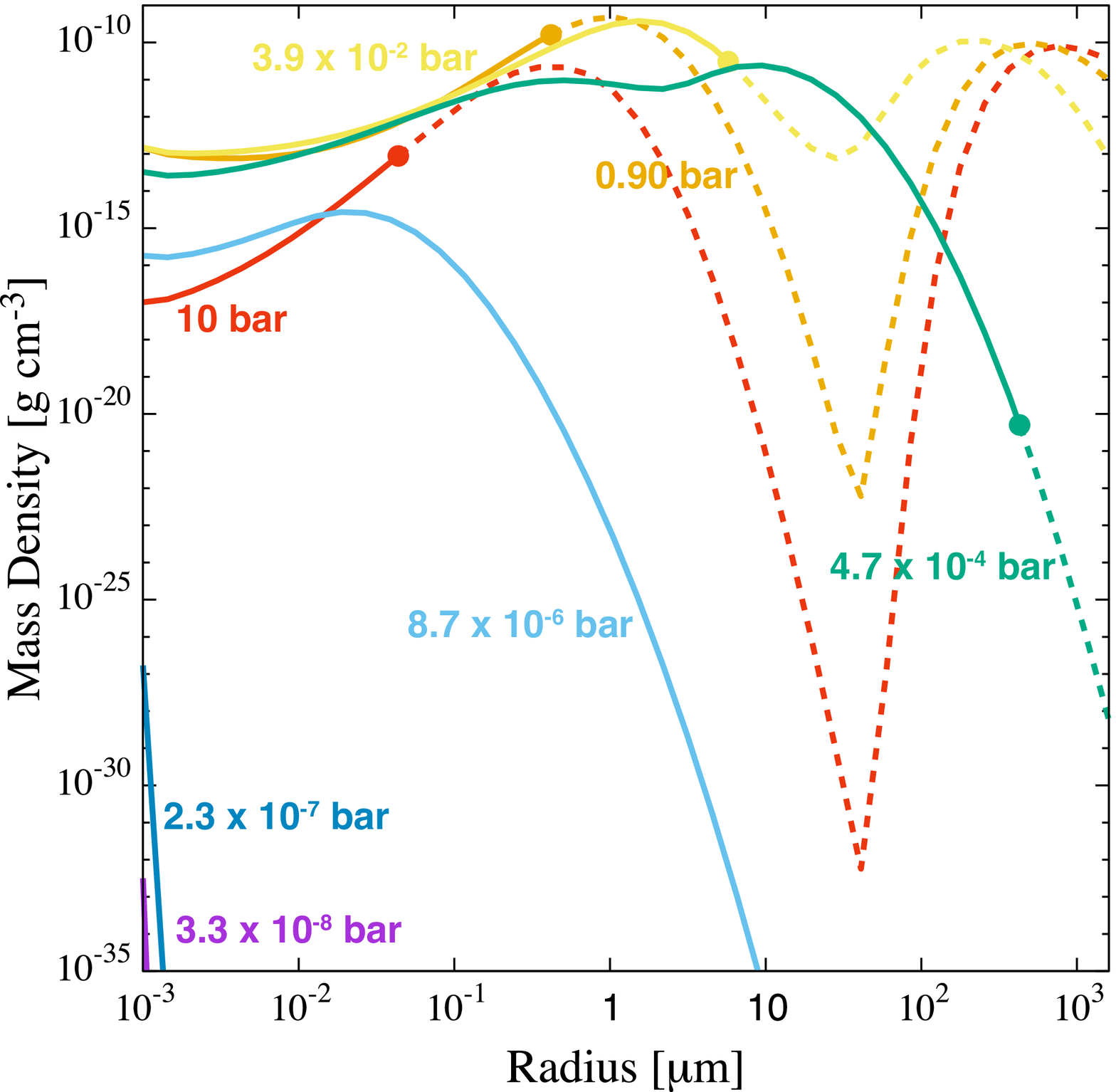}
\caption{Same as Fig.~\ref{fig_distribution-p_n_uv5} but the distribution of mass density.}
\label{fig_distribution-p_rho_uv5}
\end{figure}

In Figure~\ref{fig_distribution-p_rho_uv5}, we plot the distributions of mass density for all the size bins at the same seven different pressure levels as shown in Fig.~\ref{fig_distribution-p_n_uv5} for the case of $\beta = 10^{5}$.
In contrast to the case of $\beta = 1$ (Fig.~\ref{fig_distribution-p_rho}), the distribution is clearly bimodal for the pressure levels, $3.9 \times 10^{-2}$~bar, 0.15~bar, and 10~bar.
The distributions of mass density are qualitatively similar to those of number density (Fig.~\ref{fig_distribution-p_n_uv5}).
The obvious difference is that the two peaks of mass density are comparable in value.

\begin{figure}[ht!]
\includegraphics[width=0.48\textwidth]{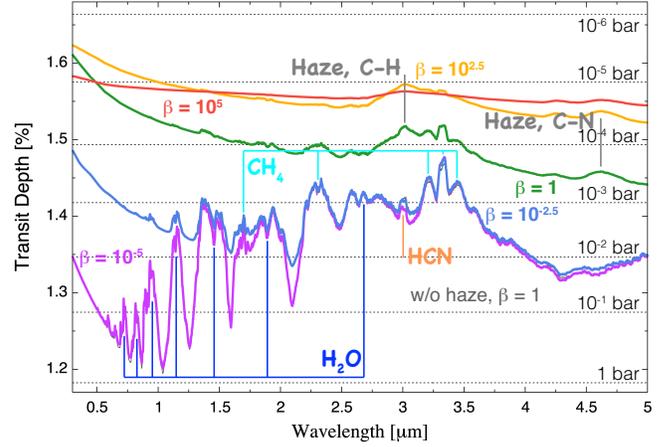}
\caption{Transmission spectrum models for the atmosphere with haze for the five cases where $\beta$ is $10^5$ (red line), $10^{2.5}$ (yellow line), $1$ (green line, same as the green line in Fig. \ref{fig_spectra}), $10^{-2.5}$ (blue line), and $10^{-5}$ (purple line). 
Transmission spectrum model for the atmosphere without haze in the case of $\beta = 1$ (black line) is also plotted, but can be hardly seen as it overlaps with that for the atmosphere with haze in the case of $\beta = 10^{-5}$ (purple line).
Same as Fig.~\ref{fig_spectra}, horizontal dotted lines represent the transit depths corresponding to the pressure levels from $1 \times 10^{-6}$~bar to $1$~bar for the atmosphere without haze in the case of $\beta = 1$.
Note that the transmission spectrum models are smoothed for clarity.}
\label{fig_dep_spectra}
\end{figure}

Figure~\ref{fig_dep_spectra} shows the transmission spectrum models for the atmosphere with haze for the five cases where $\beta$ is $10^5$ (red line), $10^{2.5}$ (yellow line), $1$ (green line, same as the green line in Fig. \ref{fig_spectra}), $10^{-2.5}$ (blue line), and $10^{-5}$ (purple line).
Transmission spectrum model for the atmosphere without haze in the case of $\beta = 1$ (black line) is also plotted, but can be hardly seen as it overlaps with that for the atmosphere with haze for $\beta = 10^{-5}$ (purple line).
Similarly to Fig.~\ref{fig_spectra}, the horizontal dotted lines represent the transit depths corresponding to the pressure levels from $1 \times 10^{-6}$~bar to $1$~bar for the atmosphere in the case of $\beta = 1$.
From this figure, we can see that the transmission spectrum varies with the value of $\beta$ significantly.
In the case of $\beta = 10^{5}$ (red line), the overall spectrum is rather flat. This is because the floating haze particles at high altitudes ($P \sim 10^{-5}$~bar) make the atmosphere so optically thick that their absorption obscures spectral absorption features due to the molecules in the lower ($P \gtrsim 10^{-5}$~bar) atmosphere.
Also, it turns out that the bimodal size distribution seen in the range of $P \gtrsim 10^{-5}$ bar (see Fig.~\ref{fig_dep_growth}) hardly affects the resultant transmission spectrum.
In the case of $\beta = 10^{2.5}$ (yellow line), some features of the haze can been seen, which include the \kawashima{spectral} slope \kawashima{due to Rayleigh scattering} in the optical and the absorption features at 3.0~$\mu\mathrm{m}$ and 4.6~$\mu\mathrm{m}$ coming from the vibrational transitions of the C-H and C$\equiv$N bonds, respectively.
As $\beta$ decreases, the overall transit depth becomes lower. This is because the altitude at which the atmosphere becomes optically thick also decreases.
In the case of $\beta = 10^{-5}$ (purple line), the spectrum is almost the same as that of the atmosphere without haze (black line).
In conclusion, these results demonstrate that the difference in monomer production rate, which relates to the UV irradiation intensity from the host star, makes the diversity of transmission spectrum: completely flat spectrum, spectrum with only extinction features of hazes (i.e., \kawashima{spectral} slope \kawashima{due to Rayleigh scattering} and absorption features of hazes), spectrum with slope \kawashima{due to Rayleigh scattering} and some molecular absorption features, and spectrum with only molecular absorption features.

\section{Validity of characteristic size approximation in Particle Growth Calculation} \label{sec:single}
When comparing theoretical transmission spectra of hazy atmospheres with high-precision observational data, the distribution of haze particles has to be determined with multiple-size growth calculations ($\S$~\ref{method_growth}).
To explore the possibility of reducing the computational cost and understand the effect of bimodality on transmission spectra, we examine the validity of characteristic size approximation quantitatively, applying the grain growth model of \cite{2014ApJ...789L..18O}.
The characteristic size approximation assumes that there are particles of a single size and monomers in the atmosphere.
This approximation is validated, at least, in the studies of the dynamics of dust grains in protoplanetary disks \citep{2011ApJ...731...95O} and proto-envelopes of gas giants \citep{2014ApJ...789L..18O}.

We assume that the haze particle size distribution at any altitude $z$ is characterized by a characteristic mass $m^*$, defined as \citep{2014ApJ...789L..18O}
\begin{equation}
m^* \equiv \frac{\int \xi \left( m \right) m \, \mathrm{d}m}{\int \xi \left( m \right) \mathrm{d}m},
\label{eq_m*}
\end{equation}                                                                                                                                     
where $\xi \left( m \right)$ is the distribution function of particles of mass $m$.
The characteristic mass $m^*$ changes by both coagulation of haze particles and production of monomers. The latter effect decreases the value of $m^*$ toward the monomer mass.
In this study, because focusing on the effect of size distribution, we neglect the gravitational collection and eddy diffusion, which are included in our particle growth module developed in $\S$~\ref{method_growth}.
Thus, we assume that coagulation occurs due to the Brownian collision only.
The gravitational collection is important when both small and large particles are abundant.
Thus, as shown in the previous section, this has a significant influence on the vertical profile of haze particles in the case of $\beta = 10^5$.
However, as also shown above because the altitude where gravitational collection becomes important is optically thick enough for transmitted radiation, the exclusion of gravitational collection has a little effect on resultant transmission spectra.
Also, as the particle transport mechanism, we take only gravitational sedimentation into account and ignore eddy diffusion.
While the eddy diffusion affects the vertical profile of haze particles in the lower atmosphere in the case of $\beta = 10^{-5}$ to some extent, we ignore the effect because we want to focus on the effect of size distribution.
The maximum differences in transit depth between spectrum models obtained from the multiple size calculations with and without two effects (gravitational collection and eddy diffusion) in the wavelength range of 0.3-5~$\mu$m are 38, 64, 43, 202, and 85ppm for $\beta = 10^5$, $10^{2.5}$, $1$, $10^{-2.5}$, and $10^{-5}$, respectively.
The relatively large difference for $\beta = 10^{-2.5}$ case comes from the eddy diffusion effect.

\begin{figure}[ht!]
\includegraphics[width=0.48\textwidth]{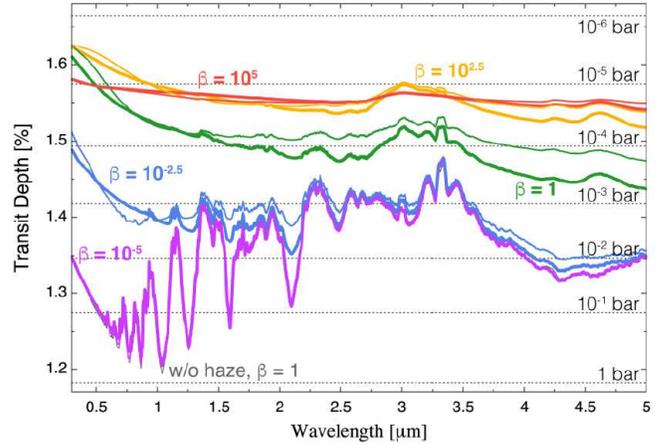}
\caption{Transmission spectrum models for the atmosphere with haze for the five cases where the haze monomer production parameter $\beta$ is $10^5$ (red lines), $10^{2.5}$ (yellow lines), $1$ (green lines), $10^{-2.5}$ (blue lines), and $10^{-5}$ (purple lines).
Models obtained from the multiple size calculations ($\S$~\ref{method_growth}) are shown with thick lines, while those calculated with the characteristic size approximation are plotted with thin lines.
Transmission spectrum model for the atmosphere without haze in the case of $\beta = 1$ (black line) is also plotted.
Same as Fig.~\ref{fig_spectra}, the horizontal dotted lines represent the transit depths corresponding to the pressure levels from $1 \times 10^{-6}$~bar to $1$~bar for the atmosphere without haze in the case of $\beta = 1$.
Note that the transmission spectra are smoothed for clarity.}
\label{fig_single_spectra}
\end{figure}

Figure~\ref{fig_single_spectra} shows the transmission spectrum models for the atmosphere with haze for the five cases where $\beta$ is $10^5$ (red line), $10^{2.5}$ (yellow line), $1$ (green line), $10^{-2.5}$ (blue line), and $10^{-5}$ (purple line).
Models obtained from the multiple size calculations ($\S$~\ref{method_growth}) are shown with thick lines, while those calculated with the characteristic size approximation are plotted with thin lines.
The model for the haze-free atmosphere for $\beta = 1$ (black line) is also plotted.
Same as Fig.~\ref{fig_spectra}, the horizontal dotted lines represent the transit depths corresponding to the pressure levels from $1 \times 10^{-6}$~bar to $1$~bar for the atmosphere without haze in the case of $\beta = 1$.
Again, we ignore the gravitational collection and eddy diffusion also in the multiple-size particle growth calculations to compare the results from those with the characteristic size approximation. 

In the case of $\beta = 10^5$ (red lines), although the size distribution is obviously bimodal in the lower atmosphere (see Figs. \ref{fig_distribution-p_n_uv5} and \ref{fig_distribution-p_rho_uv5}), the difference between the two spectrum models are very small. This is because haze particles are so abundant that the atmosphere is optically thick at low pressures ($P \sim 10^{-5}$~bar) and therefore the difference in haze particle distribution in the lower atmosphere ($P \gtrsim 10^{-5}$~bar) hardly affects the resultant spectrum.
In the case of the intermediate values of $\beta = 10^{2.5}$ (yellow lines), $1$ (green lines), and $10^{-2.5}$ (blue lines), the differences between the two models are relatively large, because the size multiplicity is important.
In the case of $\beta = 10^{-5}$ (purple lines), the difference in transit depth between the two models are relatively small because of their small abundance of haze in the atmosphere.

The maximum differences in transit depth between the two models in the wavelength range of 0.3-5~$\mu$m for $\beta = 10^5$, $10^{2.5}$, $1$, $10^{-2.5}$, and $10^{-5}$ are 87, 205, 393, 393, and 101ppm, respectively.
Precision of observed transit depths depends on properties of the planet, host star, observational instrument, and so on.
If the precision of observed transit depths is larger than the difference in transit depth between the multiple-size and characteristic-size models, the characteristic size approximation is useful.

\section{Summary and Conclusions} \label{sec:summary}
In this study, we have developed the transmission spectrum models of a close-in warm ($\sim$ 500~K) exoplanet with a hazy hydrogen-dominated atmosphere by calculating directly the creation, growth, and settling of hydrocarbon haze particles to derive the distribution of haze particles. 
More specifically, we have done photochemical calculations to derive the vertical profiles of volume mixing ratios of the haze precursors, HCN and $\mathrm{C_2H_2}$. Then, using the obtained vertical profiles of the precursors, we have calculated the growth and settling of haze particles in the atmosphere to derive the steady-state distribution of the size and number density of haze particles.
We have also modeled transmission spectra of the atmospheres with obtained properties of hazes to explore whether the recently-observed diversity of transmission spectra can be explained by the variation in the production rate of haze monomers. 

We have found that the haze particles tend to distribute in a wider region than previously assumed and consists of various sizes.
We have also found that the difference in the production rate of haze monomers, which relates to the UV irradiation intensity from the host star, yields the diversity of transmission spectra observationally suggested: completely flat spectra, spectra with only extinction features of hazes (i.e., \kawashima{spectral} slope \kawashima{due to Rayleigh scattering} and absorption features of hazes), spectra with slope \kawashima{due to Rayleigh scattering} and some molecular absorption features, and spectra with only molecular absorption features.

Also, by applying the grain growth model of \cite{2014ApJ...789L..18O}, we have examined the validity of characteristic size approximation in particle growth calculation.
We have quantified the precisions of observed transit depths beyond which the characteristic approximation suffices to be used for comparison with observation.

\ikome{In this paper, we have focused mainly on describing the methodology and demonstrating the sensitivity of transmission spectra to} haze monomer production rate.
\ikome{In our forthcoming papers, we make detailed investigation of the} dependence of \ikome{transmission spectra} on model parameters other than monomer production rate such as \ikome{atmospheric} metallicity, \kawashima{C/O ratio, }eddy diffusion coefficient, atmospheric temperature profile, and monomer size. 
\ikome{Also, we explore in detail the composition of the atmospheres of known warm exoplanets by comparing the observed spectra with our theoretical ones, taking into account other possibilities of cloud/haze constituents.} 

\acknowledgments

We would like to express special thanks to the following people.
N. Narita and A. Fukui motivated us to work on this study and gave fruitful suggestions through observational collaboration. Advice and comments from Y. Sekine and S. Okuzumi were great help in modeling the properties of haze particles. Also, we had fruitful discussions with N. Iwagami, Y. Ito, and K. Kurosaki regarding transmission spectrum modeling and with Y. Abe, H. Genda, Y. Miguel, and A. Youngblood regarding photochemical and thermochemical modeling.  We are grateful to S.-M. Tsai for kindly providing us his calculation data for model comparison. We also thank the anonymous referee for his/her careful reading and constructive comments that helped us improve this paper greatly. Y.~K. is supported by the Grant-in-Aid for JSPS Fellow (JSPS KAKENHI No.~15J08463) and Leading Graduate Course for Frontiers of Mathematical Sciences and Physics. M.~I. is also supported by the Astrobiology Center Program of National Institutes of Natural Sciences (NINS) (No.~AB291004) and JSPS Core-to-Core Program “International Network of Planetary Sciences”. This work has made use of the MUSCLES Treasury Survey High-Level Science Products.



%


\clearpage
\appendix
\section{Comparison with T\lowercase{sai et al}. (2017)} \label{tsai}
\kawashimas{
To verify our \ikomas{photochemical} model \ikomas{presented in section~{2.1}}, 
we first examine our thermochemical reaction networks. 
In this section, we attempt to reproduce the results of \cite{2017ApJS..228...20T} for two hot Jupiters, HD~189733b and HD~209458b. They considered thermochemistry and \ikomas{eddy-diffusion} transport, but \ikomas{ignored} photochemistry.
They \ikomas{then} simulated the atmospheric chemistry of these two planets to compare their model\ikomas{s} with those of \cite{2011ApJ...737...15M}.
}

\kawashimas{
For comparison, \ikomas{we adopt the same assumptions and values of input parameters that \cite{2017ApJS..228...20T} adopted:} 
\ikomas{The fluxes of} all the species \ikomas{are zero both at the} lower and upper  \ikomas{boundaries}.
\ikomas{The temperature profiles are the} dayside-average\ikomas{d ones} \ikomas{taken} from \ikomas{the} supplementary material of \cite{2011ApJ...737...15M}.
The value of eddy diffusion coefficient $K_{zz}$ \ikomas{is} $1 \times 10^9$~$\mathrm{cm^2}$~$\mathrm{s^{-1}}$ and \ikomas{the} solar elemental abundance ratios from Table~10 of \cite{2009LanB...4B...44L}.
O abundance is multiplied by a factor of 0.793 to account for the effect of oxygen sequestration \citep[see][]{2011ApJ...737...15M}.
We prepare 90 layers with thickness of 50~km and 140~km for the simulations of HD~189733b and HD~209458b, respectively, and place the lower boundary pressure at 1000~bar.
For the values of planet mass and 1000-bar radius, we use 1.15~$M_\mathrm{J}$ and 1.26~$R_\mathrm{J}$ for HD~189733b \citep{2005A&A...444L..15B}, and 0.685~$M_\mathrm{J}$ and 1.359~$R_\mathrm{J}$ for HD~209458b \citep{2008ApJ...677.1324T}.
}

\begin{figure}[ht!]
\gridline{\fig{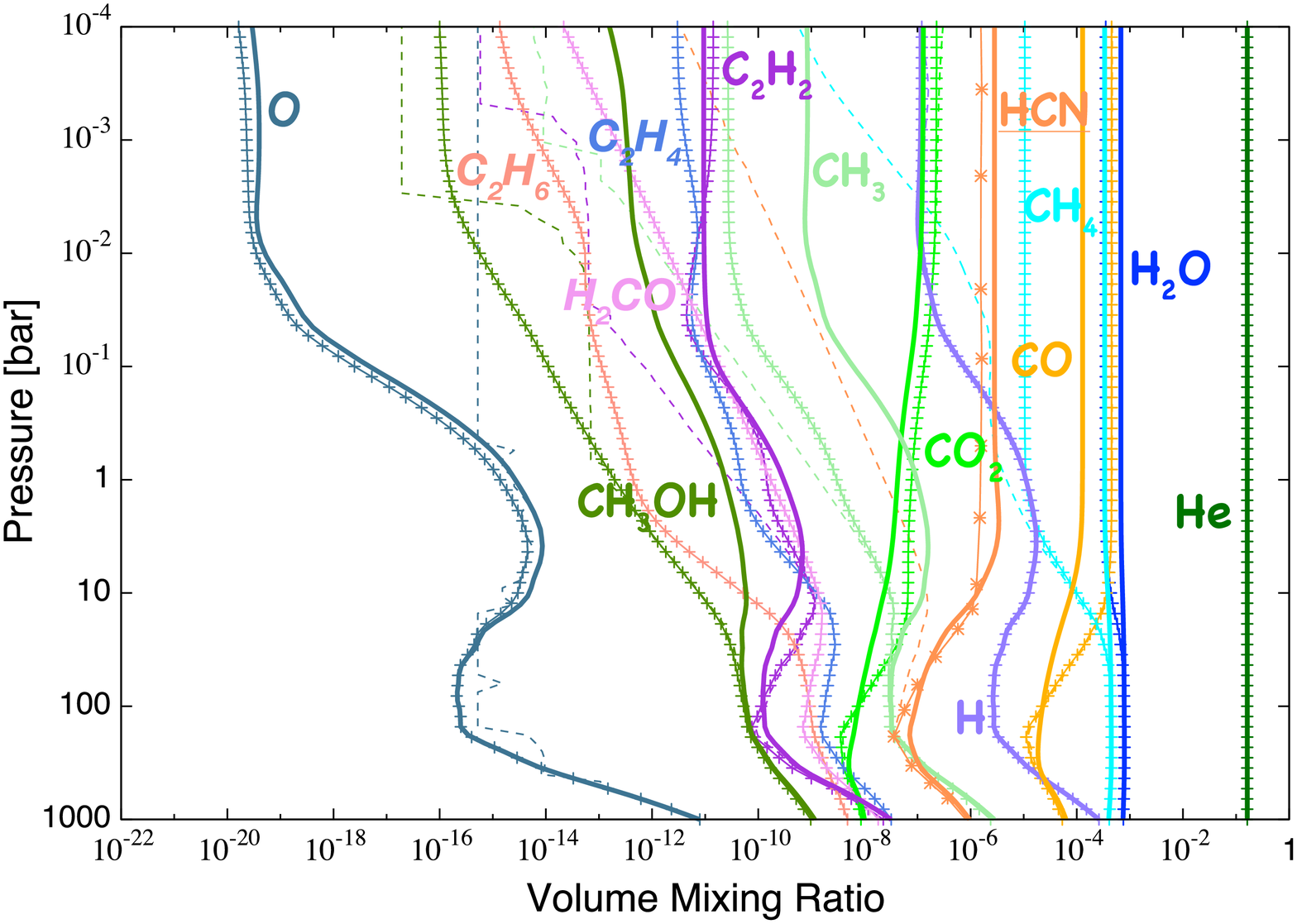}{0.5\textwidth}{(a) HD~189733b}\fig{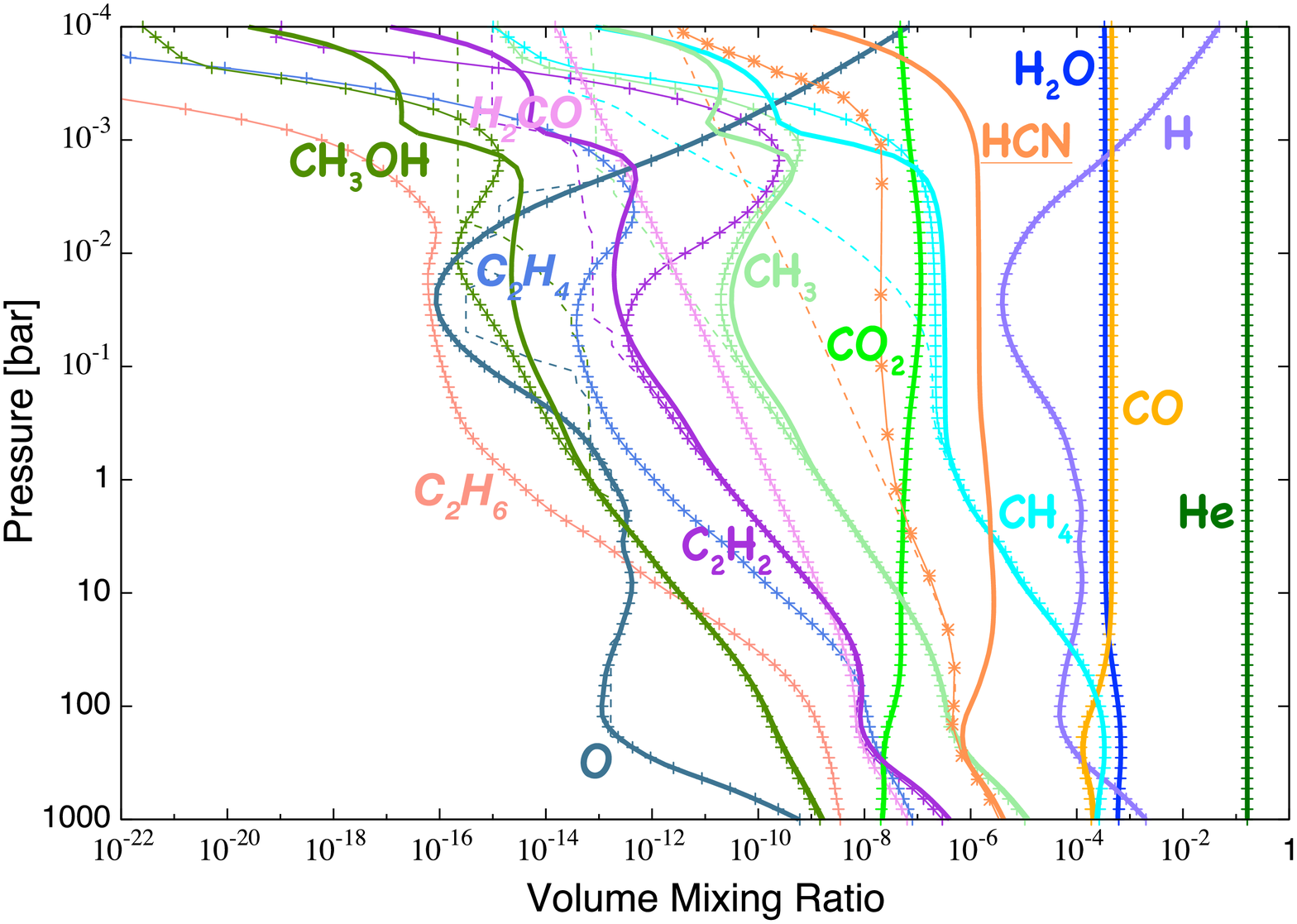}{0.5\textwidth}{(b) HD~209458b}}
\caption{Vertical distributions of gaseous species (solid lines) for the atmospheres of (a) HD~189733b and (b) HD~209458b, compared to 
those from \cite{2017ApJS..228...20T} (solid lines with crosses). HCN is not included in the model of \cite{2017ApJS..228...20T} while the molecules indicated in italics are not included in our model. 
Vertical distributions of HCN from ``no photon" models of \cite{2011ApJ...737...15M}, in which they omit photochemistry, are also shown (thin solid lines with asterisks). We take these data by tracing their Figure~3 with the use of the software, PlotDigitizer X.
We also present the thermochemical equilibrium abundances with dashed lines for reference. Note that the eddy diffusion transport is not included in the thermochemical equilibrium calculations.}
\label{fig_tsai}
\end{figure}

\kawashimas{
Figure~\ref{fig_tsai} shows the calculated vertical distributions of gaseous species (solid lines) for the atmospheres of (a)~HD~189733b and (b)~HD~209458b, \ikomas{which are} compared to the results of \cite{2017ApJS..228...20T} (thin solid lines with crosses). 
HCN is not included in the model of \cite{2017ApJS..228...20T}, 
while the molecules \ikomas{indicated} in italics are not included in our model. 
Vertical distributions of HCN from ``no photon" models of \cite{2011ApJ...737...15M}, in which they omit photochemistry, are also shown (thin solid lines with asterisks). We take these data by tracing their Figure~3 with the use of the software, PlotDigitizer X\footnote{http://www.surf.nuqe.nagoya-u.ac.jp/~nakahara/software/plotdigitizerx/index-e.html}.
We also present the thermochemical equilibrium abundances with dashed lines for reference. 
}

\kawashimas{
\ikomas{In} the case of (a)~HD~189733b first, \ikomas{the mixing ratios of ours and} \cite{2017ApJS..228...20T} 
\ikomas{differ by a factor of $\sim$ 30 for $\mathrm{CH_4}$, $\sim$ 4 for $\mathrm{CO}$, and $\sim$ 2 for} $\mathrm{H_2O}$, 
\ikomas{because quench occurs} at higher pressure \ikomas{in our model.}
\ikomas{Because of such difference in $f_\mathrm{CH_4}$,} our abundances of $\mathrm{CH_3}$ and $\mathrm{CH_3OH}$ are larger \ikomas{by} 1-2 and 1-3 orders of magnitude, respectively. 
The abundances of species in thermochemical equilibrium such as $\mathrm{CO_2}$, $\mathrm{H}$, and $\mathrm{O}$ match theirs well.
As for haze precursors, 
since HCN is not considered in their models, 
we cannot do any comparison \ikomas{regarding HCN}.
However, the ``no photon" models of \cite{2011ApJ...737...15M} (thin solid line with asterisks), in which they omit photochemistry, \ikomas{yield similar abundances to ours}.
The abundance of $\mathrm{C_2H_2}$ \ikomas{differs little between \cite{2017ApJS..228...20T}'s and ours}.
This slight difference in $\mathrm{C_2H_2}$ abundance \ikomas{never} affects our results \ikomas{regarding haze distributions and transmission spectra,} 
since the profile of the production rate of monomers is determined mainly by that of HCN abundance (see $\S$~\ref{result_photo} and \ref{result_dependence}).
}

\kawashimas{
In the case of (b)~HD~209458b, the abundances of the species, $\mathrm{CO}$, $\mathrm{H_2O}$, $\mathrm{H}$, $\mathrm{CH_4}$, $\mathrm{CO_2}$, $\mathrm{CH_3}$, $\mathrm{CH_3OH}$, and O, match theirs well.
This is because of \ikomas{higher} temperature of HD~209458b, \ikomas{for} which \ikomas{the molecules} tend to \ikomas{be closer to} thermochemical equilibrium.
As for haze precursors,
we again compare 
\ikomas{the} HCN abundance \ikomas{from our model} with \ikomas{that from} the ``no photon" models of \cite{2011ApJ...737...15M}.
In our \ikomas{model}, the abundance of HCN deviates from its thermochemical equilibrium values at higher pressure ($\sim 100$~bar) compared to \cite{2011ApJ...737...15M} ($\sim 1$~bar), and HCN results in being quenched at larger abundance in the pressure range of $1 \times 10^{-3}$~bar to \ikomas{100}~bar. 
If we use\ikomas{d} the result of \cite{2011ApJ...737...15M} \ikomas{as the distribution of the precursor molecules}, we would assume smaller monomer production at high altitude\ikomas{s} and larger at low altitude\ikomas{s}. This would hamper particle growth a little and result in less flat transmission spectra.
The abundance of $\mathrm{C_2H_2}$ \ikomas{is} larger than that of \cite{2017ApJS..228...20T} in the region where $\mathrm{C_2H_2}$ is not in thermochemical equilibrium \ikomas{(i.e.,} $P \lesssim 10^{-1}$~bar). 
However, again, this difference \ikomas{never} affect\ikomas{s} our results \ikomas{regarding haze distributions and transmission spectra,} because the profile of the production rate of monomers is determined mainly by that of HCN abundance.}

\section{Comparison with K\lowercase{opparapu et al}. (2012)} \label{kopparapu}
\kawashimas{In this section, we compare our \ikomas{photochemical} model with the model of \cite{2012ApJ...745...77K} for \ikomas{the} hot Jupiter WASP-12b, 
in which photochemistry \ikomas{is} considered in addition to thermochemistry and transport by eddy diffusion.}

\kawashimas{
For comparison, we use the same profiles of temperature and eddy diffusion coefficient by tracing the Figure~1 of \cite{2012ApJ...745...77K} \ikomas{with use of} PlotDigitizer X.
Following them, we neglect transport \ikomas{of the} short-lived species $\mathrm{O(^1D)}$ and $\mathrm{^1CH_2}$.
Since the photodissociation reactions for CO, $\mathrm{H_2}$, $\mathrm{N_2}$, and $\mathrm{CH_3OH}$ are not taken into account in their model, we exclude photochemical reactions, P7, P10, P11, and P12, from our photochemical reaction list \ikomas{used in this section}.
Also, while we consider the following photochemical reaction,
\begin{equation}
\mathrm{P6:\;} \mathrm{CH_4} \rightarrow \mathrm{CH} + \mathrm{H_2} + \mathrm{H},  \nonumber
\end{equation}
which they do not consider, they consider \ikomas{the} following photochemical reaction,
\begin{equation}
\mathrm{CH_4} \rightarrow \mathrm{CH_2} + \mathrm{H} + \mathrm{H}, \nonumber
\end{equation}
which we do not consider.
The other photochemical reactions are identical to theirs.
Following them, we use the G0V star spectrum from \cite{1998PASP..110..863P} and convert it to suit for WASP-12 by using the relation between \ikomas{the} flux at \ikomas{5556~\AA} and the visual magnitude, $V$, from \cite{1992oasp.book.....G}.
We use $V = 11.69$ \citep{2009ApJ...693.1920H}, \ikomas{427~pc as the} distance to the star \citep{2011AJ....141..179C}, and \ikomas{0.0229~AU as the} semi-major axis \citep{2009ApJ...693.1920H}.
We take \ikomas{the} solar elemental abundance ratios from Table~1 of \cite{2005ASPC..336...25A} following them. 
We prepare 100 layers with thickness of 128~km placing the lower boundary pressure at 1~bar.
For the values of planet mass and 1-bar radius, we use 1.41~$M_\mathrm{J}$ and 1.79~$R_\mathrm{J}$, respectively \citep{2009ApJ...693.1920H}.
}

\begin{figure}[ht!]
\begin{center}
\includegraphics[width=0.5\textwidth]{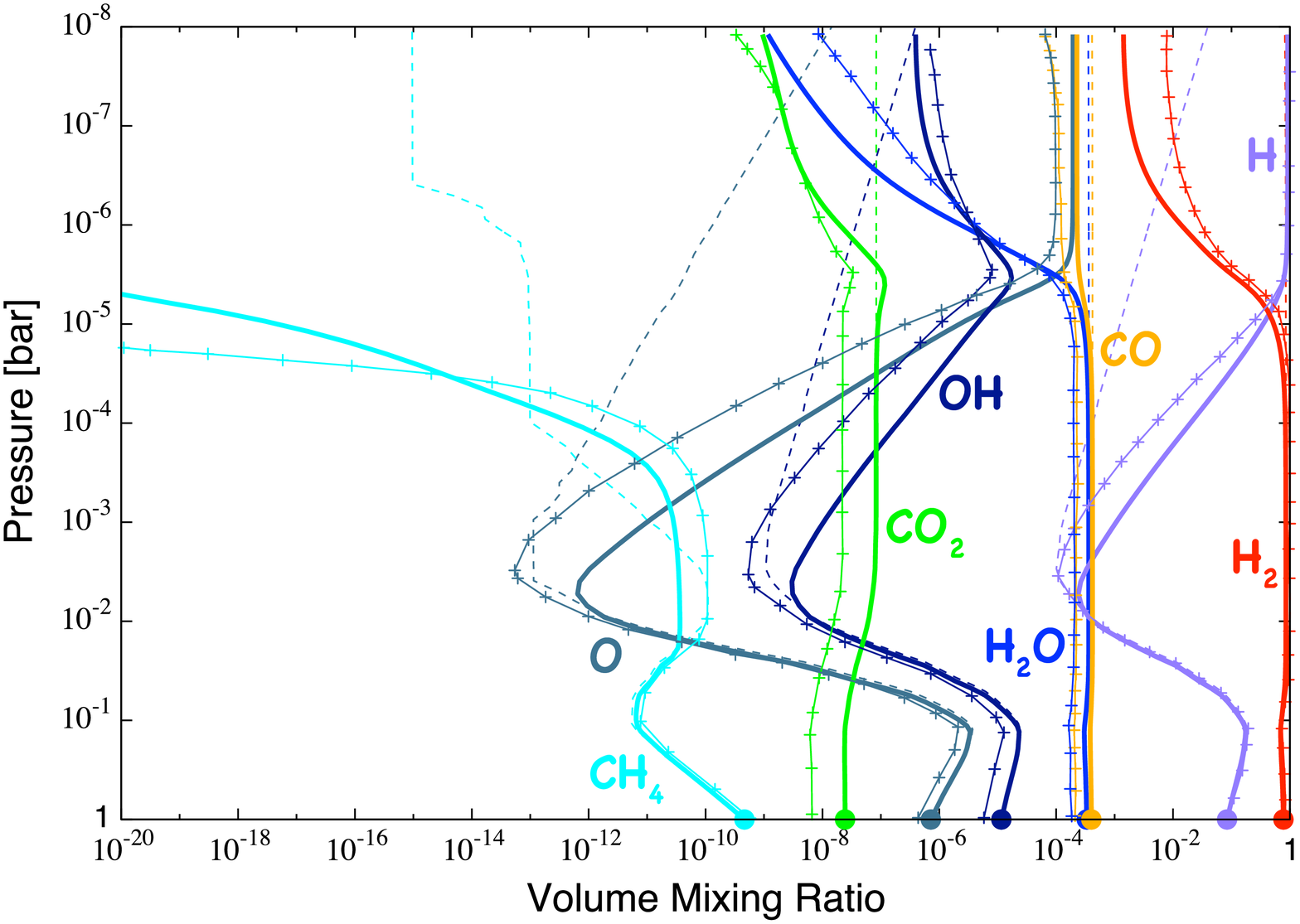}
\end{center}
\caption{Vertical distributions of gaseous species (solid lines) compared to those from \cite{2012ApJ...745...77K} (solid lines with symbols), which we take by tracing their Figure~4 with the use of PlotDigitizer X. 
Filled circles represent the thermochemical equilibrium values at the lower boundary. The thermochemical equilibrium abundances are shown with dashed lines for reference. Note that the eddy diffusion transport is not included in the thermochemical equilibrium calculations.}
\label{fig_kopparapu}
\end{figure}

\kawashimas{
Figure~\ref{fig_kopparapu} shows the calculated vertical distributions of gaseous species (solid lines)\ikomas{, which are} compared to the results of \cite{2012ApJ...745...77K} (solid lines with \ikomas{symbols}) \ikomas{that} we \ikomas{also} take by tracing \ikomas{their} Figure~4 with the use of PlotDigitizer X.
We also present the thermochemical equilibrium abundances with dashed lines for reference.
}

\kawashimas{
The abundances of \ikomas{the} major ($f_i \gtrsim 10^{-10}$) species agree with 
\ikomas{\cite{2012ApJ...745...77K}'s} within one order of magnitude.
The \ikomas{abundances at the} lower boundary \ikomas{(1~bar)} are slightly different from theirs, although we \ikomas{have used} thermochemical equilibrium values for \ikomas{the} lower boundary condition \ikomas{in the same way as they did} and \ikomas{also used} the same elemental abundance ratios. 
The differences in abundance profile \ikomas{may} come from \ikomas{those} in these lower boundary values. 
}

\section{Thermochemical Reactions}

\begin{deluxetable}{lllcllll}
\tablecaption{Thermochemical Reactions \label{tab:thermo_reactions}}
\tablehead{
\colhead{No.} & \colhead{No. Hu\tablenotemark{a}} & \colhead{Reactants} & & \colhead{Products} & \colhead{Rate \kawashima{Coefficients} \tablenotemark{b}} & \colhead{Ref.} & \colhead{Temp.\tablenotemark{c}}
}
\startdata
R1 & R1 & $\mathrm{C + CH_2}$ & $\rightarrow$ & $\mathrm{CH + CH}$ & $2.69 \times 10^{-12} \mathrm{e}^{-23573.0/T}$ & NIST & 1000-4000 \\
R2 & R2 & $\mathrm{C + CN}$ & $\rightarrow$ & $\mathrm{C_2 + N}$ & $4.98 \times 10^{-10} \mathrm{e}^{-18041.0/T}$ & NIST & 5000-8000 \\
R3 & R3 & $\mathrm{C + H_2}$ & $\rightarrow$ & $\mathrm{CH + H}$ & $6.64 \times 10^{-10} \mathrm{e}^{-11700.0/T}$ & NIST & 1520-2540 \\
R4 & R5 & $\mathrm{C + N_2}$ & $\rightarrow$ & $\mathrm{CN + N}$ & $8.7 \times 10^{-11} \mathrm{e}^{-22611.0/T}$ & NIST & 2000-5000 \\
R5 & R6 & $\mathrm{C + O_2}$ & $\rightarrow$ & $\mathrm{CO + O}$ & $5.1 \times 10^{-11} \left( T/298.0 \right)^{-0.3}$ & NIST & 15-295 \\
R6 & R14 & $\mathrm{C_2H + CH_3OH}$ & $\rightarrow$ & $\mathrm{C_2H_2 + CH_3O}$ & $2.0 \times 10^{-12}$ & NIST & 300-2500 \\
R7 & R17 & $\mathrm{C_2H_2 + CN}$ & $\rightarrow$ & $\mathrm{HCN + C_2H}$ & $2.2 \times 10^{-10}$ & NIST & 294 \\
R8 & R31 & $\mathrm{CH + CH}$ & $\rightarrow$ & $\mathrm{C_2H_2}$ & $2.0 \times 10^{-10}$ & NIST & 298 \\
R9 & R34 & $\mathrm{CH_2 + C_2H}$ & $\rightarrow$ & $\mathrm{C_2H_2 + CH}$ & $3.0 \times 10^{-11}$ & NIST & 300-2500 \\
R10 & R40 & $\mathrm{NH + OH}$ & $\rightarrow$ & $\mathrm{NH_2 + O}$ & $2.94 \times 10^{-12} \left( T/298.0 \right)^{0.1} \mathrm{e}^{5800.0/T}$ & NIST & 298-3000 \\
R11 & R42 & $\mathrm{CH_2 + CH_2}$ & $\rightarrow$ & $\mathrm{C_2H_2 + H_2}$ & $2.62 \times 10^{-9} \mathrm{e}^{-6010.0/T}$ & NIST & 1100-2700 \\
R12 & R43 & $\mathrm{CH_2 + CH_2}$ & $\rightarrow$ & $\mathrm{C_2H_2 + H + H}$ & $3.32 \times 10^{-10} \mathrm{e}^{-5530.0/T}$ & NIST & 1100-2700 \\
R13 & R48 & $\mathrm{CH_2 + CH_4}$ & $\rightarrow$ & $\mathrm{CH_3 + CH_3}$ & $7.12 \times 10^{-12} \mathrm{e}^{-5050.0/T}$ & NIST & 296-707 \\
R14 & R49 & $\mathrm{CH_2 + CH_3OH}$ & $\rightarrow$ & $\mathrm{CH_3 + CH_3O}$ & $1.12 \times 10^{-15} \left( T/298.0 \right)^{3.1} \mathrm{e}^{-3490.0/T}$ & NIST & 300-2500 \\
R15 & R50 & $\mathrm{CH_2 + HCO}$ & $\rightarrow$ & $\mathrm{CO + CH_3}$ & $3.0 \times 10^{-11}$ & NIST & 300-2500 \\
R16 & R57 & $\mathrm{CH_3 + C_2H_2}$ & $\rightarrow$ & $\mathrm{CH_4 + C_2H}$ & $3.0 \times 10^{-13} \mathrm{e}^{-8700.0/T}$ & NIST & 300-2500 \\
R17 & R69 & $\mathrm{CH_3 + CH_3OH}$ & $\rightarrow$ & $\mathrm{CH_4 + CH_3O}$ & $1.12 \times 10^{-15} \left( T/298.0 \right)^{3.1} \mathrm{e}^{-3490.0/T}$ & NIST & 300-2500 \\
R18 & R71 & $\mathrm{CH_3 + HCO}$ & $\rightarrow$ & $\mathrm{CH_4 + CO}$ & $2.0 \times 10^{-10}$ & NIST & 300-2500 \\
R19 & R91 & $\mathrm{CH_4 + C_2H}$ & $\rightarrow$ & $\mathrm{C_2H_2 + CH_3}$ & $3.0 \times 10^{-12} \mathrm{e}^{-250.0/T}$ & NIST & 300-2500 \\
R20 & R96 & $\mathrm{CH_4 + CH_3O}$ & $\rightarrow$ & $\mathrm{CH_3OH + CH_3}$ & $2.61 \times 10^{-13} \mathrm{e}^{-4450.0/T}$ & NIST & 300-2500 \\
R21 & R99 & $\mathrm{CH_4 + CN}$ & $\rightarrow$ & $\mathrm{HCN + CH_3}$ & $5.11 \times 10^{-13} \left( T/298.0 \right)^{2.64} \mathrm{e}^{150.3/T}$ & NIST & 290-1500 \\
R22 & R101 & $\mathrm{CH_3OH + CN}$ & $\rightarrow$ & $\mathrm{HCN + CH_3O}$ & $1.2 \times 10^{-10}$ & NIST & 294 \\
R23 & R108 & $\mathrm{HCO + C_2H}$ & $\rightarrow$ & $\mathrm{C_2H_2 + CO}$ & $1.0 \times 10^{-10}$ & NIST & 300-2500 \\
R24 & R113 & $\mathrm{HCO + CH_3O}$ & $\rightarrow$ & $\mathrm{CH_3OH + CO}$ & $1.5 \times 10^{-10}$ & NIST & 300-2500 \\
R25 & R116 & $\mathrm{HCO + CN}$ & $\rightarrow$ & $\mathrm{HCN + CO}$ & $1.0 \times 10^{-10}$ & NIST & 500-2500 \\
R26 & R120 & $\mathrm{CO + C_2H_2}$ & $\rightarrow$ & $\mathrm{C_2H + HCO}$ & $8.0 \times 10^{-10} \mathrm{e}^{-53641.4/T}$ & NIST & 300-2500 \\
R27 & R122 & $\mathrm{CO + CH_3}$ & $\rightarrow$ & $\mathrm{C_2H_2 + OH}$ & $6.3 \times 10^{-11} \mathrm{e}^{-30428.9/T}$ & NIST & 1500-1900 \\
R28 & R127 & $\mathrm{H + C_2H}$ & $\rightarrow$ & $\mathrm{C_2H_2}$ & $3.0 \times 10^{-10}$ & NIST & 300-2500 \\
R29 & R128 & $\mathrm{H + C_2H}$ & $\rightarrow$ & $\mathrm{H_2 + C_2}$ & $6.0 \times 10^{-11} \mathrm{e}^{-14192.0/T}$ & NIST & 300-2500 \\
R30 & R129 & $\mathrm{H + C_2H_2}$ & $\rightarrow$ & $\mathrm{C_2H + H_2}$ & $1.0 \times 10^{-10} \mathrm{e}^{-11200.0/T}$ & NIST & 300-2500 \\
R31 & R145 & $\mathrm{H + CH}$ & $\rightarrow$ & $\mathrm{C + H_2}$ & $1.31 \times 10^{-10} \mathrm{e}^{-85.6/T}$ & NIST & 300-2000 \\
R32 & R146 & $\mathrm{H + CH_2}$ & $\rightarrow$ & $\mathrm{CH + H_2}$ & $1.0 \times 10^{-11} \mathrm{e}^{900.0/T}$ & NIST & 300-3000 \\
R33 & R149 & $\mathrm{H + CH_3}$ & $\rightarrow$ & $\mathrm{CH_2 + H_2}$ & $1.0 \times 10^{-10} \mathrm{e}^{-7600.0/T}$ & NIST & 300-2500 \\
R34 & R152 & $\mathrm{H + CH_3O}$ & $\rightarrow$ & $\mathrm{CH_3OH}$ & $2.89 \times 10^{-10} \left( T/298.0 \right)^{0.04}$ & NIST & 300-2500 \\
R35 & R153 & $\mathrm{H + CH_3O}$ & $\rightarrow$ & $\mathrm{CH_3 + OH}$ & $1.6 \times 10^{-10}$ & NIST & 300-2500 \\
R36 & R157 & $\mathrm{H + CH_4}$ & $\rightarrow$ & $\mathrm{CH_3 + H_2}$ & $5.83 \times 10^{-13} \left( T/298.0 \right)^{3.0} \mathrm{e}^{-4040.0/T}$ & NIST & 300-2500 \\
R37 & R158 & $\mathrm{H + CH_3OH}$ & $\rightarrow$ & $\mathrm{CH_3 + H_2O}$ & $3.32 \times 10^{-10} \mathrm{e}^{-2670.0/T}$ & NIST & 1370-1840 \\
R38 & R159 & $\mathrm{H + CH_3OH}$ & $\rightarrow$ & $\mathrm{CH_3O + H_2}$ & $2.42 \times 10^{-12} \left( T/298.0 \right)^{2.0} \mathrm{e}^{-2270.0/T}$ & NIST & 300-2500 \\
R39 & R162 & $\mathrm{H + HCO}$ & $\rightarrow$ & $\mathrm{CO + H_2}$ & $1.50 \times 10^{-10}$ & NIST & 300-2500 \\
R40 & R165 & $\mathrm{H + CO_2}$ & $\rightarrow$ & $\mathrm{CO + OH}$ & $2.51 \times 10^{-10} \mathrm{e}^{-13350.0/T}$ & NIST & 300-2500 \\
R41 & R166 & $\mathrm{H + H_2O}$ & $\rightarrow$ & $\mathrm{H_2 + OH}$ & $6.82 \times 10^{-12} \left( T/298.0 \right)^{1.6} \mathrm{e}^{-9720.0/T}$ & NIST & 300-2500 \\
R42 & R184 & $\mathrm{H + NH}$ & $\rightarrow$ & $\mathrm{H_2 + N}$ & $1.69 \times 10^{-11}$ & NIST & 1500-2500 \\
R43 & R185 & $\mathrm{H + NH_2}$ & $\rightarrow$ & $\mathrm{H_2 + NH}$ & $1.05 \times 10^{-10} \mathrm{e}^{-4450.1/T}$ & NIST & 1100-1800 \\
R44 & R186 & $\mathrm{H + NH_3}$ & $\rightarrow$ & $\mathrm{H_2 + NH_2}$ & $7.80 \times 10^{-13} \left( T/298.0 \right)^{2.4} \mathrm{e}^{-4990.1/T}$ & NIST & 490-1780 \\
R45 & R191 & $\mathrm{H + O_2}$ & $\rightarrow$ & $\mathrm{O + OH}$ & $6.73 \times 10^{-10} \left( T/298.0 \right)^{-0.59} \mathrm{e}^{-8152.0/T}$ & NIST & 800-3500 \\
R46 & R193 & $\mathrm{H_2 + C}$ & $\rightarrow$ & $\mathrm{CH + H}$ & $6.64 \times 10^{-10} \mathrm{e}^{-11700.0/T}$ & NIST & 1520-2540 \\
R47 & R194 & $\mathrm{H_2 + C_2}$ & $\rightarrow$ & $\mathrm{C_2H + H}$ & $1.1 \times 10^{-10} \mathrm{e}^{-4000.0/T}$ & NIST & 2580-4650 \\
R48 & R195 & $\mathrm{H_2 + C_2H}$ & $\rightarrow$ & $\mathrm{C_2H_2 + H}$ & $8.95 \times 10^{-13} \left( T/298.0 \right)^{2.57} \mathrm{e}^{-130.0/T}$ & NIST & 200-2000 \\
R49 & R202 & $\mathrm{H_2 + CH}$ & $\rightarrow$ & $\mathrm{CH_2 + H}$ & $3.75 \times 10^{-10} \mathrm{e}^{-1660.0/T}$ & NIST & 327-397 \\
R50 & R203 & $\mathrm{H_2 + CH_3}$ & $\rightarrow$ & $\mathrm{CH_4 + H}$ & $6.86 \times 10^{-14} \left( T/298.0 \right)^{2.74} \mathrm{e}^{-4740.0/T}$ & NIST & 300-2500 \\
R51 & R204 & $\mathrm{H_2 + CH_3O}$ & $\rightarrow$ & $\mathrm{CH_3OH + H}$ & $9.96 \times 10^{-14} \left( T/298.0 \right)^2 \mathrm{e}^{-6720.0/T}$ & NIST & 300-2500 \\
R52 & R207 & $\mathrm{H_2 + CN}$ & $\rightarrow$ & $\mathrm{HCN + H}$ & $5.65 \times 10^{-13} \left( T/298.0 \right)^{2.45} \mathrm{e}^{-1131.0/T}$ & NIST & 300-2500 \\
R53 & R211 & $\mathrm{H_2 + NH}$ & $\rightarrow$ & $\mathrm{NH_2 + H}$ & $3.5 \times 10^{-11} \mathrm{e}^{-7758.0/T}$ & NIST & 833-1432 \\
R54 & R212 & $\mathrm{H_2 + NH_2}$ & $\rightarrow$ & $\mathrm{NH_3 + H}$ & $6.75 \times 10^{-14} \left( T/298.0 \right)^{2.6} \mathrm{e}^{-3006.8/T}$ & NIST & 400-2200 \\
R55 & R214 & $\mathrm{H_2O + C}$ & $\rightarrow$ & $\mathrm{CH + OH}$ & $1.3 \times 10^{-12} \mathrm{e}^{-19845.0/T}$ & NIST & 1000-4000 \\
R56 & R215 & $\mathrm{H_2O + C_2H}$ & $\rightarrow$ & $\mathrm{C_2H_2 + OH}$ & $7.74 \times 10^{-14} \left( T/298.0 \right)^{3.05} \mathrm{e}^{-376.0/T}$ & NIST & 300-2000 \\
R57 & R218 & $\mathrm{H_2O + CH}$ & $\rightarrow$ & $\mathrm{CH_3O}$ & $9.48 \times 10^{-12} \mathrm{e}^{380.0/T}$ & NIST & 298-669 \\
R58 & R219 & $\mathrm{H_2O + CN}$ & $\rightarrow$ & $\mathrm{HCN + OH}$ & $1.3 \times 10^{-11} \mathrm{e}^{-3760.0/T}$ & NIST & 500-2500 \\
R59 & R276 & $\mathrm{N + C_2}$ & $\rightarrow$ & $\mathrm{CN + C}$ & $2.8 \times 10^{-11}$ & NIST & 298 \\
R60 & R280 & $\mathrm{N + CH}$ & $\rightarrow$ & $\mathrm{C + NH}$ & $3.0 \times 10^{-11} \left( T/298.0 \right)^{0.65} \mathrm{e}^{-1203.0/T}$ & NIST & 990-1010 \\
R61 & R281 & $\mathrm{N + CH}$ & $\rightarrow$ & $\mathrm{CN + H}$ & $1.66 \times 10^{-10} \left( T/298.0 \right)^{-0.09}$ & NIST & 216-584 \\
R62 & R282 & $\mathrm{N + CH_3}$ & $\rightarrow$ & $\mathrm{H_2 + HCN}$ & $4.3 \times 10^{-10} \mathrm{e}^{-420.0/T}$ & NIST & 200-423 \\
R63 & R284 & $\mathrm{N + CN}$ & $\rightarrow$ & $\mathrm{C + N_2}$ & $3.0 \times 10^{-10}$ & NIST & 300-2500 \\
R64 & R287 & $\mathrm{N + H_2O}$ & $\rightarrow$ & $\mathrm{OH + NH}$ & $6.03 \times 10^{-11} \left( T/298.0 \right)^{1.2} \mathrm{e}^{-19243.6/T}$ & NIST & 800-3000 \\
R65 & R289 & $\mathrm{N + NH}$ & $\rightarrow$ & $\mathrm{N_2 + H}$ & $1.95 \times 10^{-11} \left( T/298.0 \right)^{0.51} \mathrm{e}^{-9.6/T}$ & NIST & 300-2500 \\
R66 & R290 & $\mathrm{N + NH_2}$ & $\rightarrow$ & $\mathrm{NH + NH}$ & $3.0 \times 10^{-13} \mathrm{e}^{-7600.0/T}$ & NIST & 1000-4000 \\
R67 & R300 & $\mathrm{NH + NH_3}$ & $\rightarrow$ & $\mathrm{NH_2 + NH_2}$ & $5.25 \times 10^{-10} \mathrm{e}^{-13470.0/T}$ & NIST & 1300-1700 \\
R68 & R305 & $\mathrm{NH + O}$ & $\rightarrow$ & $\mathrm{OH + N}$ & $1.16 \times 10^{-11}$ & NIST & 250-3000 \\
R69 & R309 & $\mathrm{NH + OH}$ & $\rightarrow$ & $\mathrm{H_2O + N}$ & $3.1 \times 10^{-12} \left( T/298.0 \right)^{1.2}$ & NIST & 298-3000 \\
R70 & R310 & $\mathrm{NH_2 + C}$ & $\rightarrow$ & $\mathrm{CH + NH}$ & $9.61 \times 10^{-13} \mathrm{e}^{-10500.0/T}$ & NIST & 1000-4000 \\
R71 & R311 & $\mathrm{NH_2 + C_2H_2}$ & $\rightarrow$ & $\mathrm{C_2H + NH_3}$ & $8.2 \times 10^{-13} \mathrm{e}^{-2780.0/T}$ & NIST & 340-510 \\
R72 & R316 & $\mathrm{NH_2 + CH_3}$ & $\rightarrow$ & $\mathrm{CH_4 + NH}$ & $8.4 \times 10^{-10} \mathrm{e}^{-4834.9/T}$ & NIST & 300-2000 \\
R73 & R317 & $\mathrm{NH_2 + CH_4}$ & $\rightarrow$ & $\mathrm{CH_3 + NH_3}$ & $8.77 \times 10^{-15} \left( T/298.0 \right)^3 \mathrm{e}^{-2130.0/T}$ & NIST & 300-2000 \\
R74 & R319 & $\mathrm{NH_2 + H_2O}$ & $\rightarrow$ & $\mathrm{OH + NH_3}$ & $2.1 \times 10^{-13} \left( T/298.0 \right)^{1.9} \mathrm{e}^{-5725.0/T}$ & NIST & 300-3000 \\
R75 & R323 & $\mathrm{NH_2 + O}$ & $\rightarrow$ & $\mathrm{OH + NH}$ & $1.16 \times 10^{-11}$ & NIST & 298-3000 \\
R76 & R326 & $\mathrm{NH_2 + OH}$ & $\rightarrow$ & $\mathrm{H_2O + NH}$ & $7.69 \times 10^{-13} \left( T/298.0 \right)^{1.5} \mathrm{e}^{-230.0/T}$ & NIST & 250-3000 \\
R77 & R327 & $\mathrm{NH_3 + CH}$ & $\rightarrow$ & $\mathrm{HCN + H_2 + H}$ & $7.24 \times 10^{-11} \mathrm{e}^{317.0/T}$ & NIST & 300-1300 \\
R78 & R328 & $\mathrm{NH_3 + CH_3}$ & $\rightarrow$ & $\mathrm{CH_4 + NH_2}$ & $9.55 \times 10^{-14} \mathrm{e}^{-4895.0/T}$ & NIST & 350-600 \\
R79 & R329 & $\mathrm{NH_3 + CN}$ & $\rightarrow$ & $\mathrm{HCN + NH_2}$ & $1.66 \times 10^{-11}$ & NIST & 300-700 \\
R80 & R388 & $\mathrm{O + C_2}$ & $\rightarrow$ & $\mathrm{CO + C}$ & $6.0 \times 10^{-10}$ & NIST & 8000 \\
R81 & R389 & $\mathrm{O + C_2H}$ & $\rightarrow$ & $\mathrm{CO + CH}$ & $1.7 \times 10^{-11}$ & NIST & 300-2500 \\
R82 & R391 & $\mathrm{O + C_2H_2}$ & $\rightarrow$ & $\mathrm{CO + CH_2}$ & $3.49 \times 10^{-12} \left( T/298.0 \right)^{1.5} \mathrm{e}^{-850.0/T}$ & NIST & 300-2500 \\
R83 & R407 & $\mathrm{O + CH}$ & $\rightarrow$ & $\mathrm{OH + C}$ & $2.52 \times 10^{-11} \mathrm{e}^{-2380.0/T}$ & NIST & 10-6000 \\
R84 & R408 & $\mathrm{O + CH}$ & $\rightarrow$ & $\mathrm{CO + H}$ & $6.6 \times 10^{-11}$ & NIST & 300-2000 \\
R85 & R409 & $\mathrm{O + CH_2}$ & $\rightarrow$ & $\mathrm{CH + OH}$ & $7.2 \times 10^{-12}$ & NIST & 300-2500 \\
R86 & R410 & $\mathrm{O + CH_2}$ & $\rightarrow$ & $\mathrm{HCO + H}$ & $5.0 \times 10^{-11}$ & NIST & 1200-1800 \\
R87 & R411 & $\mathrm{O + CH_2}$ & $\rightarrow$ & $\mathrm{CO + H + H}$ & $1.2 \times 10^{-10}$ & NIST & 300-2500 \\
R88 & R412 & $\mathrm{O + CH_2}$ & $\rightarrow$ & $\mathrm{CO + H_2}$ & $7.3 \times 10^{-11}$ & NIST & 300-2500 \\
R89 & R414 & $\mathrm{O + CH_3}$ & $\rightarrow$ & $\mathrm{CH_3O}$ & $7.51 \times 10^{-14} \left( T/298.0 \right)^{-2.12} \mathrm{e}^{-314.0/T}$ & NIST & 300-2500 \\
R90 & R416 & $\mathrm{O + CH_3}$ & $\rightarrow$ & $\mathrm{CO + H_2 + H}$ & $5.72 \times 10^{-11}$ & NIST & 290-900 \\
R91 & R420 & $\mathrm{O + CH_3O}$ & $\rightarrow$ & $\mathrm{CH_3 + O_2}$ & $2.5 \times 10^{-11}$ & NIST & 298 \\
R92 & R423 & $\mathrm{O + CH_4}$ & $\rightarrow$ & $\mathrm{CH_3 + OH}$ & $2.26 \times 10^{-12} \left( T/298.0 \right)^{2.2} \mathrm{e}^{-3820.0/T}$ & NIST & 420-1520 \\
R93 & R424 & $\mathrm{O + CH_3OH}$ & $\rightarrow$ & $\mathrm{CH_3O + OH}$ & $1.66 \times 10^{-11} \mathrm{e}^{-2360.0/T}$ & NIST & 300-1000 \\
R94 & R426 & $\mathrm{O + HCO}$ & $\rightarrow$ & $\mathrm{CO + OH}$ & $5.0 \times 10^{-11}$ & NIST & 300-2500 \\
R95 & R427 & $\mathrm{O + HCO}$ & $\rightarrow$ & $\mathrm{CO_2 + H}$ & $5.0 \times 10^{-11}$ & NIST & 300-2500 \\
R96 & R430 & $\mathrm{O + CN}$ & $\rightarrow$ & $\mathrm{CO + N}$ & $3.4 \times 10^{-11} \mathrm{e}^{-210.0/T}$ & NIST & 500-2500 \\
R97 & R433 & $\mathrm{O + H_2}$ & $\rightarrow$ & $\mathrm{H + OH}$ & $3.44 \times 10^{-13} \left( T/298.0 \right)^{2.67} \mathrm{e}^{-3160.0/T}$ & NIST & 300-2500 \\
R98 & R436 & $\mathrm{O + HCN}$ & $\rightarrow$ & $\mathrm{CO + NH}$ & $3.0 \times 10^{-12} \mathrm{e}^{-4000.0/T}$ & JPL & 470-900 \\
R99 & R445 & $\mathrm{O + OH}$ & $\rightarrow$ & $\mathrm{O_2 + H}$ & $2.2 \times 10^{-11} \mathrm{e}^{120.0/T}$ & JPL & 200-300 \\
R100 & R449 & $\mathrm{O(^1D) + CH_4}$ & $\rightarrow$ & $\mathrm{CH_3O + H}$ & $3.5 \times 10^{-11}$ & JPL & 200-300 \\
R101 & R450 & $\mathrm{O(^1D) + CH_4}$ & $\rightarrow$ & $\mathrm{CH_3 + OH}$ & $1.31 \times 10^{-10}$ & JPL & 200-300 \\
R102 & R453 & $\mathrm{O(^1D) + CH_3OH}$ & $\rightarrow$ & $\mathrm{CH_3O + OH}$ & $4.2 \times 10^{-10}$ & NIST & 300 \\
R103 & R454 & $\mathrm{O(^1D) + CO_2}$ & $\rightarrow$ & $\mathrm{CO_2 + O}$ & $7.5 \times 10^{-11} \mathrm{e}^{115.0/T}$ & JPL & 200-300 \\
R104 & R455 & $\mathrm{O(^1D) + H_2}$ & $\rightarrow$ & $\mathrm{H + OH}$ & $1.2 \times 10^{-10}$ & JPL & 200-300 \\
R105 & R456 & $\mathrm{O(^1D) + H_2O}$ & $\rightarrow$ & $\mathrm{OH + OH}$ & $1.63 \times 10^{-10} \mathrm{e}^{60.0/T}$ & JPL & 200-300 \\
R106 & R458 & $\mathrm{O(^1D) + N_2}$ & $\rightarrow$ & $\mathrm{O + N_2}$ & $2.15 \times 10^{-11} \mathrm{e}^{110.0/T}$ & JPL & 200-300 \\
R107 & R461 & $\mathrm{O(^1D) + NH_3}$ & $\rightarrow$ & $\mathrm{OH + NH_2}$ & $2.5 \times 10^{-10}$ & JPL & 200-300 \\
R108 & R464 & $\mathrm{O(^1D) + O_2}$ & $\rightarrow$ & $\mathrm{O + O_2}$ & $3.3 \times 10^{-11} \mathrm{e}^{55.0/T}$ & JPL & 200-300 \\
R109 & R476 & $\mathrm{OH + C_2}$ & $\rightarrow$ & $\mathrm{CO + CH}$ & $8.3 \times 10^{-12}$ & NIST & 2200 \\
R110 & R477 & $\mathrm{OH + C_2H}$ & $\rightarrow$ & $\mathrm{CO + CH_2}$ & $3.0 \times 10^{-11}$ & NIST & 300-2500 \\
R111 & R478 & $\mathrm{OH + C_2H}$ & $\rightarrow$ & $\mathrm{C_2H_2 + O}$ & $3.0 \times 10^{-11}$ & NIST & 300-2500 \\
R112 & R480 & $\mathrm{OH + C_2H_2}$ & $\rightarrow$ & $\mathrm{C_2H + H_2O}$ & $1.03 \times 10^{-13} \left( T/298.0 \right)^{2.68} \mathrm{e}^{-6060.0/T}$ & NIST & 300-2500 \\
R113 & R482 & $\mathrm{OH + C_2H_2}$ & $\rightarrow$ & $\mathrm{CO + CH_3}$ & $6.34 \times 10^{-18} \left( T/298.0 \right)^{4.0} \mathrm{e}^{1010.0/T}$ & NIST & 500-2500 \\
R114 & R495 & $\mathrm{OH + CH_3}$ & $\rightarrow$ & $\mathrm{CH_3O + H}$ & $6.45 \times 10^{-13} \left( T/298.0 \right)^1 \mathrm{e}^{-6012.0/T}$ & NIST & 300-3000 \\
R115 & R496 & $\mathrm{OH + CH_3}$ & $\rightarrow$ & $\mathrm{CH_2 + H_2O}$ & $1.2 \times 10^{-10} \mathrm{e}^{-1400.0/T}$ & NIST & 300-1000 \\
R116 & R501 & $\mathrm{OH + CH_4}$ & $\rightarrow$ & $\mathrm{CH_3 + H_2O}$ & $2.45 \times 10^{-12} \mathrm{e}^{-1775.0/T}$ & JPL & 200-300 \\
R117 & R502 & $\mathrm{OH + CH_3OH}$ & $\rightarrow$ & $\mathrm{CH_3O + H_2O}$ & $2.9 \times 10^{-12} \mathrm{e}^{-345.0/T}$ & JPL & 200-300 \\
R118 & R504 & $\mathrm{OH + HCO}$ & $\rightarrow$ & $\mathrm{CO + H_2O}$ & $1.69 \times 10^{-10}$ & NIST & 300-2500 \\
R119 & R507 & $\mathrm{OH + CN}$ & $\rightarrow$ & $\mathrm{O + HCN}$ & $1.0 \times 10^{-11} \mathrm{e}^{-1000.0/T}$ & NIST & 500-2500 \\
R120 & R511 & $\mathrm{OH + CO}$ & $\rightarrow$ & $\mathrm{CO_2 + H}$ & $5.4 \times 10^{-14} \left( T/298.0 \right)^{1.5} \mathrm{e}^{250.0/T}$ & NIST & 300-2000 \\
R121 & R512 & $\mathrm{OH + H_2}$ & $\rightarrow$ & $\mathrm{H_2O + H}$ & $2.8 \times 10^{-12} \mathrm{e}^{-1800.0/T}$ & JPL & 200-300 \\
R122 & R515 & $\mathrm{OH + HCN}$ & $\rightarrow$ & $\mathrm{CO + NH_2}$ & $1.1 \times 10^{-13} \mathrm{e}^{-5890.0/T}$ & NIST & 500-2500 \\
R123 & R516 & $\mathrm{OH + HCN}$ & $\rightarrow$ & $\mathrm{CN + H_2O}$ & $1.84 \times 10^{-13} \left( T/298.0 \right)^{1.5} \mathrm{e}^{-3890.0/T}$ & NIST & 298-2840 \\
R124 & R523 & $\mathrm{OH + NH_3}$ & $\rightarrow$ & $\mathrm{H_2O + NH_2}$ & $1.7 \times 10^{-12} \mathrm{e}^{-710.0/T}$ & JPL & 200-300 \\
R125 & R526 & $\mathrm{OH + OH}$ & $\rightarrow$ & $\mathrm{H_2O + O}$ & $1.8 \times 10^{-12}$ & JPL & 200-300 \\
R126 & R596 & $\mathrm{^1 CH_2 + H_2}$ & $\rightarrow$ & $\mathrm{CH_2 + H_2}$ & $1.26 \times 10^{-11}$ & YD99\tablenotemark{d} &  \\
R127 & R597 & $\mathrm{^1 CH_2 + H_2}$ & $\rightarrow$ & $\mathrm{CH_3 + H}$ & $9.24 \times 10^{-11}$ & YD99\tablenotemark{d} &  \\
R128 & R598 & $\mathrm{^1 CH_2 + CH_4}$ & $\rightarrow$ & $\mathrm{CH_2 + CH_4}$ & $1.2 \times 10^{-11}$ & YD99\tablenotemark{d} &  \\
R129 & R599 & $\mathrm{^1 CH_2 + CH_4}$ & $\rightarrow$ & $\mathrm{CH_3 + CH_3}$ & $5.9 \times 10^{-11}$ & YD99\tablenotemark{d} &  \\
R130 & R608 & $\mathrm{C_2 + CH_4}$ & $\rightarrow$ & $\mathrm{C_2H + CH_3}$ & $5.05 \times 10^{-11} \mathrm{e}^{-297.0/T}$ & YD99\tablenotemark{d} &  \\
R131 & R644 & $\mathrm{NH_2 + OH}$ & $\rightarrow$ & $\mathrm{NH_3 + O}$ & $3.32 \times 10^{-13} \left( T/298.0 \right)^{0.4} \mathrm{e}^{-250.2/T}$ & NIST & 250-3000 \\
R132 & M1 & $\mathrm{C + C}$ & $\rightarrow$ & $\mathrm{C_2}$ & $5.46 \times 10^{-31} \left( T/298.0 \right)^{-1.6} \times M$ & NIST & 5000-6000 \\
R133 & M2 & $\mathrm{C + H_2}$ & $\rightarrow$ & $\mathrm{CH_2}$ & $6.89 \times 10^{-32} \times M$ & NIST & 300 \\
R134 & M11 & $\mathrm{H + CN}$ & $\rightarrow$ & $\mathrm{HCN}$ & $9.35 \times 10^{-30} \left( T/298.0 \right)^{-2.0} \mathrm{e}^{-521.0/T} \times M$ & NIST & 500-2500 \\
R135 & M12 & $\mathrm{H + CO}$ & $\rightarrow$ & $\mathrm{HCO}$ & $5.29 \times 10^{-34} \mathrm{e}^{-370.0/T} \times M$ & NIST & 300-2500 \\
R136 & M13 & $\mathrm{H + H}$ & $\rightarrow$ & $\mathrm{H_2}$ & $6.04 \times 10^{-33 } \left( T/298.0 \right)^{-1.0} \times M$ & NIST & 300-2500 \\
R137 & M14 & $\mathrm{H + NH_2}$ & $\rightarrow$ & $\mathrm{NH_3}$ & $3.0 \times 10^{-30} \times M$ & NIST & 298 \\
R138 & M16 & $\mathrm{H + O}$ & $\rightarrow$ & $\mathrm{OH}$ & $4.36 \times 10^{-32} \left( T/298.0 \right)^{-1.0} \times M$ & NIST & 300-2500 \\
R139 & M18 & $\mathrm{H + OH}$ & $\rightarrow$ & $\mathrm{H_2O}$ & $6.87 \times 10^{-31} \left( T/298.0 \right)^{-2.0} \times M$ & NIST & 300-3000 \\
R140 & M22 & $\mathrm{N + C}$ & $\rightarrow$ & $\mathrm{CN}$ & $9.4 \times 10^{-33} \times M$ & NIST & 298 \\
R141 & M23 & $\mathrm{N + H}$ & $\rightarrow$ & $\mathrm{NH}$ & $5.0 \times 10^{-32} \times M$ & NIST & 298 \\
R142 & M24 & $\mathrm{N + H_2}$ & $\rightarrow$ & $\mathrm{NH_2}$ & $1.0 \times 10^{-36} \times M$ & NIST & 298 \\
R143 & M25 & $\mathrm{N + N}$ & $\rightarrow$ & $\mathrm{N_2}$ & $1.38 \times 10^{-33} \mathrm{e}^{502.7/T} \times M$ & JPL & 90-6400 \\
R144 & M30 & $\mathrm{O + C}$ & $\rightarrow$ & $\mathrm{CO}$ & $2.0 \times 10^{-34} \times M$ & NIST & 8000 \\
R145 & M31 & $\mathrm{O + CO}$ & $\rightarrow$ & $\mathrm{CO_2}$ & $1.7 \times 10^{-33} \mathrm{e}^{-1509.0/T} \times M$ & NIST & 300-2500 \\
R146 & M34 & $\mathrm{O + O}$ & $\rightarrow$ & $\mathrm{O_2}$ & $5.21 \times 10^{-35} \mathrm{e}^{900.0/T} \times M$ & NIST & 200-4000 \\
R147 & M55 & $\mathrm{H + CH_2}$ & $\rightarrow$ & $\mathrm{CH_3}$ & $k_0\tablenotemark{e} = 5.8 \times 10^{-30} \mathrm{e}^{355.0/T}$ & YD99\tablenotemark{d} &  \\
& & & & & $k_{\infty}\tablenotemark{e} = 2.37 \times 10^{-12} \mathrm{e}^{523.0/T}$ & & \\
R148 & M56 & $\mathrm{H + CH_3}$ & $\rightarrow$ & $\mathrm{CH_4}$ & $6.2 \times 10^{-29} \left( T/298.0 \right)^{-1.8} \times M$ & NIST & 300-1000 \\
R149 & M72 & $\mathrm{CH + H_2}$ & $\rightarrow$ & $\mathrm{CH_3}$ & $k_0\tablenotemark{e} = 5.8 \times 10^{-30} \mathrm{e}^{355.0/T}$\tablenotemark{e} & YD99\tablenotemark{d} &  \\
& & & & & $k_{\infty}\tablenotemark{e} = 2.37 \times 10^{-12} \mathrm{e}^{523.0/T}$ & & \\
R150 & T19 & $\mathrm{CH_3OH}$ & $\rightarrow$ & $\mathrm{CH_3O + H}$ & $2.16 \times 10^{-8} \mathrm{e}^{-33556.0/T} \times M$ & NIST & 1400-2500 \\
R151 & T20 & $\mathrm{CH_3OH}$ & $\rightarrow$ & $\mathrm{CH_3 + OH}$ & $1.1 \times 10^{-7} \mathrm{e}^{-33075.0/T} \times M$ & NIST & 1000-2000 \\
R152 & T22 & $\mathrm{CH_3OH}$ & $\rightarrow$ & $\mathrm{CH_2 + H_2O}$ & $9.51 \times 10^{15} \left( T/298.0 \right)^{-1.02} \mathrm{e}^{-46185.0/T}$ & NIST & 1000-3000 \\
R153 & T46 & $\mathrm{HCO}$ & $\rightarrow$ & $\mathrm{CO + H}$ & $6.0 \times 10^{-11} \mathrm{e}^{-7721.0/T} \times M$ & NIST & 298-1229 \\
R154 & T57 & $\mathrm{HCN}$ & $\rightarrow$ & $\mathrm{H + CN}$ & $1.93 \times 10^{-4} \left( T/298.0 \right)^{-2.44} \mathrm{e}^{-62782.1/T} \times M$ & NIST & 1800-5000 \\
\enddata
\tablenotetext{a}{Reaction number of \cite{2012ApJ...761..166H}}
\tablenotetext{b}{Unit of $\mathrm{cm^3 s^{-1}}$ for 2-body reactions and $\mathrm{cm^6 s^{-2}}$ for 3-body reactions}
\tablenotetext{c}{Unit of $\mathrm{K}$}
\tablenotetext{d}{\cite{1999ppa..conf.....Y}}
\tablenotetext{e}{\kawashima{R}ate \kawashima{coefficient} $k$: $k = \left( \frac{k_0 M}{1 + \frac{k_0 M}{k_\infty}} \right) 0.6^{\left[ 1 + \left( \log_{10} \frac{k_0 M}{k_{\infty}} \right)^2 \right]^{-1}}$}
\tablecomments{Thermochemical Reactions used in our photochemical model. $M$ refers to the number density of background atmosphere (unit of $\mathrm{cm^{-3}}$). We assume $M$ equals to the total number density.}
\end{deluxetable}

\section{Photochemical Reactions}

\begin{deluxetable}{lllcll}
\tablecaption{Photochemical Reactions \label{tab:photo_reactions}}
\tablehead{
\colhead{No.} & \colhead{No. Hu\tablenotemark{a}} & \colhead{Reactants} & & \colhead{Products} & \colhead{Quantum Yields}
}
\startdata
P1 & 1 & $\mathrm{O_2}$ & $\rightarrow$ & $\mathrm{O + O}$ & $\lambda < 139$ nm: 0.5 \\
& & & & & $139$ nm $\leq \lambda < 175$ nm: 0 \\
& & & & & $\lambda \geq 175$ nm: 1.0 \\
\hline
P2 & 2 & $\mathrm{O_2}$ & $\rightarrow$ & $\mathrm{O + O(^1D)}$ & $\lambda < 139$ nm: 0.5 \\
& & & & & $139$ nm $\leq \lambda < 175$ nm: 1.0 \\
& & & & & $\lambda \geq 175$ nm: 0 \\
\hline
P3 & 6 & $\mathrm{H_2O}$ & $\rightarrow$ & $\mathrm{H + OH}$ & $1.0$ \\
\hline
P4 & 32 & $\mathrm{CH_4}$ & $\rightarrow$ & $\mathrm{CH_3 + H}$ & 0.41 \citep{1999JGR...104.1873S} \\
\hline
P5 & 33 & $\mathrm{CH_4}$ & $\rightarrow$ & $\mathrm{CH_2^1 + H_2}$ & 0.53 \citep{1999JGR...104.1873S} \\
\hline
P6 & 34 & $\mathrm{CH_4}$ & $\rightarrow$ & $\mathrm{CH + H_2 + H}$ & 0.06 \citep{1999JGR...104.1873S} \\
\hline
P7 & 35 & $\mathrm{CO}$ & $\rightarrow$ & $\mathrm{C + O}$ & $\lambda < 111$ nm: 1.0 \\
& & & & & $\lambda \geq111$ nm: 0 \\
\hline
P8 & 36 & $\mathrm{CO_2}$ & $\rightarrow$ & $\mathrm{CO + O}$ & $\lambda < 167$ nm: 0 \\
& & & & & $167$ nm $\leq \lambda < 205$ nm: 1.0 \\
& & & & & $\lambda \geq 205$ nm: 0 \\
\hline
P9 & 37 & $\mathrm{CO_2}$ & $\rightarrow$ & $\mathrm{CO + O(^1D)}$ & $\lambda < 167$ nm: 1.0 \\
& & & & & $\lambda \geq 167$ nm: 0 \\
\hline
P10 & 38 & $\mathrm{H_2}$ & $\rightarrow$ & $\mathrm{H + H}$ & $\lambda < 80$ nm: 0.1 \citep{1970JChPh..52.5641M} \\
& & & & & $80$ nm $\leq \lambda < 85$ nm: 1.0 \\
& & & & & $\lambda \geq 85$ nm: 0 \\
\hline
P11 & 39 & $\mathrm{N_2}$ & $\rightarrow$ & $\mathrm{N + N}$ & 1.0 \\
\hline
P12 & 40 & $\mathrm{CH_3OH}$ & $\rightarrow$ & $\mathrm{CH_3O + H}$ & 1.0 \\
\hline
P13 & 41 & $\mathrm{HCN}$ & $\rightarrow$ & $\mathrm{H + CN}$ & 1.0 \\
\hline
P14 & 42 & $\mathrm{NH_3}$ & $\rightarrow$ & $\mathrm{NH_2 + H}$ & $\lambda < 106$ nm: 0.3 \citep{Lilly:1974ji} \\
& & & & & $106$ nm $\leq \lambda < 165$ nm: Linear interpolation \\
& & & & & $\lambda \geq 165$ nm: 1.0 \\
\hline
P15 & 43 & $\mathrm{NH_3}$ & $\rightarrow$ & $\mathrm{NH + H_2}$ & $\lambda < 106$ nm: 0.7 \\
& & & & & $106$ nm $\leq \lambda < 165$ nm: Linear interpolation \\
& & & & & $\lambda \geq 165$ nm: 1.0 \\
\hline
P16 & 55 & $\mathrm{C_2H_2}$ & $\rightarrow$ & $\mathrm{C_2H + H}$ & $\lambda < 217$ nm: 1.0 \citep{Lauter:2002ka} \\
& & & & & $\lambda \geq 217$ nm: 0 \\
\enddata
\tablenotetext{a}{Reaction number of \cite{2012ApJ...761..166H}}
\tablenotetext{b}{We use the polynomial expansion written in page 4D-3 of \citet{Sander2011}.}
\tablecomments{Photochemical Reactions used in our photochemical model and values of quantum yields.}
\end{deluxetable}

\section{UV Cross Sections}

\begin{deluxetable}{llll}
\tablecaption{UV Cross Sections \label{tab:photo_cross-sections}}
\tablehead{
\colhead{Species} & \colhead{Wavelength} & \colhead{Cross Sections} & \colhead{T\tablenotemark{a}}
}
\startdata
$\mathrm{O_2}$ & $4.13$ nm $\leq \lambda \leq 103.00$ nm & \citet{BRION:1979jy} & N \\
& $108.75$ nm $\leq \lambda \leq 129.60$ nm & \citet{1975CaJPh..53.1845O} & N \\
& $129.62$ nm $\leq \lambda \leq 172.53$ nm & \citet{2005JMoSp.229..238Y} & N \\
& $176.8$ nm $\leq \lambda \leq 202.6$ nm & \citet{Kockarts:1976uz} & N \\
& $205$ nm $\leq \lambda \leq 245$ nm & \citet{Sander2011} & N \\
\hline
$\mathrm{H_2O}$ & $6.20$ nm $\leq \lambda \leq 118.08$ nm & \citet{Chan:1993kj} & N \\
& $121.00$ nm $\leq \lambda \leq 198.00$ nm & \citet{Sander2011} & N \\
& $198.00$ nm $\leq \lambda \leq 240$ nm & Extrapolation & \\
\hline
$\mathrm{CH_4}$ & $52.054$ nm $\leq \lambda \leq 124.629$ nm & \citet{Kameta:2002bf} & N \\
& $125$ nm $\leq \lambda \leq 141$ nm & \citet{Chen:2004ia} & N \\
& $142$ nm $\leq \lambda \leq 152$ nm & \citet{Lee:2001eo} & N \\
\hline
$\mathrm{CO}$ & $6.199$ nm $\leq \lambda \leq 177$ nm & \citet{Chan:1993ij} & N \\
\hline
$\mathrm{CO_2}$ & $35.0000$ nm $\leq \lambda \leq 197.6950$ nm & \citet{2010DPS....42.4813H} & N \\
& $197.70$ nm $\leq \lambda \leq 270.15$ nm & \citet{2008CPL...462...31I} & N \\
\hline
$\mathrm{H_2}$ & $18$ nm $\leq \lambda \leq 70$ nm & \citet{1976JQSRT..16..873L} & N \\
& $77.00$ nm $\leq \lambda \leq 86.88$ nm & \citet{Cook:1964ty} & N \\
& $88.6$ nm $\leq \lambda \leq 124$ nm & \citet{Backx:1976hx} & N \\
\hline
$\mathrm{N_2}$ & $6.199$ nm $\leq \lambda \leq 113$ nm &\citet{Chan:1993ht} & N \\
\hline
$\mathrm{CH_3OH}$ & $15.5$ nm $\leq \lambda \leq 103$ nm & \citet{Burton:1992ho} & N \\
& $106.50$ nm $\leq \lambda \leq 165.00$ nm & \citet{Nee:1985jq} & N \\
& $165.5$ nm $\leq \lambda \leq 219.5$ nm & \citet{Cheng:2002cn} & N \\
\hline
$\mathrm{HCN}$ & $133.42$ nm $\leq \lambda \leq 144.75$ nm & \citet{Macpherson:1978jn} & N \\
\hline
$\mathrm{NH_3}$ & $8.0$ nm $\leq \lambda \leq 105.0$ nm & \citet{1987JChPh..87.6416S}& N \\
& $106.00$ nm $\leq \lambda \leq 139.98$ nm & \citet{2007JChPh.127o4311W} & N \\
& $140.00$ nm $\leq \lambda \leq 230.00$ nm & \citet{Cheng:2006ws} & N \\
\hline
$\mathrm{C_2H_2}$ & $6.20$ nm $\leq \lambda \leq 131$ nm & \citet{Cooper:1995kr} & N \\
& $136.90378$ nm $\leq \lambda \leq 185.62863$ nm & \citet{Smith:1991jy} & Y \\
& $185.63$ nm $\leq \lambda \leq 236.290$ nm & \cite{Benilan:2000fu} & Y \\
\enddata
\tablenotetext{a}{Temperature dependence: Y and N indicate whether temperature dependence is taken into account or not.}
\tablenotetext{b}{We use the expression written in page 4D-2 of \citet{Sander2011}.}
\end{deluxetable}

\bibliography{Untitled_final,ykawashima-RefList,plus,ads}


\listofchanges

\end{document}